\documentclass[a4paper,12pt]{article}
\usepackage[utf8]{inputenc}
\usepackage{bm,amssymb,cite,color,verbatim,graphicx,multicol}
\usepackage{amsmath}
\usepackage{psfrag}
\usepackage{tikz}
\usepackage[breaklinks = true]{hyperref}
\usepackage{caption, subcaption}
\allowdisplaybreaks

\newcommand{\ket}[1]{\left| #1 \right\rangle}
\newcommand{\bra}[1]{\left\langle #1 \right|}
\newcommand{\braket}[2]{\left\langle #1 | #2 \right\rangle}

\newlength{\mylength}
\def\ii{{\rm i}}

\def\xRmxe#1{\vcenter{\hbox{\includegraphics[width=#1em]{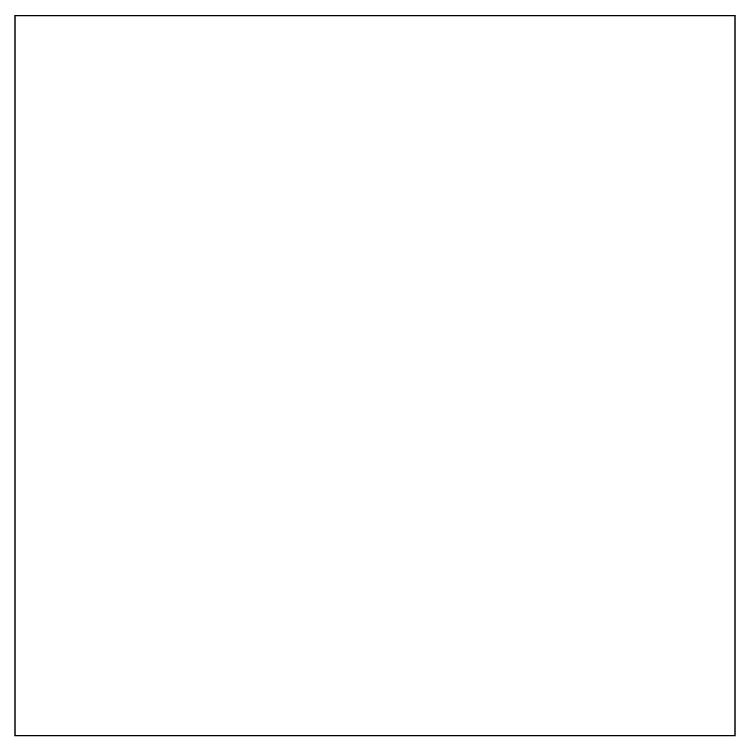}}}}
\def\Rmxe{{
  \mathchoice
    {\xRmxe2}%
    {\xRmxe2}%
    {\xRmxe\defaultscriptratio}%
    {\xRmxe\defaultscriptscriptratio}}}
    
\def\xRmxtl#1{\vcenter{\hbox{\includegraphics[width=#1em]{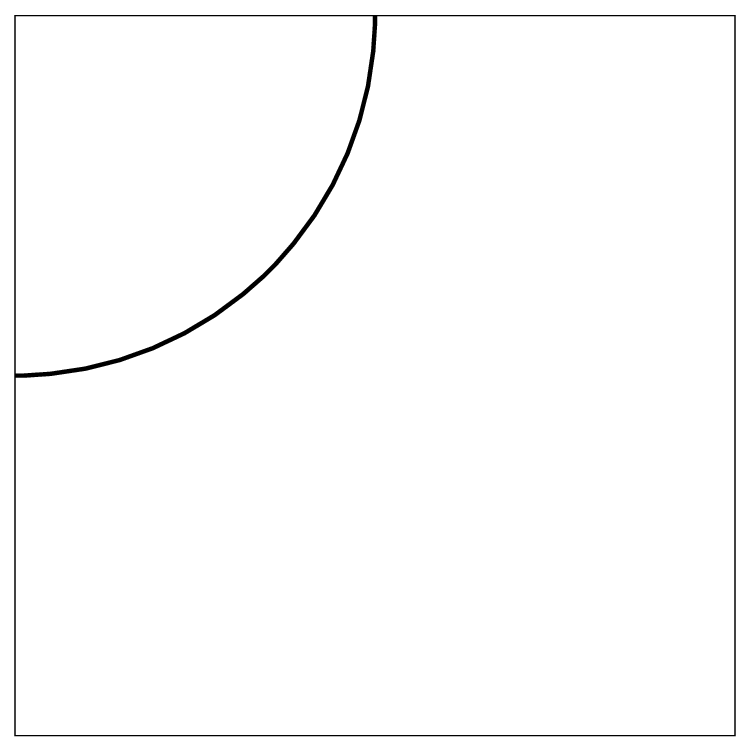}}}}
\def\Rmxtl{{
  \mathchoice
    {\xRmxtl2}%
    {\xRmxtl2}%
    {\xRmxtl\defaultscriptratio}%
    {\xRmxtl\defaultscriptscriptratio}}}
    
\def\xRmxtr#1{\vcenter{\hbox{\includegraphics[width=#1em]{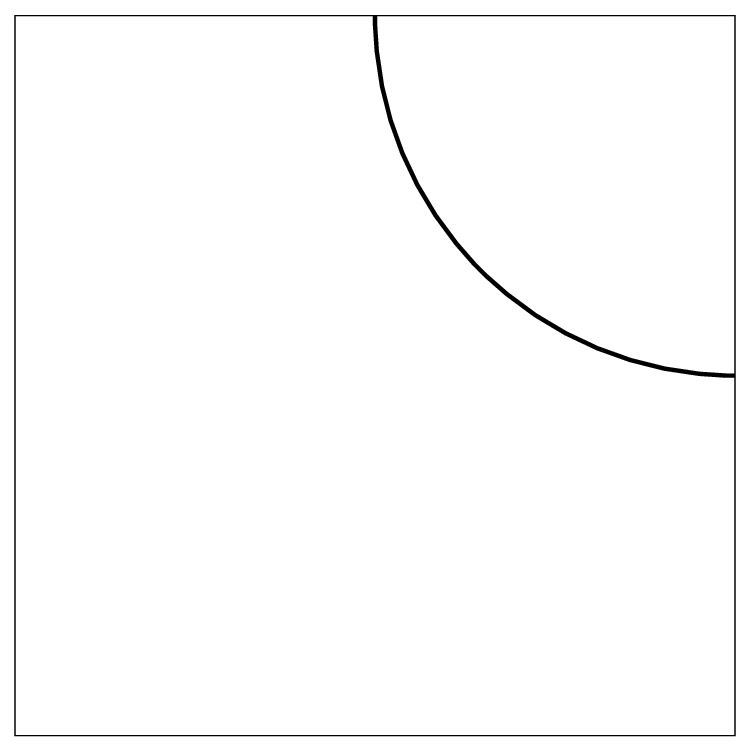}}}}
\def\Rmxtr{{
  \mathchoice
    {\xRmxtr2}%
    {\xRmxtr2}%
    {\xRmxtr\defaultscriptratio}%
    {\xRmxtr\defaultscriptscriptratio}}}
    
\def\xRmxbl#1{\vcenter{\hbox{\includegraphics[width=#1em]{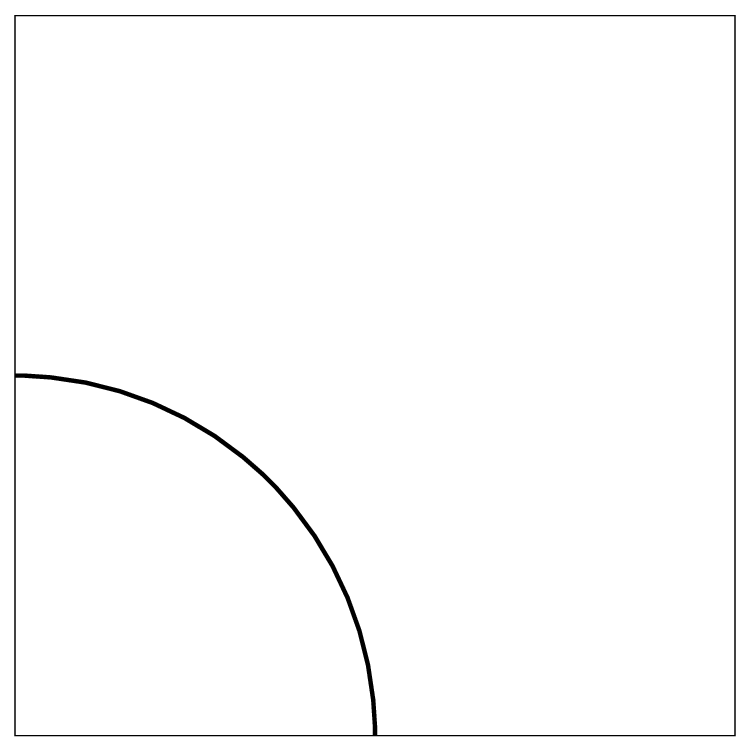}}}}
\def\Rmxbl{{
  \mathchoice
    {\xRmxbl2}%
    {\xRmxbl2}%
    {\xRmxbl\defaultscriptratio}%
    {\xRmxbl\defaultscriptscriptratio}}}

\def\xRmxbr#1{\vcenter{\hbox{\includegraphics[width=#1em]{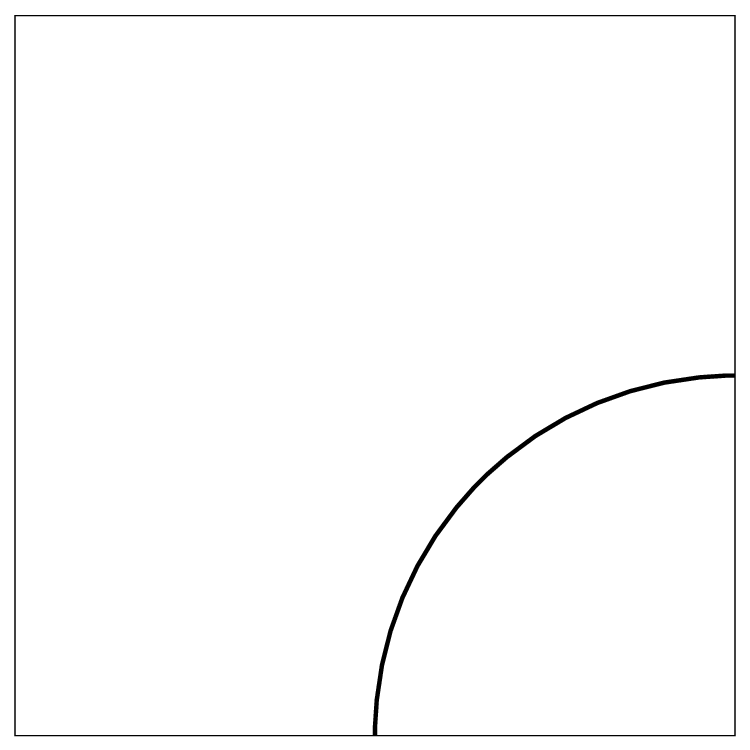}}}}
\def\Rmxbr{{
  \mathchoice
    {\xRmxbr2}%
    {\xRmxbr2}%
    {\xRmxbr\defaultscriptratio}%
    {\xRmxbr\defaultscriptscriptratio}}}
    
\def\xRmxtb#1{\vcenter{\hbox{\includegraphics[width=#1em]{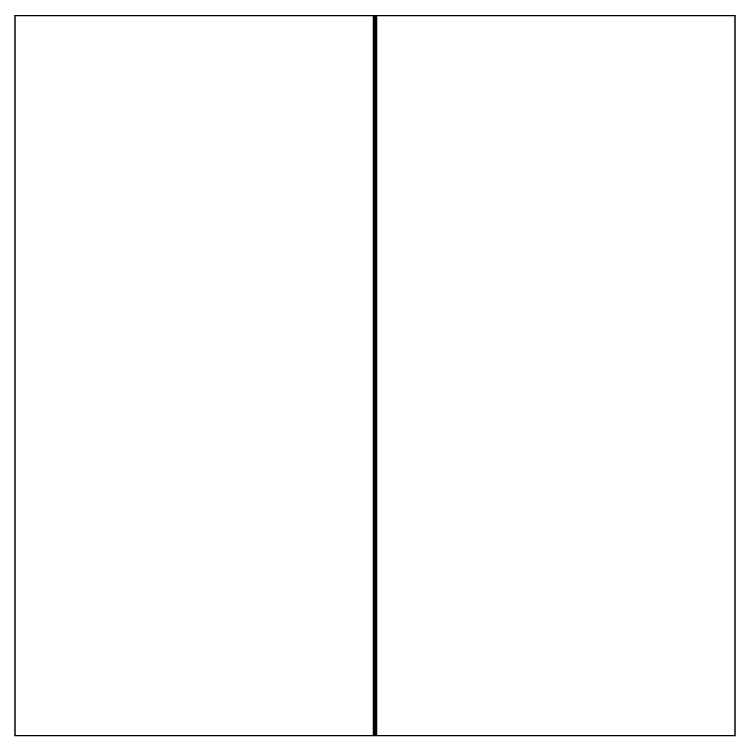}}}}
\def\Rmxtb{{
  \mathchoice
    {\xRmxtb2}%
    {\xRmxtb2}%
    {\xRmxtb\defaultscriptratio}%
    {\xRmxtb\defaultscriptscriptratio}}}
    
\def\xRmxlr#1{\vcenter{\hbox{\includegraphics[width=#1em]{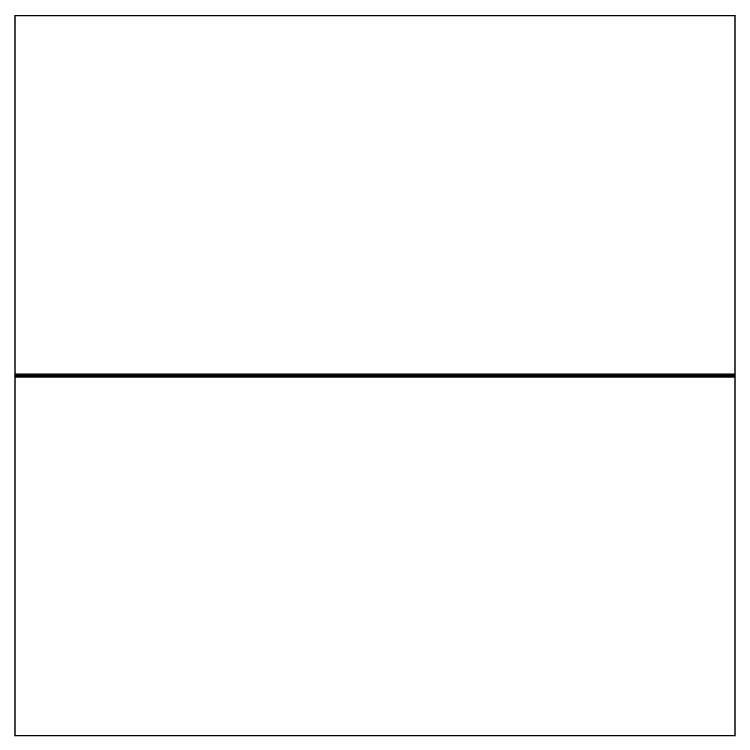}}}}
\def\Rmxlr{{
  \mathchoice
    {\xRmxlr2}%
    {\xRmxlr2}%
    {\xRmxlr\defaultscriptratio}%
    {\xRmxlr\defaultscriptscriptratio}}}
    
\def\xRmxtlbr#1{\vcenter{\hbox{\includegraphics[width=#1em]{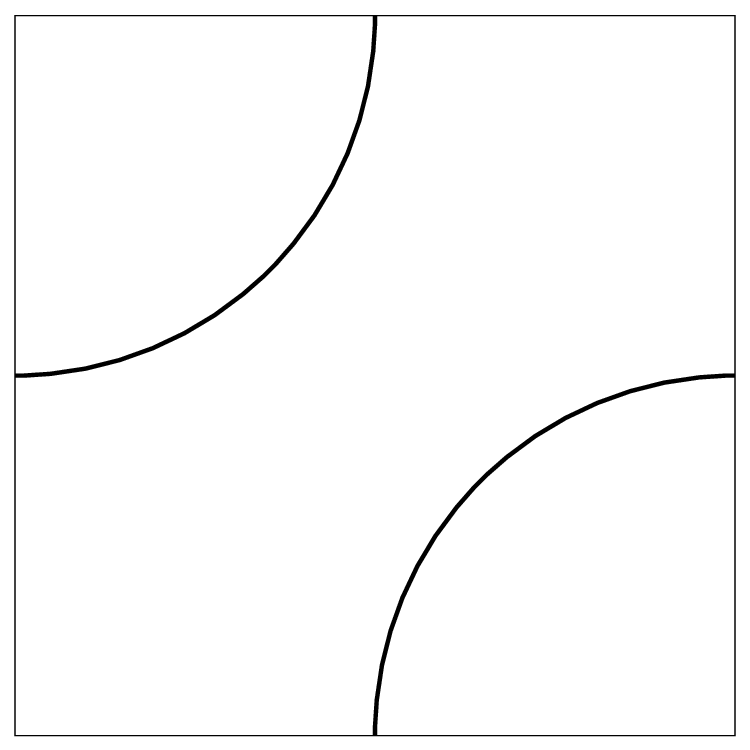}}}}
\def\Rmxtlbr{{
  \mathchoice
    {\xRmxtlbr2}%
    {\xRmxtlbr2}%
    {\xRmxtlbr\defaultscriptratio}%
    {\xRmxtlbr\defaultscriptscriptratio}}}
    
\def\xRmxtrbl#1{\vcenter{\hbox{\includegraphics[width=#1em]{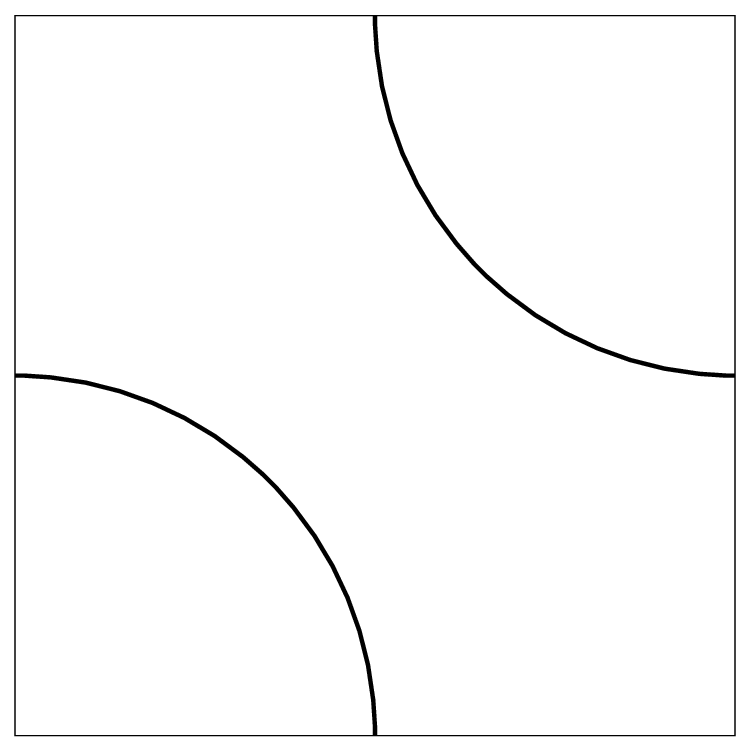}}}}
\def\Rmxtrbl{{
  \mathchoice
    {\xRmxtrbl2}%
    {\xRmxtrbl2}%
    {\xRmxtrbl\defaultscriptratio}%
    {\xRmxtrbl\defaultscriptscriptratio}}}

\def\xRmxrapii#1{\vcenter{\hbox{\psfrag{z}{$z$}\psfrag{w}{$w$}\includegraphics[width=#1em]{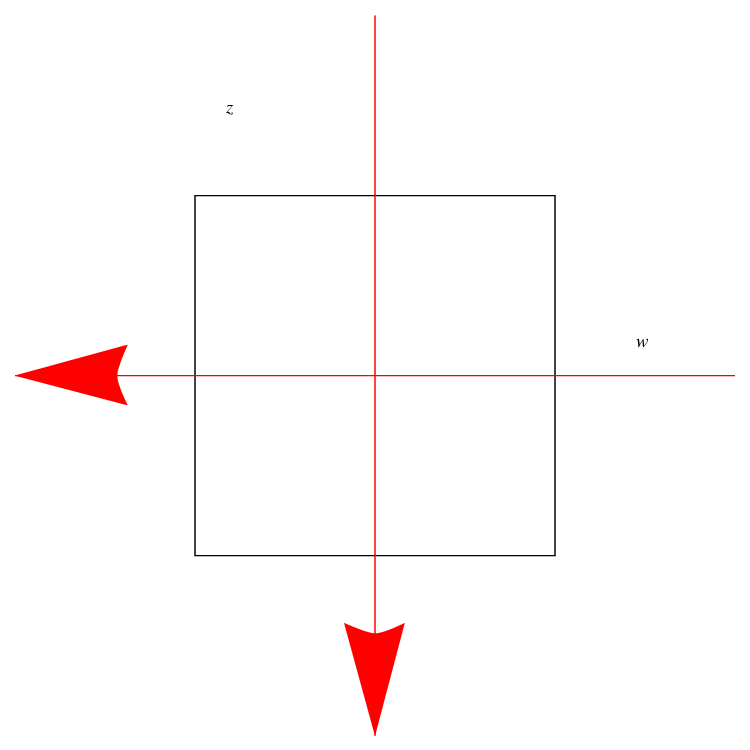}}}}
\def\Rmxrapii{{
  \mathchoice
    {\xRmxrapii4}%
    {\xRmxrapii4}%
    {\xRmxrapii\defaultscriptratio}%
    {\xRmxrapii\defaultscriptscriptratio}}}

\def\xKlmxrapii#1{\vcenter{\hbox{\psfrag{a}{$z$}\psfrag{b}{$z^{-1}$}\psfrag{c}{$z_B$}\includegraphics[height=#1em]{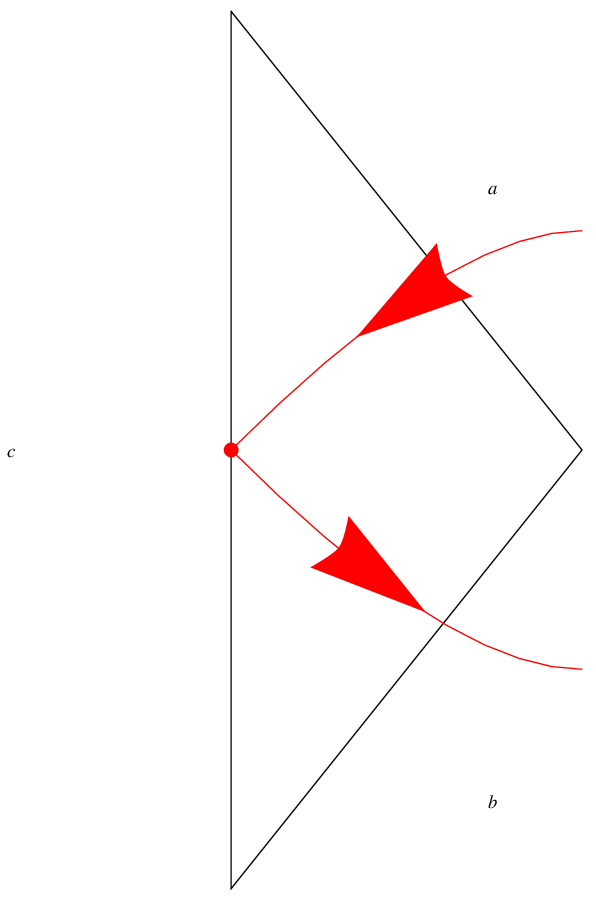}}}}
\def\Klmxrapii{{
  \mathchoice
    {\xKlmxrapii5}%
    {\xKlmxrapii5}%
    {\xKlmxrapii\defaultscriptratio}%
    {\xKlmxrapii\defaultscriptscriptratio}}}    
    
\def\xKlmxe#1{\vcenter{\hbox{\includegraphics[width=#1em]{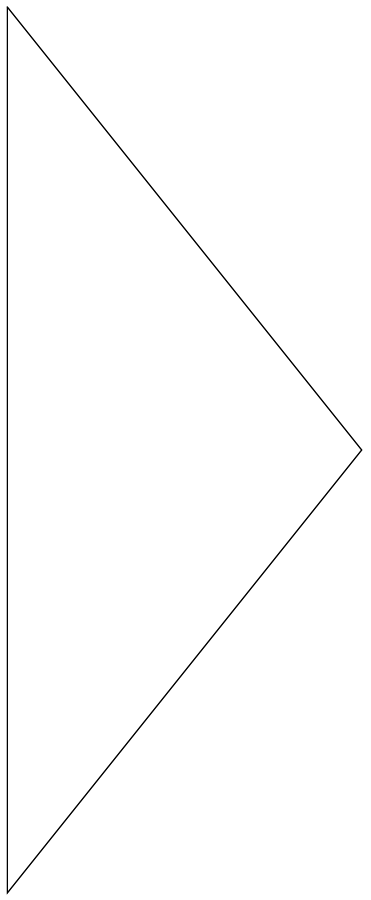}}}}
\def\Klmxe{{
  \mathchoice
    {\xKlmxe2}%
    {\xKlmxe2}%
    {\xKlmxe\defaultscriptratio}%
    {\xKlmxe\defaultscriptscriptratio}}}
    
\def\xKlmxl#1{\vcenter{\hbox{\includegraphics[width=#1em]{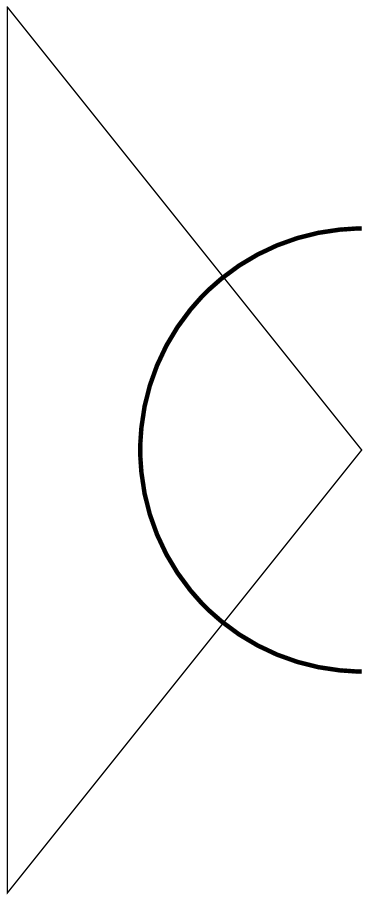}}}}
\def\Klmxl{{
  \mathchoice
    {\xKlmxl2}%
    {\xKlmxl1}%
    {\xKlmxl\defaultscriptratio}%
    {\xKlmxl\defaultscriptscriptratio}}}
    
\def\xKlmxt#1{\vcenter{\hbox{\includegraphics[width=#1em]{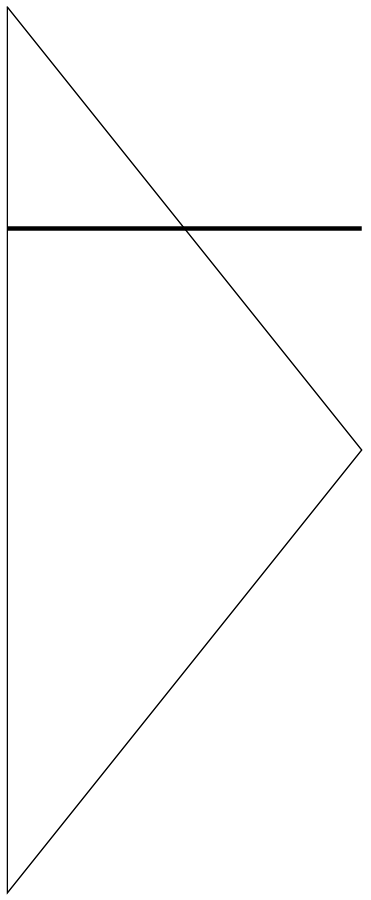}}}}
\def\Klmxt{{
  \mathchoice
    {\xKlmxt2}%
    {\xKlmxt2}%
    {\xKlmxt\defaultscriptratio}%
    {\xKlmxt\defaultscriptscriptratio}}}
    
\def\xKlmxb#1{\vcenter{\hbox{\includegraphics[width=#1em]{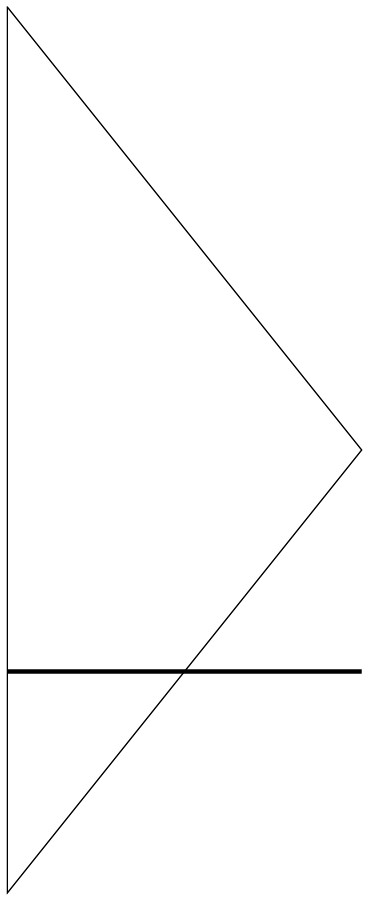}}}}
\def\Klmxb{{
  \mathchoice
    {\xKlmxb2}%
    {\xKlmxb2}%
    {\xKlmxb\defaultscriptratio}%
    {\xKlmxb\defaultscriptscriptratio}}}
    
\def\xKlmxm#1{\vcenter{\hbox{\includegraphics[width=#1em]{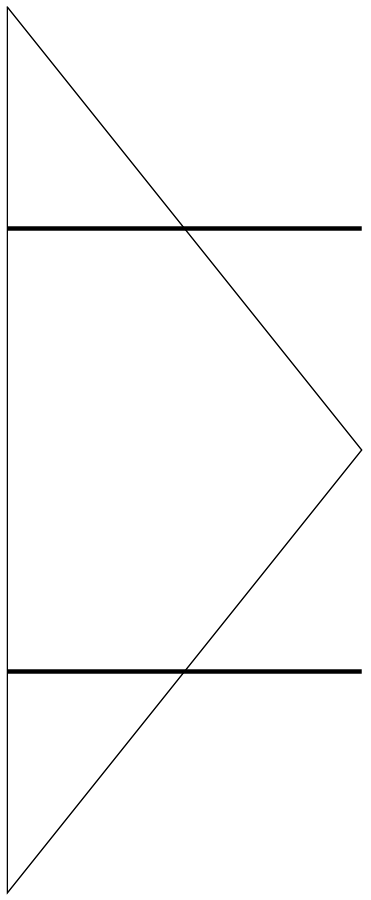}}}}
\def\Klmxm{{
  \mathchoice
    {\xKlmxm2}%
    {\xKlmxm1}%
    {\xKlmxm\defaultscriptratio}%
    {\xKlmxm\defaultscriptscriptratio}}}

\def\xKrmxrapii#1{\vcenter{\hbox{\psfrag{a}{$z$}\psfrag{b}{$z^{-1}$}\psfrag{c}{$z_B$}\includegraphics[height=#1em]{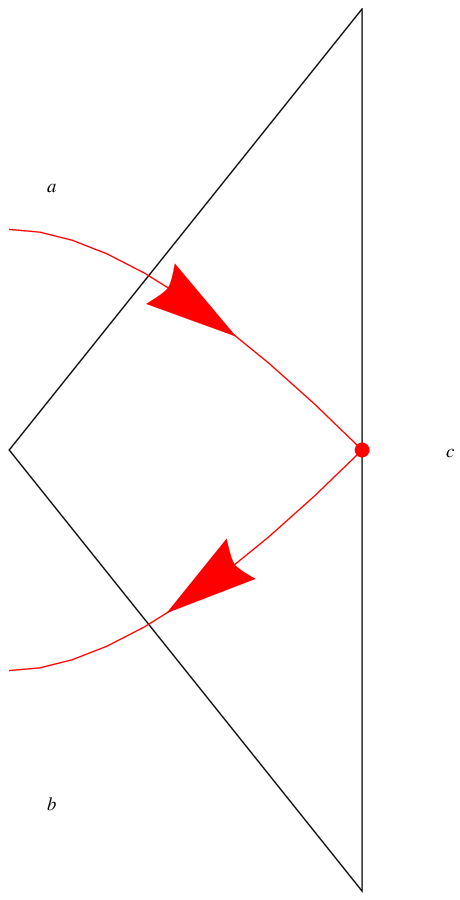}}}}
\def\Krmxrapii{{
  \mathchoice
    {\xKrmxrapii5}%
    {\xKrmxrapii5}%
    {\xKrmxrapii\defaultscriptratio}%
    {\xKrmxrapii\defaultscriptscriptratio}}} 
    
\def\xKrmxe#1{\vcenter{\hbox{\includegraphics[width=#1em]{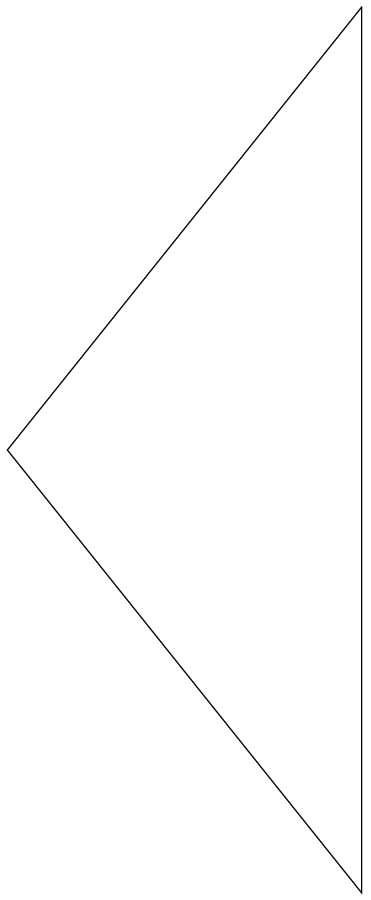}}}}
\def\Krmxe{{
  \mathchoice
    {\xKrmxe2}%
    {\xKrmxe2}%
    {\xKrmxe\defaultscriptratio}%
    {\xKrmxe\defaultscriptscriptratio}}}
                
\def\xKrmxl#1{\vcenter{\hbox{\includegraphics[width=#1em]{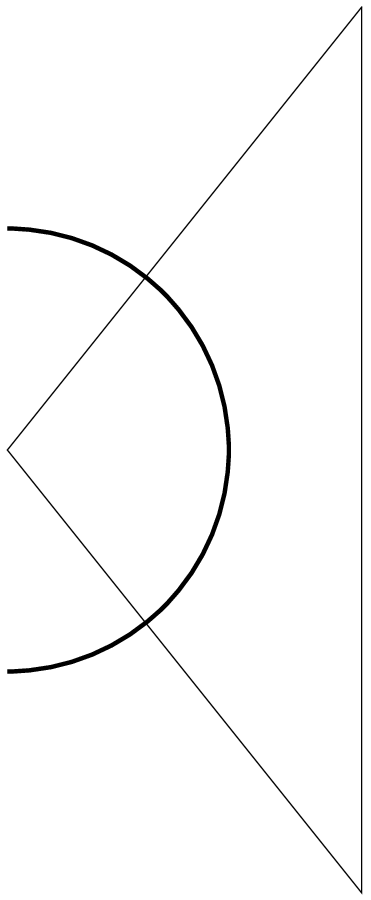}}}}
\def\Krmxl{{
  \mathchoice
    {\xKrmxl2}%
    {\xKrmxl2}%
    {\xKrmxl\defaultscriptratio}%
    {\xKrmxl\defaultscriptscriptratio}}}
    
\def\xKrmxt#1{\vcenter{\hbox{\includegraphics[width=#1em]{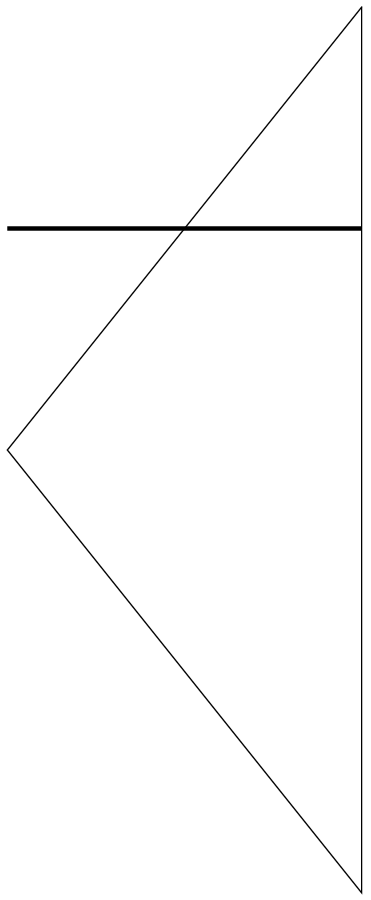}}}}
\def\Krmxt{{
  \mathchoice
    {\xKrmxt2}%
    {\xKrmxt2}%
    {\xKrmxt\defaultscriptratio}%
    {\xKrmxt\defaultscriptscriptratio}}}
        
\def\xKrmxb#1{\vcenter{\hbox{\includegraphics[width=#1em]{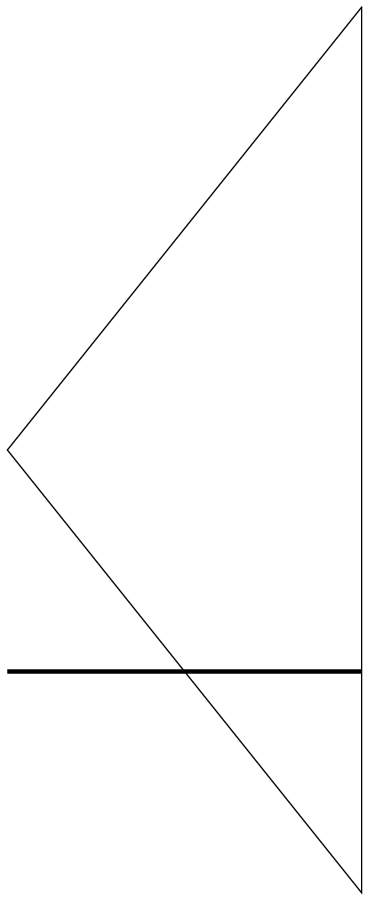}}}}
\def\Krmxb{{
  \mathchoice
    {\xKrmxb2}%
    {\xKrmxb2}%
    {\xKrmxb\defaultscriptratio}%
    {\xKrmxb\defaultscriptscriptratio}}}

\def\xKrmxm#1{\vcenter{\hbox{\includegraphics[width=#1em]{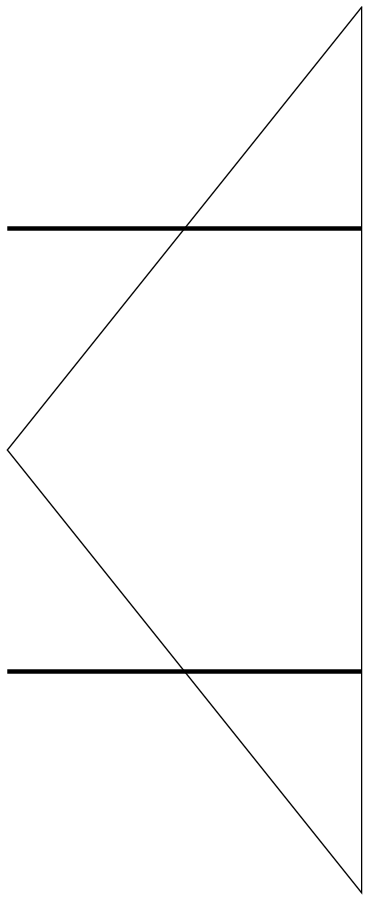}}}}
\def\Krmxm{{
  \mathchoice
    {\xKrmxm2}%
    {\xKrmxm2}%
    {\xKrmxm\defaultscriptratio}%
    {\xKrmxm\defaultscriptscriptratio}}}
    
\def\xSmxe#1{\vcenter{\hbox{\includegraphics[width=#1em]{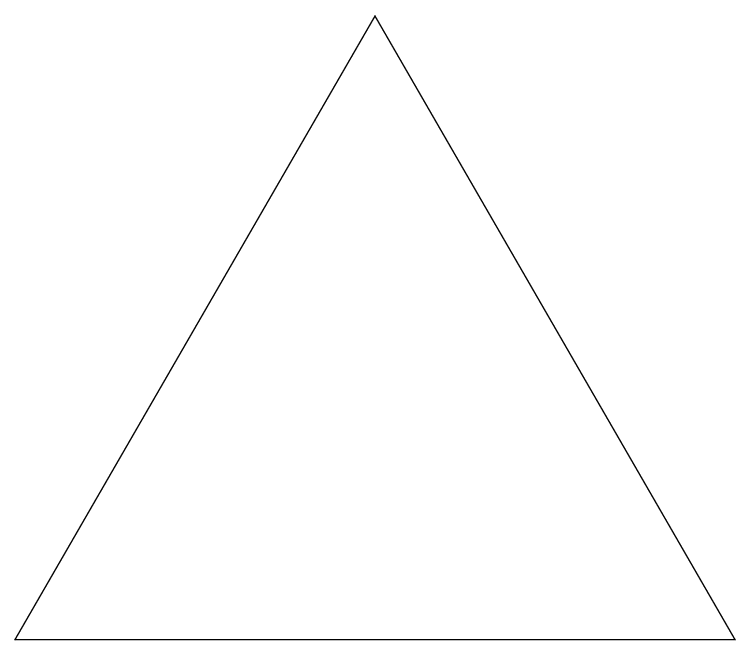}}}}
\def\Smxe{{
  \mathchoice
    {\xSmxe{1.732}}%
    {\xSmxe{1.732}}%
    {\xSmxe\defaultscriptratio}%
    {\xSmxe\defaultscriptscriptratio}}}
    
\def\xSmxlr#1{\vcenter{\hbox{\includegraphics[width=#1em]{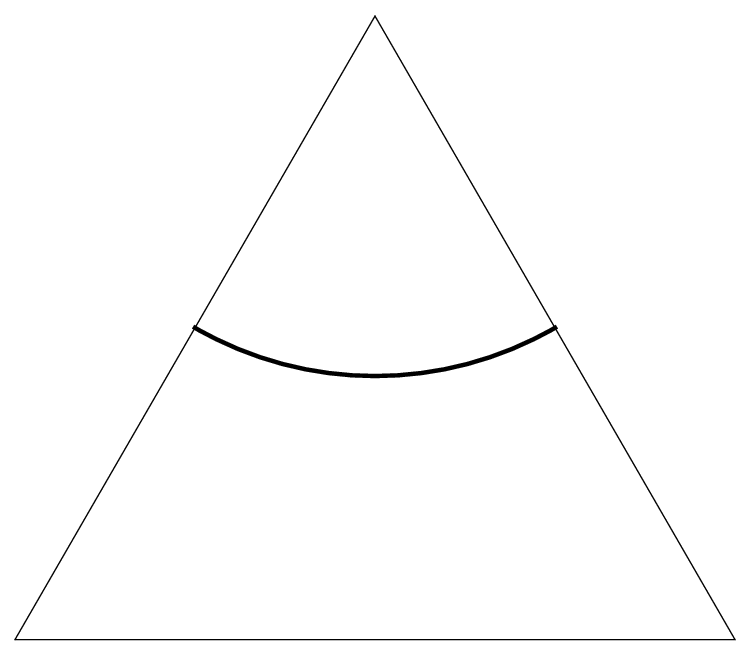}}}}
\def\Smxlr{{
  \mathchoice
    {\xSmxlr{1.732}}%
    {\xSmxlr{1.732}}%
    {\xSmxlr\defaultscriptratio}%
    {\xSmxlr\defaultscriptscriptratio}}}
    
\def\xSmxlb#1{\vcenter{\hbox{\includegraphics[width=#1em]{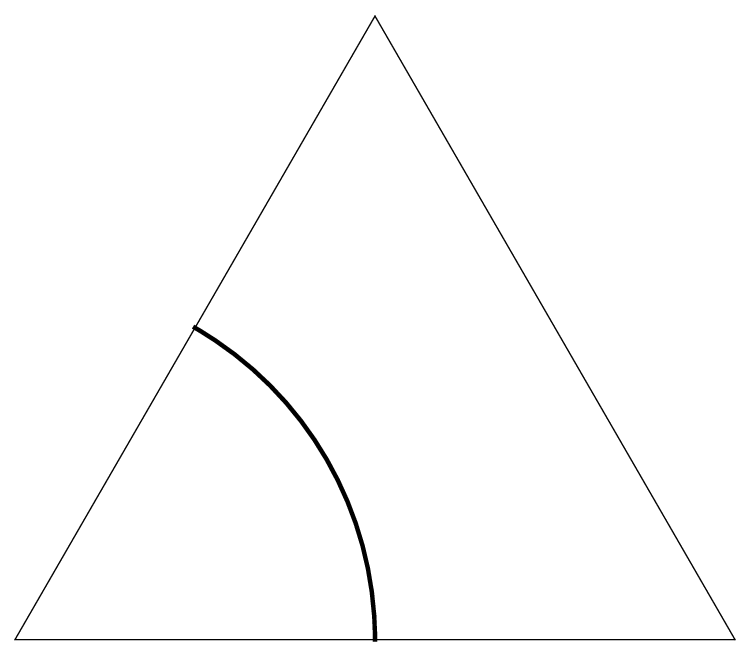}}}}
\def\Smxlb{{
  \mathchoice
    {\xSmxlb{1.732}}%
    {\xSmxlb{1.732}}%
    {\xSmxlb\defaultscriptratio}%
    {\xSmxlb\defaultscriptscriptratio}}}
    
\def\xSmxrb#1{\vcenter{\hbox{\includegraphics[width=#1em]{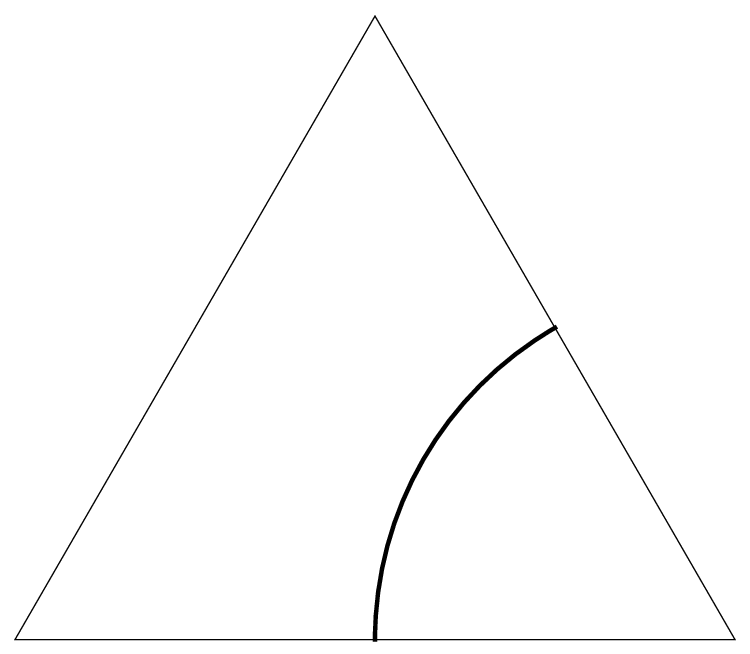}}}}
\def\Smxrb{{
  \mathchoice
    {\xSmxrb{1.732}}%
    {\xSmxrb{1.732}}%
    {\xSmxrb\defaultscriptratio}%
    {\xSmxrb\defaultscriptscriptratio}}}
    
\def\xMmxe#1{\vcenter{\hbox{\includegraphics[width=#1em]{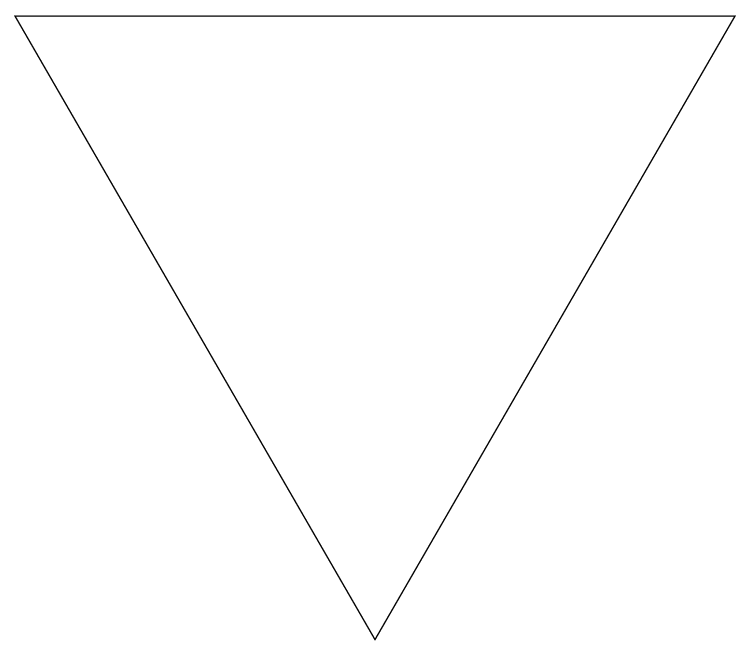}}}}
\def\Mmxe{{
  \mathchoice
    {\xMmxe{1.732}}%
    {\xMmxe{1.732}}%
    {\xMmxe\defaultscriptratio}%
    {\xMmxe\defaultscriptscriptratio}}}
    
    \def\xMmxlr#1{\vcenter{\hbox{\includegraphics[width=#1em]{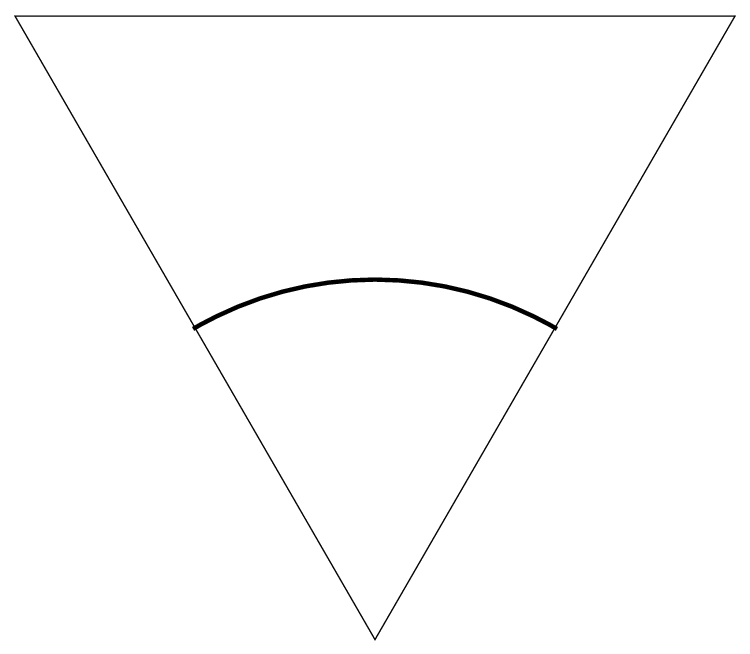}}}}
\def\Mmxlr{{
  \mathchoice
    {\xMmxlr{1.732}}%
    {\xMmxlr{1.732}}%
    {\xMmxlr\defaultscriptratio}%
    {\xMmxlr\defaultscriptscriptratio}}}
    
    \def\xMmxtr#1{\vcenter{\hbox{\includegraphics[width=#1em]{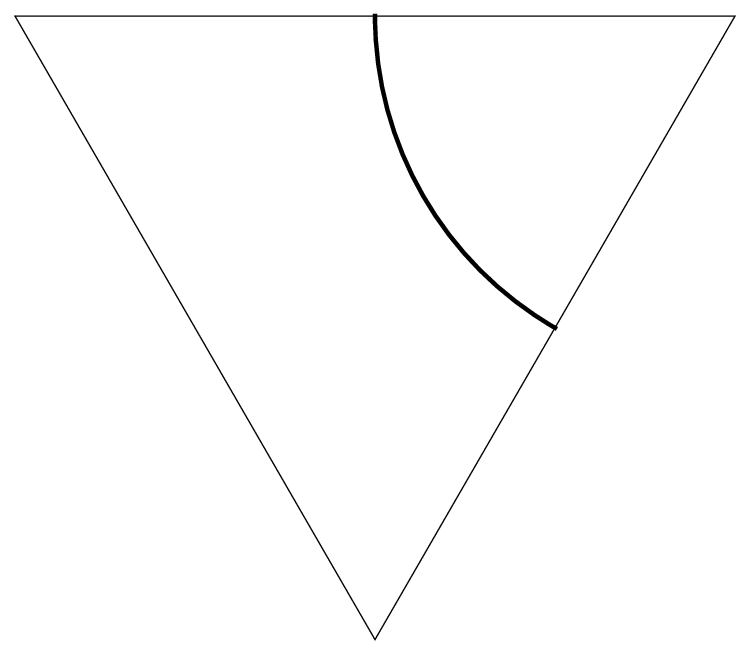}}}}
\def\Mmxtr{{
  \mathchoice
    {\xMmxtr{1.732}}%
    {\xMmxtr{1.732}}%
    {\xMmxtr\defaultscriptratio}%
    {\xMmxtr\defaultscriptscriptratio}}}
    
    \def\xMmxtl#1{\vcenter{\hbox{\includegraphics[width=#1em]{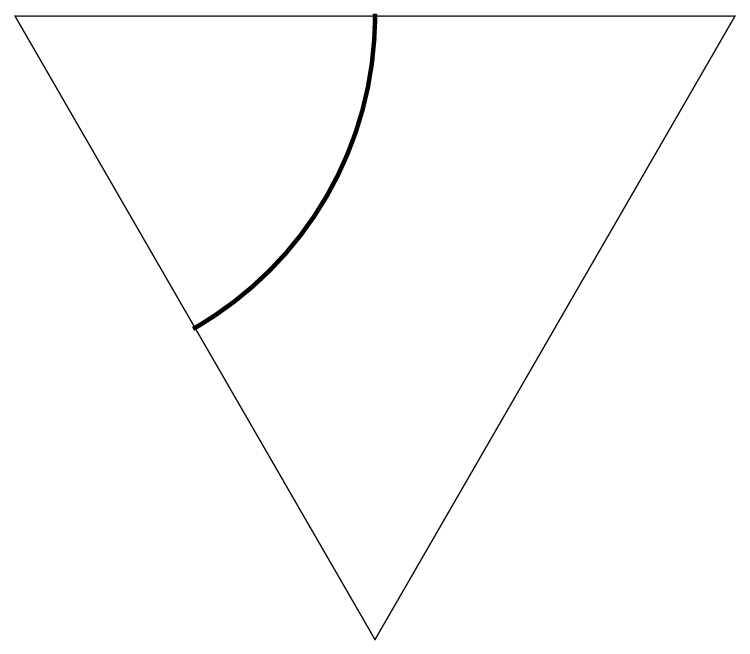}}}}
\def\Mmxtl{{
  \mathchoice
    {\xMmxtl{1.732}}%
    {\xMmxtl{1.732}}%
    {\xMmxtl\defaultscriptratio}%
    {\xMmxtl\defaultscriptscriptratio}}}

    \def\xUmxe#1{\vcenter{\hbox{\includegraphics[width=#1em]{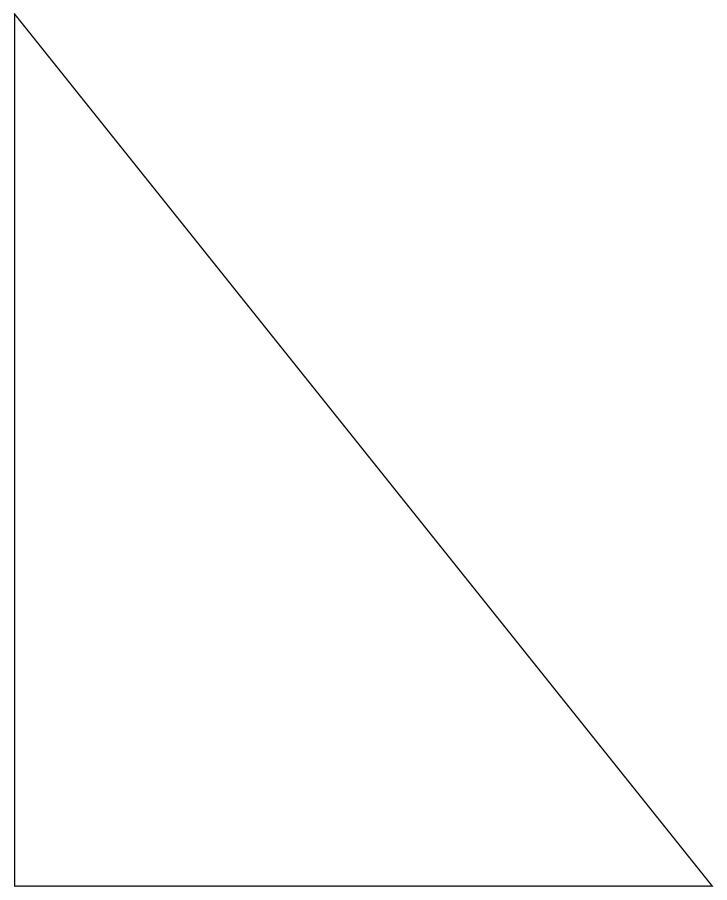}}}}
\def\Umxe{{
  \mathchoice
    {\xUmxe2}%
    {\xUmxe2}%
    {\xUmxe\defaultscriptratio}%
    {\xUmxe\defaultscriptscriptratio}}}
    
    \def\xUmxl#1{\vcenter{\hbox{\includegraphics[width=#1em]{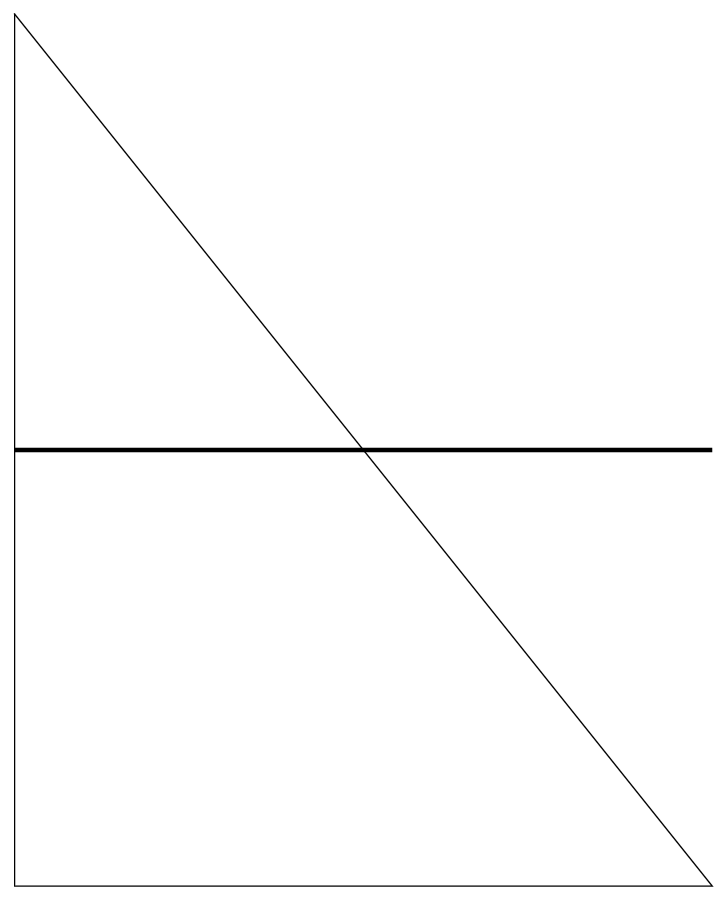}}}}
\def\Umxl{{
  \mathchoice
    {\xUmxl2}%
    {\xUmxl2}%
    {\xUmxl\defaultscriptratio}%
    {\xUmxl\defaultscriptscriptratio}}}

    \def\xLmxe#1{\vcenter{\hbox{\includegraphics[width=#1em]{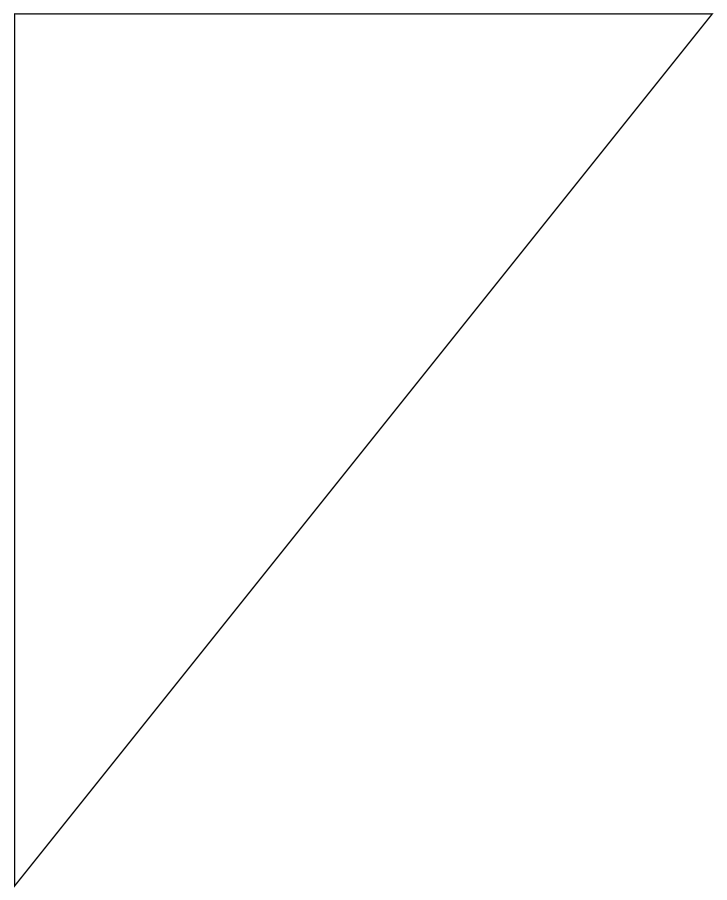}}}}
\def\Lmxe{{
  \mathchoice
    {\xLmxe2}%
    {\xLmxe2}%
    {\xLmxe\defaultscriptratio}%
    {\xLmxe\defaultscriptscriptratio}}}
    
    \def\xLmxl#1{\vcenter{\hbox{\includegraphics[width=#1em]{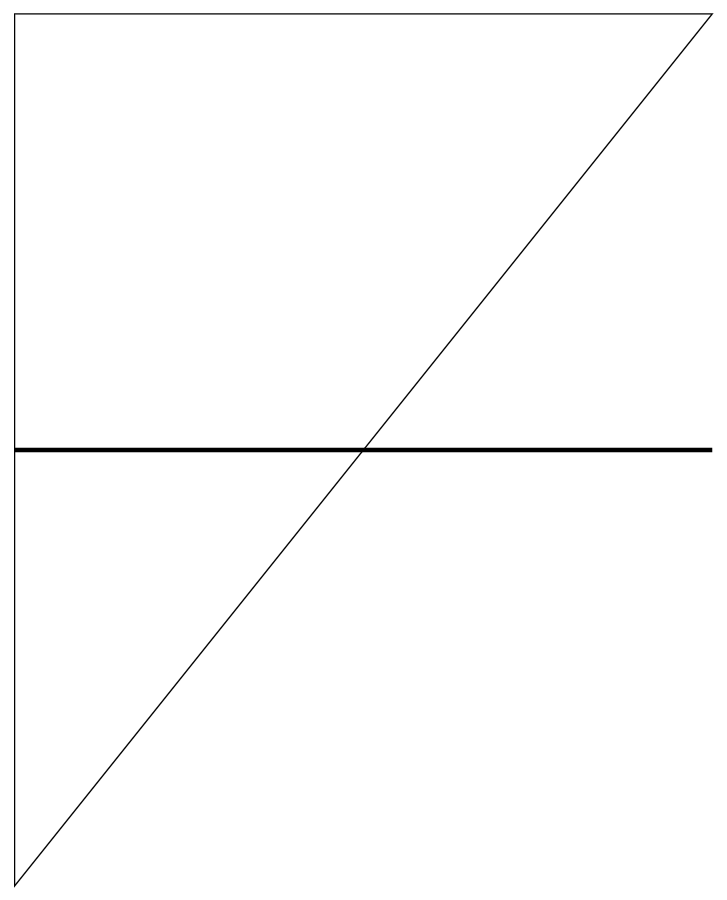}}}}
\def\Lmxl{{
  \mathchoice
    {\xLmxl2}%
    {\xLmxl2}%
    {\xLmxl\defaultscriptratio}%
    {\xLmxl\defaultscriptscriptratio}}}

\title{Currents in the dilute $O(n=1)$ loop model}
\author{G.Z. Feh\'er$^{1,2}$, B. Nienhuis$^{1}$ \bigskip \\
        \parbox{0.6\textwidth}{\center
        \small $^1$\textit{Instituut voor Theoretische Fysica, \\
        University of Amsterdam, \\
        P.O.Box 94485, 1090 GL Amsterdam, The Netherlands}
\medskip\\
$^2$\textit{Department of Theoretical Physics,\\ Budapest University
of Technology and Economics,\\ 1111 Budapest, Budafoki \'ut 8, Hungary
        }}
        \bigskip}
\small\date{\small\today}

\begin{document}

\maketitle

\begin{abstract}
In the framework of an inhomogeneous solvable lattice model, we derive exact expressions for a 
boundary-to-boundary current on a lattice of finite width. 
The model we use is the dilute O($n$=1) 
loop model, related to the Izergin-Korepin spin~$\!-1$ chain and the critical site percolation on the triangular lattice. Our expressions are derived based on solutions of the $q$-Knizhnik-Zamolodchikov equations, and recursion 
relations. 
\end{abstract}

\let\thefootnote\relax\footnotetext{Email: \texttt{g.feher@eik.bme.hu\, B.Nienhuis@uva.nl }}

\tableofcontents
\setlength{\parindent}{0pt}
\setlength{\parskip}{1ex}
 
\section{Introduction}

In the last decades, a growing interest surrounds integrable loop models, which is fueled by connections to the distinct 
fields of combinatorics, correlation functions and discrete holomorphicity.\\
Loop models were originally introduced as high temperature expansions
for spin models, or $n$-vector models. From the symmetry group in these models, they are now known as $O(n)$ loop models \cite{nienhuis1982exact}. 
Later an integrable version of the model was identified \cite{nienhuis1990critical1, nienhuis1990critical2}. In \cite{di2005around} a set of
finite difference equation, the --loosely called -- quantum Knizhnik-Zamolodchikov equations were introduced in order
to find the partition sum and the ground state elements of the dense $O(1)$ model with periodic boundary conditions. This approach results in inhomogeneous, polynomial, finite size expressions.  
The method was used to extend the calculations to the ground state of dense $O(1)$ loop model 
on a strip with various boundary conditions \cite{di2005inhomogeneous, cantini2009qkz, de2009exact}. 
Later, this approach was used to compute quantities not directly related to the ground state of the transfer-matrix, e.g. 
left passage probability in percolations \cite{ikhlef2012finite}, current \cite{de2010exact},
correlation functions \cite{cantini2012finite} (on finite temperature \cite{aufgebauer2012finite}). Observables have been computed in the dense O($n$=1) model on a cylinder, for the homogeneous case \cite{mitra2004exact1,mitra2004exact2}. There was an 
attempt to prove the Razumov-Stroganov conjecture \cite{razumov2001spin1, razumov2001spin2, razumov2004combinatorial} 
(see also \cite{batchelor2001quantum}) by this method \cite{di2005around, di2006sum}, which was finally proven 
by purely combinatorial method \cite{cantini2011proof}. In~\cite{gorin2015asymptotics}, homogenous, continuum limits of some of these finite size size expressions were computed. In \cite{kasatani2007polynomials}, the $O(n)$ models are related to the ground state of the quantum Hall-effect. \\
The motivation for this work is coming from several direction. The integrability of finite size lattice models
provides a good ground to compute complicated quantities (e.g. correlations) exactly. \\
In this paper, we deal with a discretely holomorphic, parafermionic observable, 
the spin-1 boundary to boundary current (for further connection between integrability and discrete holomorphicity, see e.g. 
\cite{batchelor2013surprising, de2013discrete}). Discretely
holomorphic observables on the lattice necessarily turn into holomorphic observables of the CFT, in the scaling limit. \\
\begin{figure}

\centering

\begin{subfigure}{.5\textwidth}
  \centering
     \begin{tikzpicture}[scale=5]
     \draw[fill=lightgray,lightgray] (0,0) rectangle (1,1);
     \draw[fill=gray, gray] (-.05,0) rectangle (0,1);
     \draw[ultra thick] (0,0)--(0,1); 
     \draw[fill=gray, gray] (1,0) rectangle (1.05,1);
     \draw[ultra thick] (1,0)--(1,1); 

     \draw[thick] (0, .75)--(.125, .85)--(.25, .82)--(.37, .63)--(.24, .49)--(.36,.376)--(.4, .387)--(.49, .51)--(.54, .497)--(.8, .67)--(.81, .58)--(.92, .583)--(.87, .33)--(1, .375);

     \draw[fill] (.625,.625) circle [radius=0.01];
     \draw[fill] (.625,.5) circle [radius=0.01];

     \node[above right] at (.625,.625) {$x_1$};
     \node[right] at (.625,.5) {$x_2$};

     \end{tikzpicture}
  \caption{Path with $+1$ contribution}
  \label{fig:PathExamplePlus}
\end{subfigure}%
\begin{subfigure}{.5\textwidth}
  \centering
     \begin{tikzpicture}[scale=5]
     \draw[fill=lightgray,lightgray] (0,0) rectangle (1,1);
     \draw[fill=gray, gray] (-.05,0) rectangle (0,1);
     \draw[ultra thick] (0,0)--(0,1); 
     \draw[fill=gray, gray] (1,0) rectangle (1.05,1);
     \draw[ultra thick] (1,0)--(1,1); 

     \draw[thick] (0, .375)--(.1, .5)--(.22, .41)--(.24, .55)--(.375, .6)--(.382,
.54)--(.45, .73)--(.55, .75)--(.61, .72)--(.755, .715)--(.756, .555)--(.5, .54)--(.38, .47)--(.62, .37)--(.74, .362)--(.81, .26)--(.9,.17)--(.88, .09)--(1, .125);

     \draw[fill] (.625,.625) circle [radius=0.01];
     \draw[fill] (.625,.5) circle [radius=0.01];

     \node[right] at (.625,.625) {$x_1$};
     \node[below right] at (.625,.5) {$x_2$};

     \end{tikzpicture}
  \caption{Path with $-1$ contribution}
  \label{fig:PathExampleMinus}
\end{subfigure}
\caption{The spin-1 property of the current: different paths contribute with different signs.}
\label{fig:PathExamples}
\end{figure}
%
%
%
%
%
%
We apply these ideas to the dilute loop model, with unit loop weight, also called the O($n$=1) loop model. 
We will study the statistical ensemble of non-intersecting paths on a strip of finite width 
and infinite height, in the framework of dilute loop model which we define in Section \ref{sec:modelDef}.
The paths may form closed loops, 
or terminate on the boundary. We assume, the paths connecting the two boundaries carry equal 
unit of current from the left boundary to the right boundary. The closed loops, and paths connected 
only to one of the boundaries do not carry any current. In this paper, we are computing the mean current density 
induced by the statistics of the paths. Introduce the observable $F^{\left( x_1, x_2 \right)}$, 
which is the mean current between points $x_1$ and $x_2$:
\begin{equation}
\label{eq:Fdef}
 F^{( x_1, x_2 )}= \sum_{C \in \Gamma} P(C) N_C^{\left( x_1, x_2 \right)} \text{sign}_C^{( x_1, x_2 )} \;.
\end{equation}
Here $\Gamma$ is the set of all configurations, $N_C^{\left( x_1, x_2 \right)}$ is the number of paths passing in between
points  $x_1$ and $x_2$, and running form the left to the right boundary, $P(C)$ is the ensemble probability of configuration $C$ 
and $\text{sign}_C^{\left( x_1, x_2 \right)}$ is $+1$ if $x_1$ lies in the region above the paths, and $-1$ if it lies below (Fig.~\ref{fig:PathExamples}). 
The observable $F$ is antisymmetric, and additive:
\begin{align}
 F^{\left( x_1, x_2 \right)} &=-F^{\left( x_2, x_1 \right)} \\
 F^{\left( x_1, x_3\right)}  &= F^{\left( x_1, x_2 \right)} + F^{\left( x_2, x_3 \right)}
\end{align}
Up to a phase factor, $F$ is the $s=1$ special case of the more general, arbitrary $s$ spin case:
\begin{equation}
 \tilde{F}^{s,( x_1, x_2 )}= \sum_{C \in \Gamma} P(C) N_C^{\left( x_1, x_2 \right)} e^{\ii s \phi(C)}\;,
\end{equation}
where $\phi(C)$ is the winding angle from the initial direction until the crossing of the path with the $x_1$, $x_2$ line.

The dilute $O(n)$ model is related to the site percolation on a triangular lattice (Appendix~\ref{app:Perco}) and the Izergin-Korepin type 19-vertex model \cite{izergin1981inverse, tarasov1988algebraic, garbali2014domain}. 
\\
This work is a direct continuation of \cite{garbali2017dilute1, garbali2017dilute2}, where the ground state elements and partition sum has been computed.
Our result is also related to \cite{de2010exact}, where the same current was computed for the dense $O(1)$ loop model.
\\
The structure of the paper is as follows. In Section~\ref{sec:modelDef}, we define the model, in Section~\ref{sec:current}, 
we present our main result, in Section~\ref{sec:RS1},~\ref{sec:Sym}, we build up the necessary tools for our statements, 
and in Section~\ref{sec:Proof}, --under a technical assumption-- we
prove the main result. Some specific calculations are deferred to appendices.

\section{The square lattice dilute loop model on a strip}
\label{sec:modelDef}

Consider a square lattice of width $L$ and infinite height. 
Each square of the lattice is decorated randomly by one of the nine plaquettes:
\par
\vspace{10pt}
  \begin{minipage}{\linewidth}
            \centering
	    \psfrag{a}{$b_1$}
	    \psfrag{b}{$b_2$}
	    \psfrag{c}{$b_3$}
	    \psfrag{d}{$b_4$}
	    \psfrag{e}{$b_5$}
	    \psfrag{f}{$b_6$}
	    \psfrag{g}{$b_7$}
	    \psfrag{h}{$b_8$}
	    \psfrag{i}{$b_9$}
	    \includegraphics[scale=0.35]{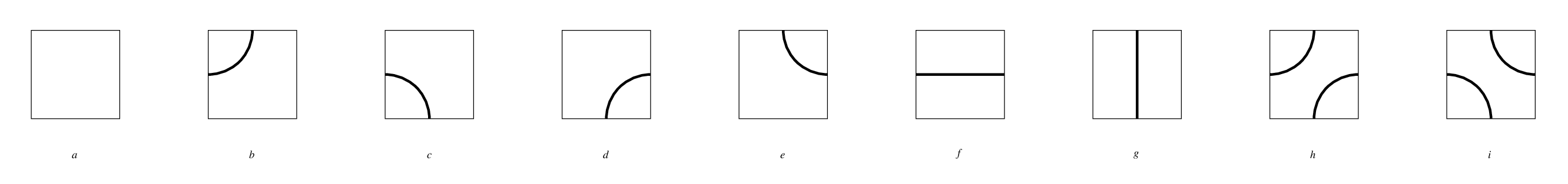}
        \end{minipage}
       \vspace{20pt}

The decoration is subject to the restriction, 
that each path built up by the decoration of the plaquettes has to be continuous, 
so it either ends on the boundaries, or forms a closed loop. Because certain configurations are not allowed, we can not associate 
\emph{independent probabilities} with the plaquettes, we can only associate \emph{statistical weights} with them.
To describe the interaction with the left and right boundary, we introduce different kind of plaquettes, respectively: 
\par
\vspace{10pt}
  \begin{minipage}{\linewidth}
            \centering
	    \psfrag{a}{$l_1$}
	    \psfrag{b}{$l_2$}
	    \psfrag{c}{$l_3$}
	    \psfrag{d}{$l_4$}
	    \psfrag{e}{$l_5$}
	    \psfrag{f}{$r_1$}
	    \psfrag{g}{$r_2$}
	    \psfrag{h}{$r_3$}
	    \psfrag{i}{$r_4$}
	    \psfrag{j}{$r_5$}
	    \includegraphics[scale=0.5]{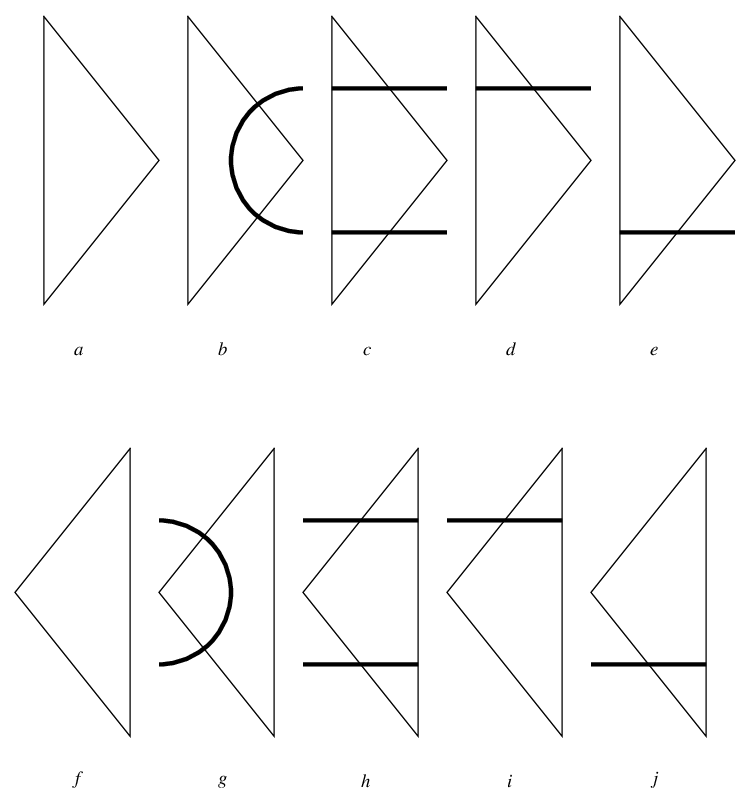}
        \end{minipage}
       \vspace{20pt}
\begin{figure}
\begin{center}

 \includegraphics[scale=0.5]{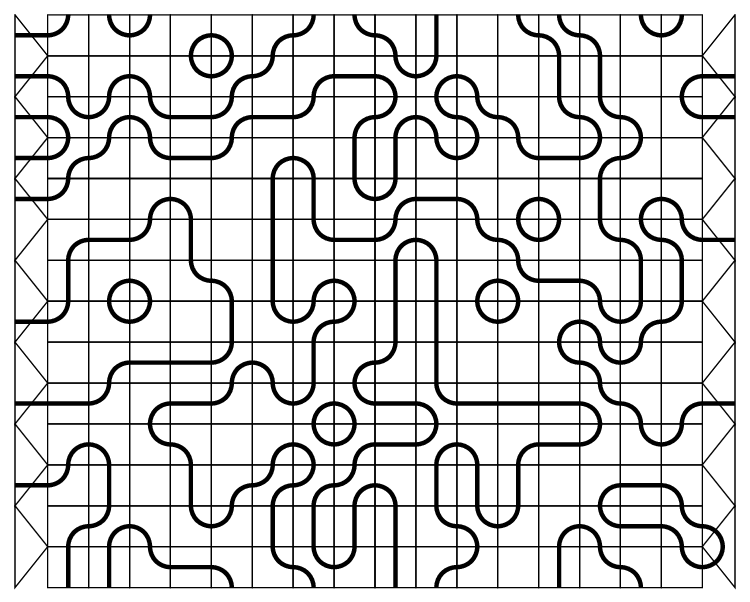}
  \caption{A typical configuration at $L=16$}
  \label{fig:TypicalConfig}
\end{center}
\end{figure}

\begin{figure}
\begin{center}

 \includegraphics[scale=0.5]{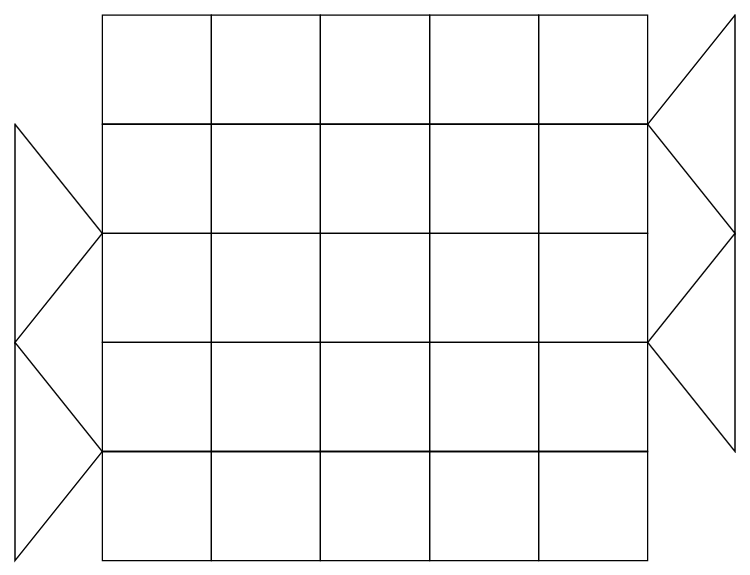}
  \caption{Zigzag boundary conditions}
  \label{fig:ZigZagBC}
\end{center}
\end{figure}

A typical configuration is shown in Fig.~\ref{fig:TypicalConfig}.
To assign probabilities with arbitrary finite or infinite configuration, we use the following definition:
Each closed loop in the bulk carries the weight $n$, and each path attached to the boundaries also carries the weight $n$.
Also each plaquette carries a weight, given by the label in the pictures
above.
The statistical weight of a configuration $C$ is given by the product of the weights of the constituent plaquettes and the weight 
of the loops:
\begin{multline}
\label{eq:ProbDef}
 P(C)=\left( \prod_{i=1}^9 b_i^{\text{\# of $b_i$ plaquettes}} \right) \left( \prod_{i=1}^5 l_i^{\text{\# of $l_i$ plaquettes}} \right) \\
      \left( \prod_{i=1}^5 r_i^{\text{\# of $r_i$ plaquettes}} \right) n^{\text{\# of loops}} \;.
\end{multline}
We will deal only with the case $n=1$, i.e. we can ignore the number of loops. \\
By this, we have defined the \emph{homogeneous dilute O(n=1) loop model} \cite{nienhuis1990critical1, nienhuis1990critical2}, with open boundary conditions. \\
There are some other subtleties, to define the model differently, we can distinguish the 
lines connecting the two boundaries from the ones connected to only one \cite{de2013discrete},
we can introduce 'zigzag' boundary conditions, as in Fig.~\ref{fig:ZigZagBC}, we can introduce other boundary conditions by
restricting the set of boundary plaquettes, we can introduce mixed
boundary conditions by distinguishing the allowed plaquettes on the two
boundaries, we can assign different weights to loops in bulk and paths
connected to the boundaries.  In this paper we always choose weight
$n=1$ for the loops as well as for the paths terminating on the
boundary, aside from the boundary weight that can be attributed locally.

\subsection{Baxterization, inhomogeneous weights}

\begin{figure}
\begin{center}
 \psfrag{a}{$z_1$} \psfrag{b}{$z_2$} 
 \psfrag{e}{$z_{L-1}$} \psfrag{F}{$z_L$}
 
 \psfrag{f}{$w_1$} \psfrag{g}{$w_1^{-1}$} \psfrag{h}{$w_2$}
  \psfrag{i}{$w_2^{-1}$} \psfrag{j}{$w_3$} \psfrag{k}{$w_3^{-1}$}
  \psfrag{A}{$z_0$}
  \psfrag{B}{$z_{L+1}$}
 \includegraphics[scale=0.7]{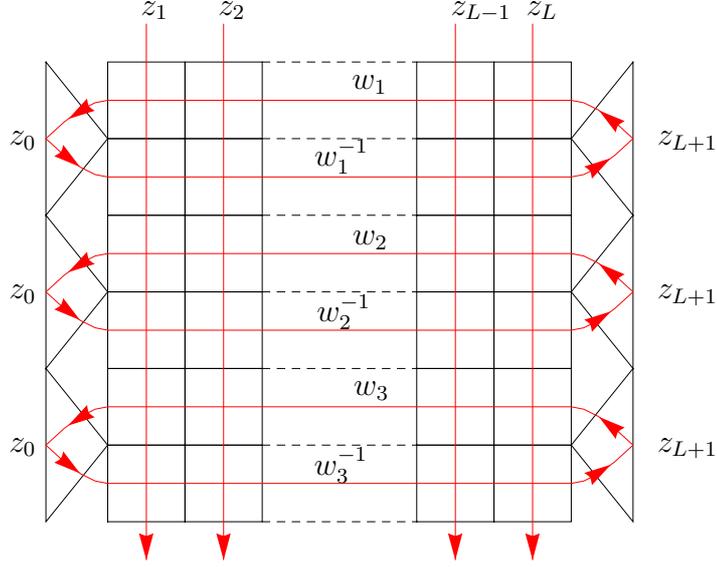}
  \caption{Rapidities: $z_1, \ldots, z_L$ and $w_i$ are the
    rapidities. The $z_0$ and $z_{L+1}$ are further parameters of the
    left and right $K$-matrices. As we see later, for many occasions we
    can treat them the same as the rapidities in the bulk; hence we will call them boundary rapidities.}
  \label{fig:Rapidities}
\end{center}
\end{figure}

In this section we define the \emph{inhomogeneous} version of our model. 
To make our model accessible for the toolbox of integrability, we introduce inhomogeneous weights both for the bulk and boundary 
plaquettes which satisfy the Yang-Baxter \cite{baxter1982exactly} and the reflection equations \cite{sklyanin1988boundary}. 
To do so, we introduce \emph{rapidities} -- sometimes called \emph{spectral parameters}-- flowing through the sites, as in Fig.~\ref{fig:Rapidities}. 
We make the statistical weight of a plaquette to be the function of the two rapidities crossing it, which process is sometimes called Baxterization. 
To make it explicit, we introduce the \emph{$R$-matrix}:
%
\begin{multline}
 R(z,w)=\Rmxrapii = W_{1}(z,w) \left(\, \Rmxe + \Rmxtb + \Rmxlr \, \right) + \\
 + W_{t}(z,w) \left( \, \Rmxtl + \Rmxtr + \Rmxbl + \Rmxbr \, \right) + W_{2} (z,w)\, \Rmxtlbr +W_{m} (z,w)\, \Rmxtrbl \;,
\end{multline}
with the following weights:
\def\omega{q}
\begin{subequations}
\begin{align}
 W_1 (z,w) &= -1+\left( \frac{w}{z} \right)^2 \; ,\\
 W_t (z,w) &= \left(\omega+\omega^2 \right) \frac{w}{z} \; ,\\
 W_2 (z,w) &=  \omega^2+\omega  \left( \frac{w}{z} \right)^2 \; ,\\
 W_m (z,w) &= -\omega-\omega^2  \left( \frac{w}{z} \right)^2 \;.
\end{align}
\end{subequations}
%
%
Here, and in the rest of the paper, $\omega$ satisfies $\omega - \omega^2 = 1$, which sets its value to $\omega=e^{\pm \ii \pi/3}$. 
The $R$-matrix can be regarded in two different ways: as the statistical weight of the plaquette, or as an operator. 
We need to introduce some more objects to understand the $R$-matrix as an operator, so, we postpone this for later.
The $R$-matrix is normalized by the factor
\begin{equation}
 W_R (z, w)=-1 - \frac{w}{z} + 2 \omega \frac{w}{z} + \frac{w^2}{z^2} \; ,
\end{equation}
resulting in a stochastic matrix,  so that the elements of $R(z,w)/W_R(z,w)$ are probabilities. The $R$-matrix in fact depends only on the ratio $\frac{w}{z}$. \\
In the same way, we can introduce the left and right $K$-matrices, which describes the interaction with the boundaries:
\begin{align}
  K_l (z,z_B) & = \Klmxrapii  =  K_{id}^l (z,z_B) \left( \, \Klmxe+ \Klmxl \, \right) +\\
	\nonumber
  & K_m^l (z,z_B)  \left(  \, \Klmxe + \Klmxm  \, \right) + K_1^l (z,z_B) \left( \, \Klmxb+ \Klmxt \, \right) \; , \\
  K_r (z,z_B) & = \Krmxrapii  \quad =  K_{id}^r (z,z_B) \left( \, \Krmxe+ \Krmxl  \, \right) +\\
	\nonumber
  & K_m^r (z,z_B)  \left( \, \Krmxe + \Krmxm \, \right) + K_1^r (z,z_B) \left( \,  \Krmxb+ \Krmxt \, \right) \;,
\end{align}
with the following weights:
\begin{subequations}
\begin{align}
 K_{id}^l (z,z_B) & = k^2 \left( z^{-1} \right) x^2 \left( z_B \right) -1 \; , \\
 K_m^l (z,z_B) &= x^2 (z_B) k \left( z^{-1} \right) \left( k(z)-k\left( z^{-1} \right) \right) \; ,\\
 K_1^l (z,z_B) &= x (z_B) \left( k(z)- k\left( z^{-1} \right) \right) \; , \\ \nonumber \\
 K_{id}^r (z,z_B) &= 1- k^2 \left( z \right) x^2 \left( z_B \right) \; , \\
 K_m^r (z,z_B) &= x^2 \left( z_B \right) k(z) \left( k\left(z \right) - k \left( z^{-1} \right) \right)  \; , \\
 K_1^r (z,z_B) &= x (z_B) \left( k(z)- k\left( z^{-1} \right) \right) \; .
\end{align}
\end{subequations}
Here, $z_B$ is a free parameter, which we call \emph{boundary rapidity}.
The left and right $K$-matrix is normalized by
\begin{align}
 W_{K_l} \left( z, z_B \right) &= \left( k(z) x(z_B) - 1 \right) \left( 1 + k\left(z^{-1} \right) x \left(z_B \right) \right)\;  , \\ \nonumber \\
 W_{K_r} \left( z, z_B \right) &= \left( 1 - k \left( z^{-1} \right) x \left( z_B \right) \right) \left(1 + k \left( z \right) x \left( z_B \right) \right) 
\end{align}
respectively, and we use the auxiliary functions
\begin{equation}
 k(z) = \omega z - z^{-1} \qquad\mbox{and}\qquad x(z) = \omega \frac{z}{z^2-1} \, . 
\end{equation}
By these definitions, the weight of a given configuration is a function of the 
rapidities involved in the $R$- and $K$-matrices. 

\subsection{Vector space of link patterns}

\begin{figure}
 \begin{center}
  \includegraphics[scale=0.3]{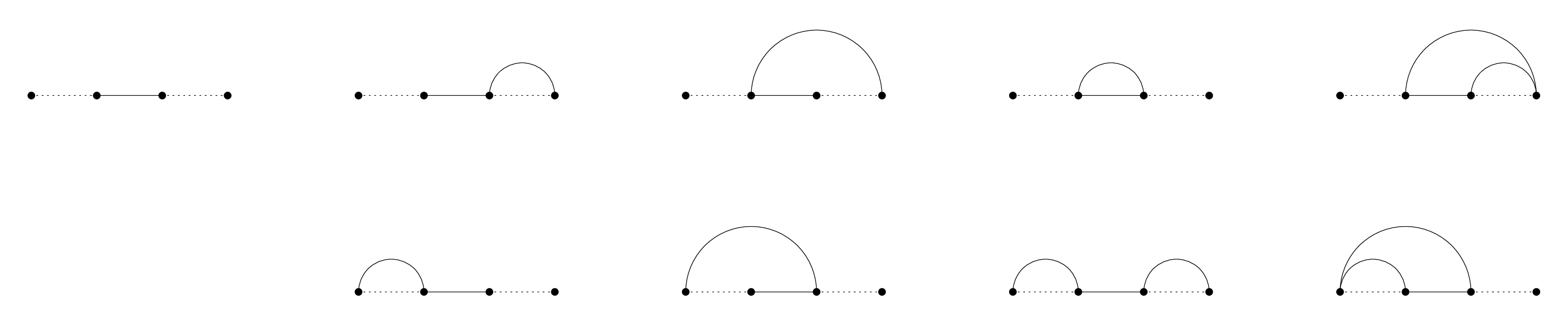}
  \caption{The elements of $dLP_{L=2}$. The two outermost points represent the boundaries, the inner ones the sites. The top row: $\ket{\bullet\bullet}$ (empty element), $\ket{\bullet(}$, $\ket{(\bullet}$, $\ket{()}$, $\ket{((}$.
  The bottom row: $\ket{)\bullet}$, $\ket{\bullet)}$, $\ket{)(}$, $\ket{))}$.}
  \label{fig:dLP2}
 \end{center}
\end{figure}

In this section we introduce the vector space of link patterns, in which the $R$ and $K$-matrices act as stochastic operators. Based on this description, we present the equations, which are satisfied by the inhomogeneous model. The $R$ and $K$-matrices act on the vector space as elements of the two boundary dilute Temperley-Lieb algebra (for a brief overview, see \cite{grimm1996dilute}).\\
Consider the dilute $O(n=1)$ loop model on a half-infinite strip, finite in width, infinite upward, and with a bottom edge. 
We are interested in the 
connectivity configurations on the bottom edge. 
\begin{figure}
 \begin{center}
  \includegraphics[scale=0.7]{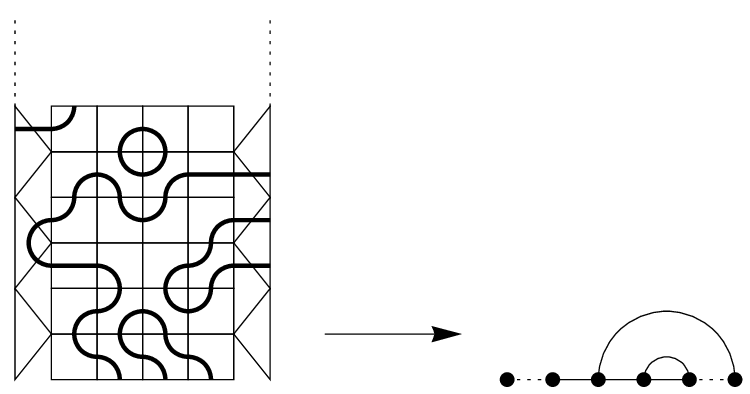}
  \caption{The mapping from loop configurations to link patterns. The two outermost points --connected with dotted line to the others-- represent the two
  boundaries. Here the image of the mapping is $\ket{\bullet(()}$}
  \label{fig:dLPElementExample}
 \end{center}
\end{figure}
Introduce $dLP_L$, the set of  \emph{dilute link patterns} of size $L$, the possible connectivities on the bottom edge of a half-infinite strip with width $L \in \mathbb{N}$. The set is built up as follows: Every site can be either occupied or empty. 
If a site is occupied, it can be connected to an other site, or to the
left or right boundary. The connectivities are such that the chords
corresponding to the connectivities are non-intersecting. Hence, every
dilute loop configuration on the $L$ size half-infinite strip
corresponds to an element of $dLP_L$ (Fig.~\ref{fig:dLPElementExample}). 
 $dLP_L$ is in bijection with $L$ long strings of the characters '(', '$\bullet$' and ')', consequently $dLP_L$ contains $3^L$ element.
We use $dLP_L$, as the formal basis for the vector space $V_L =
\text{span}(dLP_L)$. We denote the \emph{basis of link patterns} as
$dLP_L$, and the \emph{vector space of link patterns} as $V_L$. The
space $V_L$ is thus spanned by the basis vectors $\ket{\pi} \in dLP_L$. 
%
As an example, the basis elements of $dLP_{L=2}$ are in Fig.~\ref{fig:dLP2}.
\begin{figure}
 \begin{center}
  \psfrag{A}{$=$}
  \includegraphics[scale=0.3]{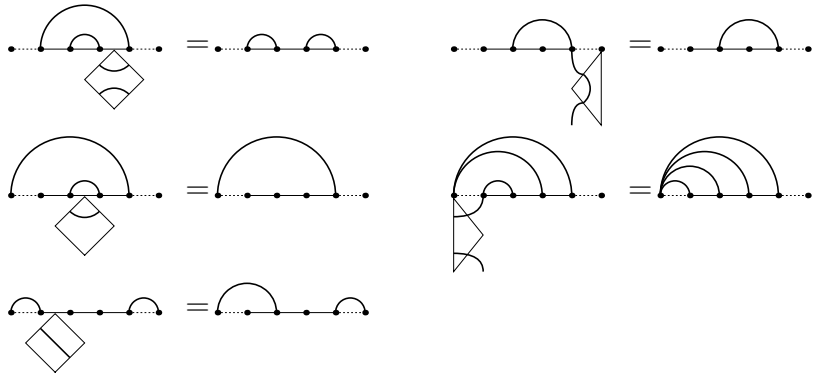}
  \caption{Some examples of $R$, and $K$-matrix elements acting on Temperley-Lieb states}
  \label{fig:qKZexamples}
 \end{center}
\end{figure}
The $R$-matrix and $K$-matrix act on the link pattern vector space as operators. The left and right $K$-matrices act as operators, as they are attached to the first and last site. The $R$-matrices act as operators, as they are rotated by $45^\circ$, and act on two consecutive sites. The image of a specific plaquette on a link pattern is the link pattern
which is formed after attaching the given plaquette to the bottom (Fig.~\ref{fig:qKZexamples}). The $R$-matrix is an element of the dilute Temperley-Lieb algebra.
The $R$-matrix and $K$-matrix act as stochastic operators over $V_L$.
When we need to specify that $R$ acts on sites $i$ and $i+1$, 
we will use the subscript $R_{i,i+1}$.
To compute the correct weights, 
$R$ and $K$-matrices are initially defined to map from the sites, where rapidities enter
to the ones, where they exit, and their orientation is taken into account respecting this rule.
Relations involving $R$ and $K$-matrices can be represented by pictures. These relations mean that the sum of all possible configurations realizing the same connectivity has the same statistical weight on the two sides of the equation. The directed red lines always represent 
rapidity flows. A crossing of two rapidity flows is an $R$-matrix, its weight is computed respecting the direction of lines.
We introduce lines for the boundary rapidities too, it is helpful to treat rapidities and boundary rapidities in a more uniform way.
By this, we give new pictorial representation for the $K$-matrices:
\vspace{15pt}
  \par
  \begin{minipage}{\linewidth}
            \centering
  \psfrag{a}{$w$}
  \psfrag{b}{$w^{-1}$}
  \psfrag{c}{$z_B$}
  \psfrag{A}{$\equiv$}
  \includegraphics[scale=0.4]{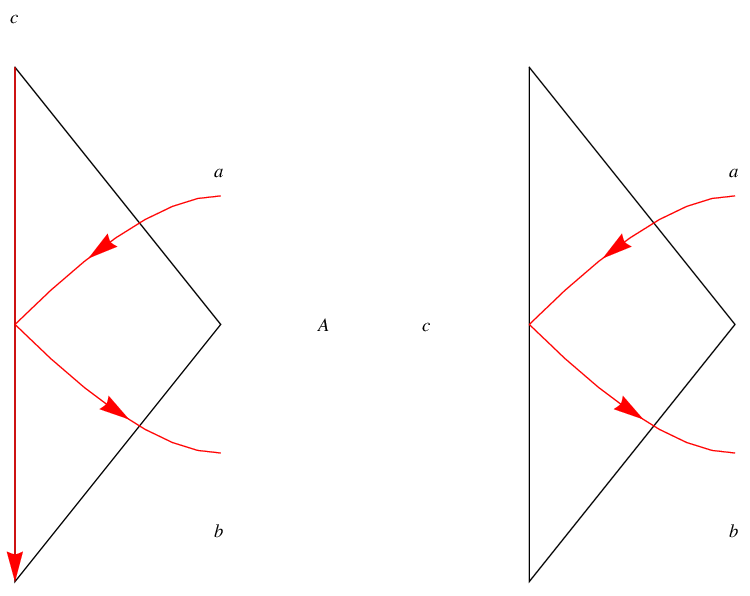}
  \hspace{20pt}
  \psfrag{a}{$w$}
  \psfrag{b}{$w^{-1}$}
  \psfrag{c}{$z_B$}
  \psfrag{A}{$\equiv$}
  \includegraphics[scale=0.4]{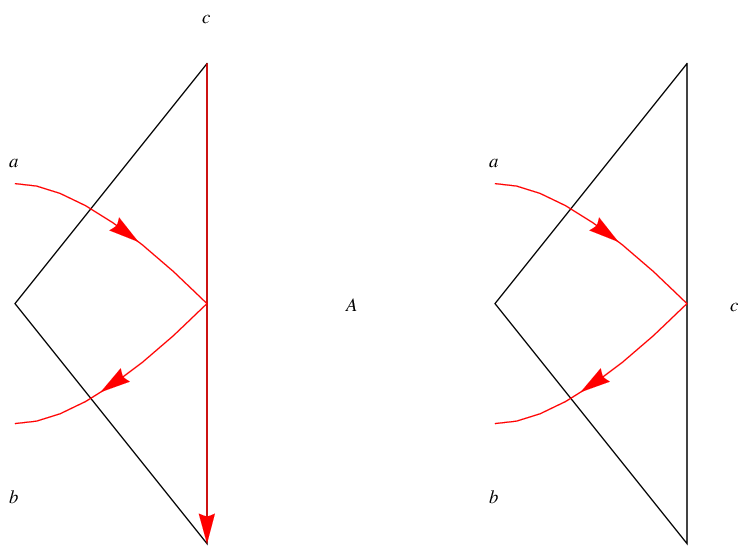}
        \end{minipage}
\vspace{15pt}

A $K$-matrix is represented by a reflection of a rapidity on a boundary. The order of the operators is prescribed by the direction of rapidities.
The tiles are only drawn on some of the pictures.


%

With the previously mentioned definition, the $R$-matrix satisfies the following equations (In the pictures, for legibility, we omit prefactors):
\begin{itemize}
  \item the inversion/unitary relation:
  \begin{equation}
	\label{eq:RReqid}
    R(z_2,z_1) R(z_1,z_2) = W_R (z_2,z_1) W_R (z_1,z_2) \cdot \text{id}
  \end{equation}
  \par
  \begin{minipage}{\linewidth}
            \centering
            \psfrag{a}{$z_1$}
	    \psfrag{b}{$z_2$}
	    \psfrag{c}{$z_1$}
	    \psfrag{d}{$z_2$}
	    \psfrag{A}{$=$}
	    \includegraphics[scale=0.4]{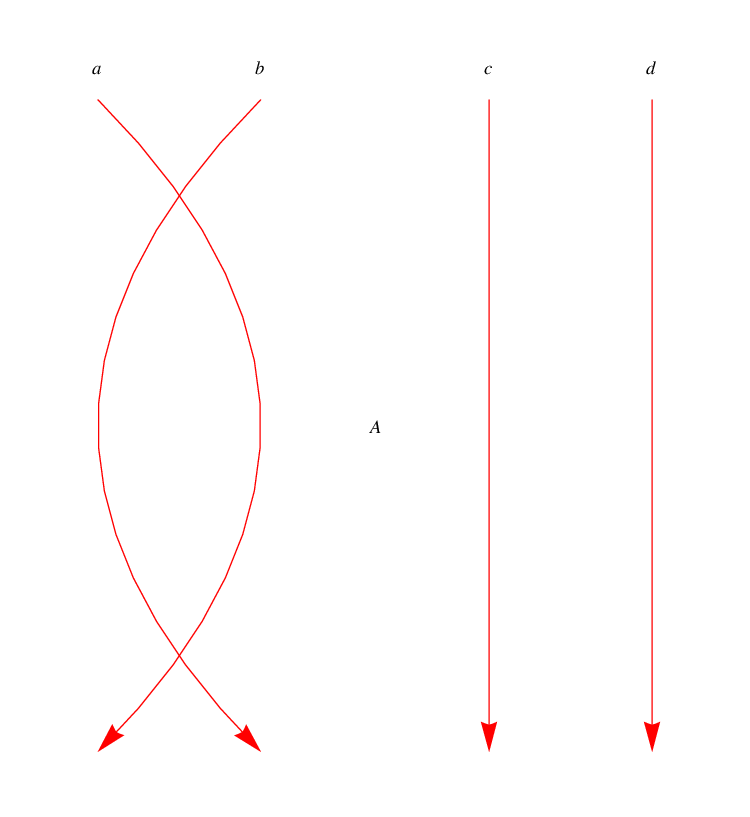}
        \end{minipage}
  \item the crossing relation: 
  \begin{equation}
    R(z ,w) = - \left( \frac{w}{z} \right) ^2 R^{rot} (-w,z)
  \end{equation}
  \par
  \begin{minipage}{\linewidth}
            \centering
            \psfrag{a}{$z$}
	    \psfrag{b}{$w$}
	    \psfrag{c}{$z$}
	    \psfrag{d}{$-w$}
	    \psfrag{A}{$=$}
	    \includegraphics[scale=0.5]{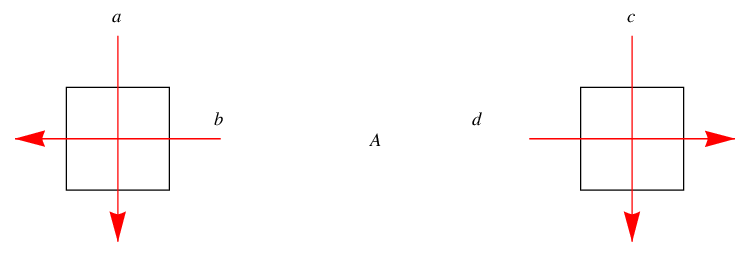}
  \end{minipage}
		As the $R$-matrix only depends on the ratio $\frac{w}{z}$, an equivalent form of the crossing relation is
	\begin{equation}
    R(z ,w) = - \left( \frac{w}{z} \right) ^2 R^{rot} (w,-z).
    \end{equation}
  \item the Yang-Baxter equation:
  \begin{equation}
   R_{23} (u,v) R_{12} (u,w) R_{23} (v,w) = R_{12} (v,w) R_{23} (u,w) R_{12} (u,v)
  \end{equation}
  \par
  \begin{minipage}{\linewidth}
            \centering
            \psfrag{a}{$u$}
	    \psfrag{b}{$v$}
	    \psfrag{c}{$w$}
	    \psfrag{A}{$=$}
	    \includegraphics[scale=0.5]{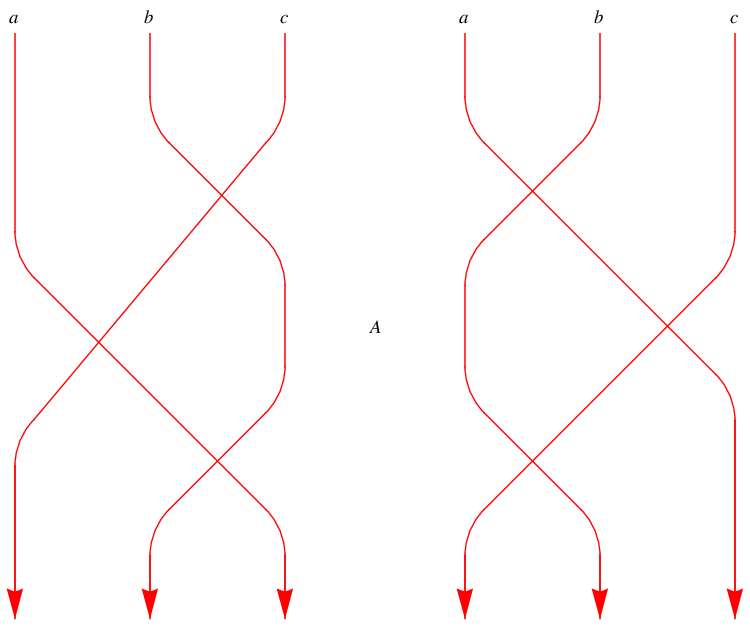}
  \end{minipage}
\end{itemize}

The $K$-matrix satisfies the following equations:
\begin{itemize}
  \item the boundary inversion/unitary relation:
  \begin{itemize}
   \item left boundary:
   \begin{equation}
	 \label{eq:leftKKid}
   K_l (w^{-1},z_B) K_l (w, z_B) =  W_{K_l} (w^{-1},z_B) W_{K_l} (w, z_B) \cdot \text{id}
  \end{equation}
  \par
  \begin{minipage}{\linewidth}
            \centering
            \psfrag{a}{$w$}
	    \psfrag{b}{$w^{-1}$}
	    \psfrag{c}{$\!\! z_B$}
	    \psfrag{A}{$=$}
	    \includegraphics[scale=0.4]{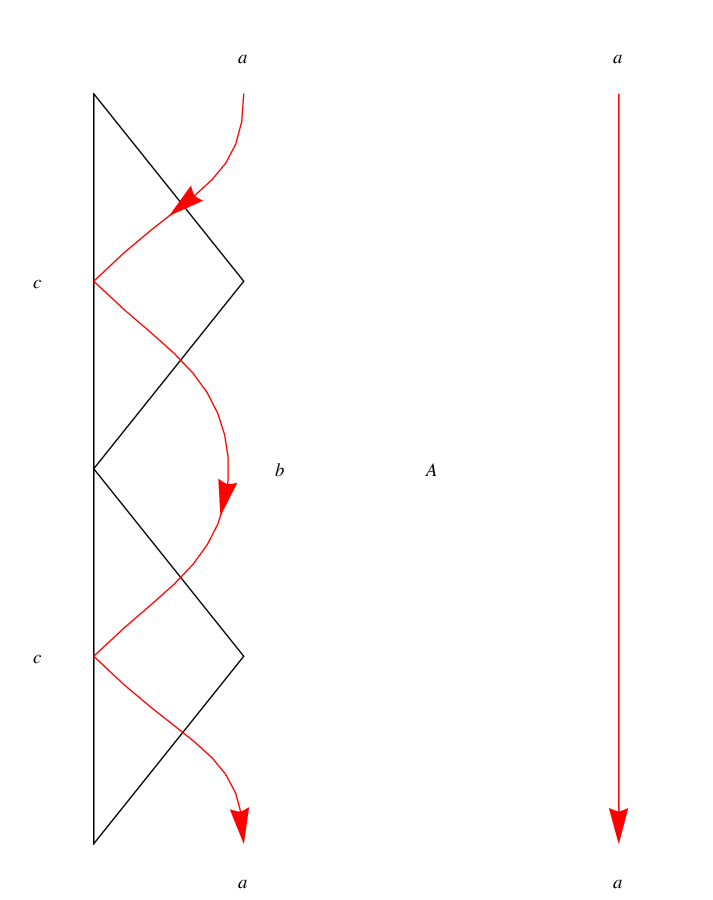}
        \end{minipage}
   \item right boundary:
   \begin{equation}
	  \label{eq:rightKKid}
   K_r (w^{-1},z_B) K_r (w, z_B) = W_{K_r} (w^{-1},z_B) W_{K_r} (w, z_B) \cdot \text{id}
  \end{equation}
  \par
  \begin{minipage}{\linewidth}
            \centering
            \psfrag{a}{$w$}
	    \psfrag{b}{$\! \! \! \! \!w^{-1}$}
	    \psfrag{c}{$z_B$}
	    \psfrag{A}{$=$}
	    \includegraphics[scale=0.4]{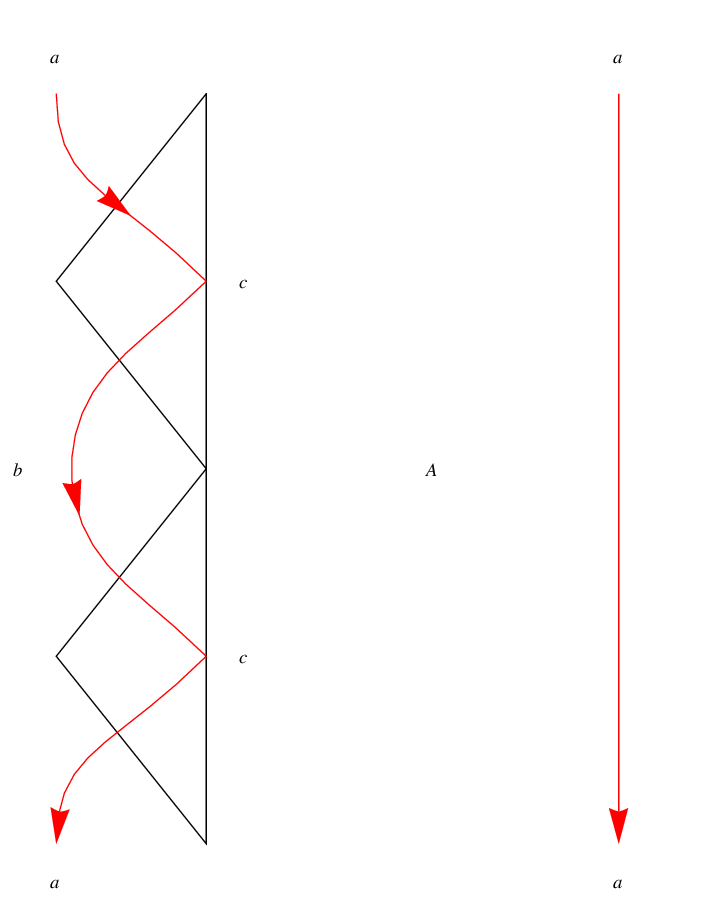}
        \end{minipage}
  \end{itemize}
  \item the boundary crossing relation:
  \begin{equation}
  K_l (w, z_B) = - K_r (-w^{-1}, -z_B)
  \end{equation}
  \par
  \begin{minipage}{\linewidth}
            \centering
            \psfrag{a}{$w$}
	    \psfrag{b}{$w^{-1}$}
	    \psfrag{c}{$-w$}
	    \psfrag{d}{$-w^{-1}$}
	    \psfrag{e}{$z_B$}
	    \psfrag{f}{$-z_B$}
	    \psfrag{A}{$=$}
	    \includegraphics[scale=0.5]{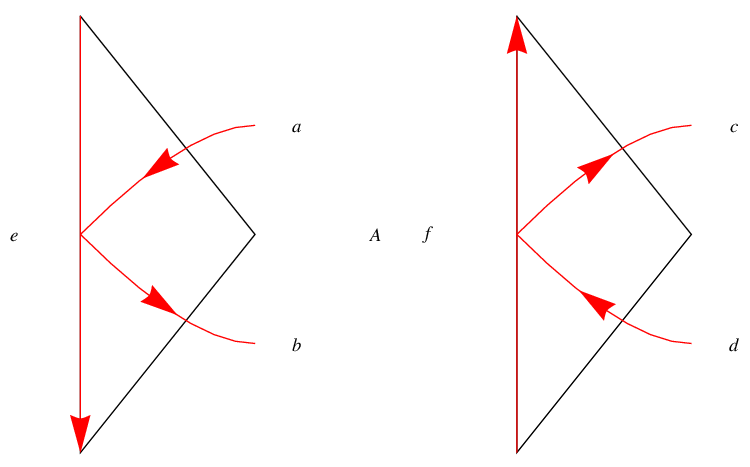}
        \end{minipage}
  \item the reflection equation:
  \begin{itemize}
   \item left boundary:
   \begin{multline}
   R(v^{-1}, u^{-1}) K_l (v,z_B) R (u^{-1},v) K_l (u,z_B)=\\=K_l (u, z_B ) R (v^{-1},u) K_l (v,z_B) R(u,v)
  \end{multline}
  \par
  \begin{minipage}{\linewidth}
      \centering
      \psfrag{a}{\small $z_B$}
	    \psfrag{b}{\small $u$}
	    \psfrag{c}{\small $v$}
	    \psfrag{d}{\small $u^{-1}$}
	    \psfrag{e}{\small $v^{-1}$}
	    \psfrag{A}{$=$}
	    \includegraphics[scale=0.4]{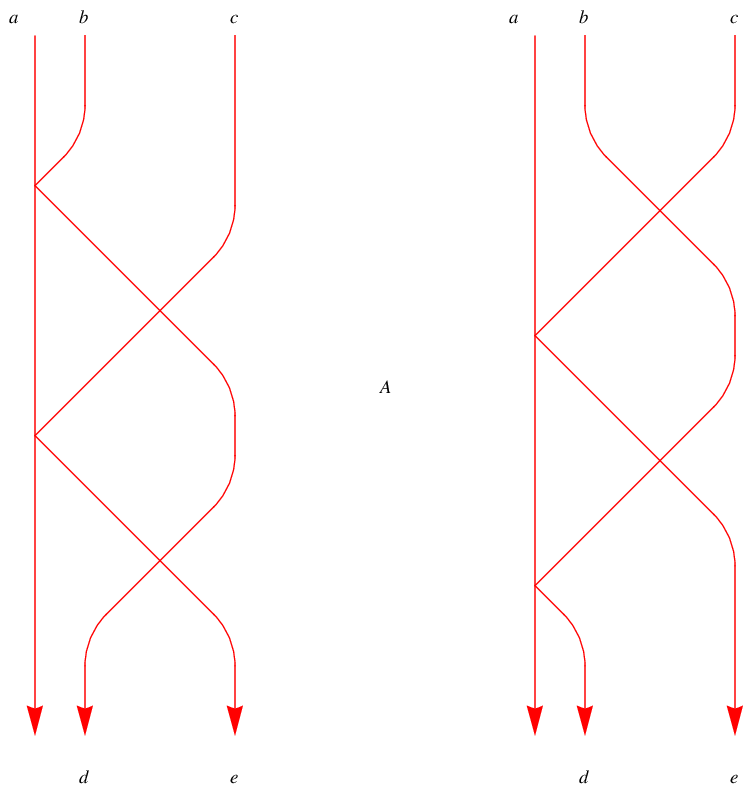}
   \end{minipage}
  \item right boundary:
    \begin{multline}
 R(v^{-1}, u^{-1}) K_r (u, z_B) R (u, v^{-1}) K_r (v, z_B) =\\= K_r (v,z_B) R (v, u^{-1}) K_r (u, z_B) R (u,v)
	\end{multline}
  \par
  \begin{minipage}{\linewidth}
      \centering
      \psfrag{a}{\small $u$}
	    \psfrag{b}{\small $v$}
	    \psfrag{c}{\small $z_B$}
	    \psfrag{e}{\small $u^{-1}$}
	    \psfrag{f}{\small $v^{-1}$}
	    \psfrag{g}{\small $z_B$}
	    \psfrag{A}{$=$}
	    \includegraphics[scale=0.4]{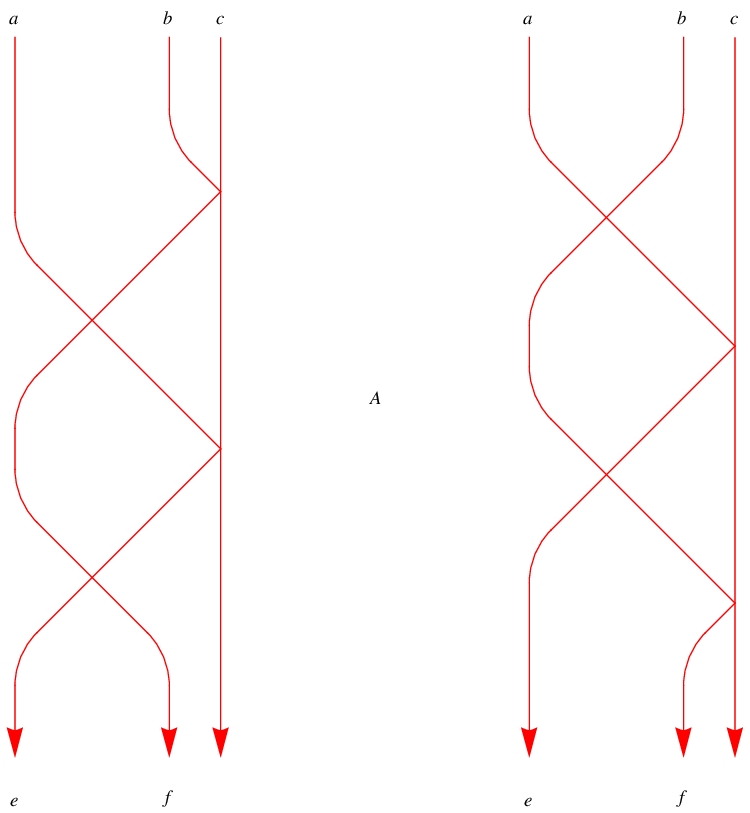}
  \end{minipage}
  \end{itemize}
\end{itemize}

%
Since there is no restriction on the orientation of the figures, the operators $K_l$ and $K_r$ are related by symmetry, and an inversion of the rapidity line.
For our purposes, it is enough to replace $K_r$  with an "upside-down" left $K_l$-matrix. 
The following reflection equation holds for this case (The interpretation as operators might seem obscure, however we can always refer to statistical interpretation.
Note the direction of rapidities and boundary rapidities):
 \begin{equation}
 \label{eq:RevReflRel}
   R (u, v^{-1} ) K_l (u,-z_B) R (v,u^{-1}) K_r (v,z_B) = K_r ( v,z_B ) R (u,v) K_l (u,-z_B) R (v, u^{-1}) \; .
 \end{equation}
  \par
  \begin{minipage}{\linewidth}
            \centering
            \psfrag{a}{$u^{-1}$}
	    \psfrag{b}{$v$}
	    \psfrag{c}{$z_B$}
	    \psfrag{d}{$-z_B$}
	    \psfrag{e}{$u$}
	    \psfrag{f}{$v^{-1}$}
	    \psfrag{A}{$=$}
	    \includegraphics[scale=0.5]{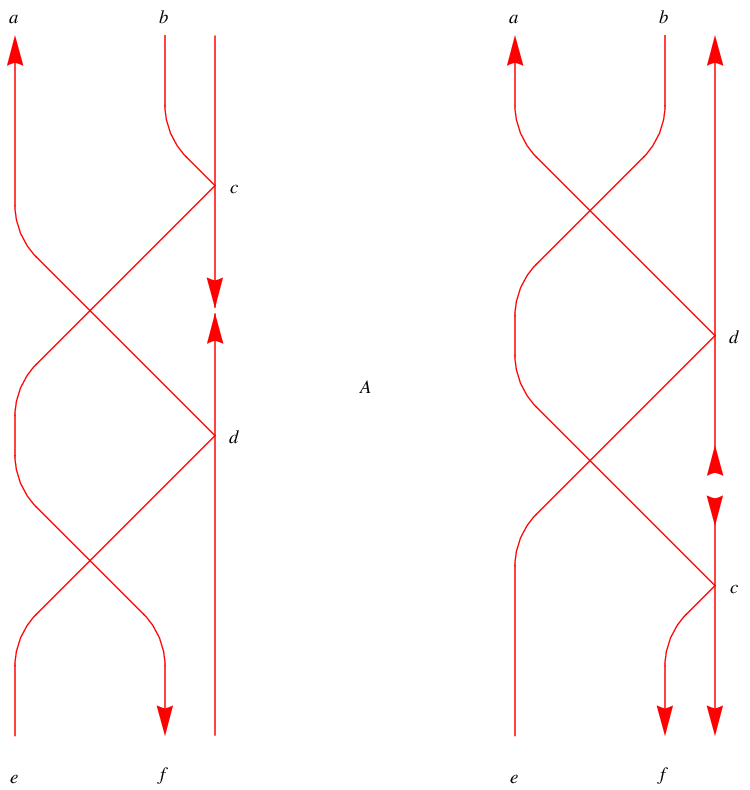}
   \end{minipage}
       
To make the boundary rapidity flows continuous, we introduce the following 'boundary-crossed' left $K$-matrix, by the following definition:
 \begin{equation}
   K_l^{b.reversed} (w,-z_B) = K_r (w, z_B) \; .
 \end{equation}
  \par
  \begin{minipage}{\linewidth}
            \centering
            \psfrag{a}{$w$}
	    \psfrag{b}{$w^{-1}$}
	    \psfrag{c}{$-z_B$}
	    \psfrag{d}{$z_B$}
	    \psfrag{A}{$=$}
	    \includegraphics[scale=0.5]{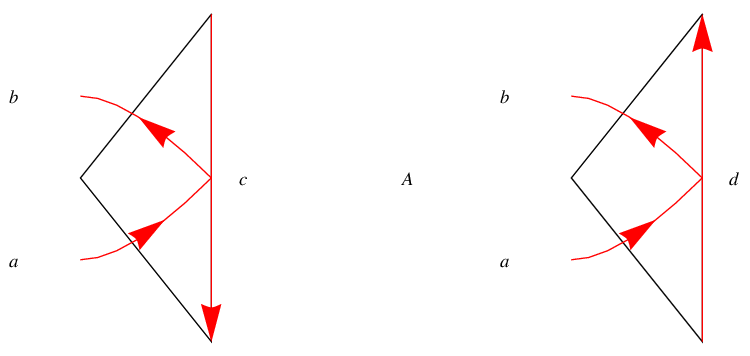}
  \end{minipage}

By this, Eq.~(\ref{eq:RevReflRel}) can be written in the following form:
 \begin{multline}
   R (u, v^{-1} ) K_l^{b.reversed} (u,z_B) R (v,u^{-1}) K_r (v,z_B) =\\= K_r ( v,z_B ) R (u,v) K_l^{b.reversed} (u,z_B) R (v, u^{-1}) \; .
 \end{multline}
  \par
  \begin{minipage}{\linewidth}
            \centering
           \psfrag{a}{$u^{-1}$}
	    \psfrag{b}{$v$}
	    \psfrag{c}{$z_B$}
	    \psfrag{d}{$z_B$}
	    \psfrag{e}{$u$}
	    \psfrag{f}{$v^{-1}$}
	    \psfrag{A}{$=$}
	    \includegraphics[scale=0.5]{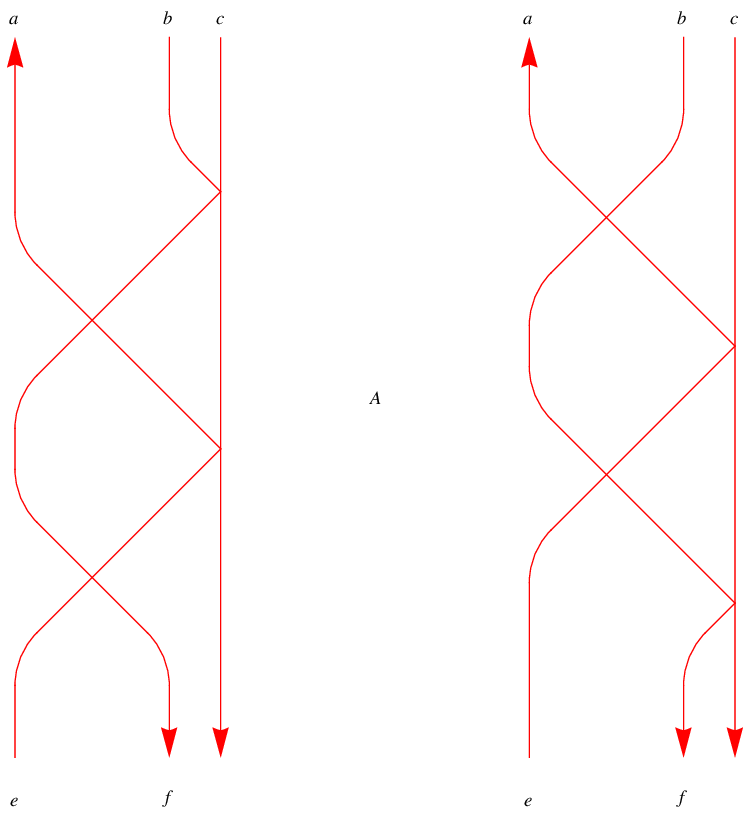}
  \end{minipage}

Similarly, we can introduce 'boundary-crossed' right $K$-matrix, what we leave for the educated reader. We introduce the 'boundary-crossed' $K$-matrices in order to be able to define the transfer matrix. 

\subsection{Double row transfer matrix}

Based on these definitions, we define the \emph{double row transfer matrix}:
\begin{equation}
	\begin{minipage}{\linewidth}
      \centering
      \psfrag{a}{$z_0$}
	    \psfrag{b}{$z_1$}
	    \psfrag{c}{$z_2$}
	    \psfrag{d}{$z_{L-1}$}
	    \psfrag{e}{$z_L$}
	    \psfrag{f}{$z_{L+1}$}
	    \psfrag{g}{$w$}
	    \psfrag{h}{$w^{-1}$}
	    \psfrag{A}{$T_L (w, z_0, z_1, \ldots, z_L, z_{L+1})=$}
	    \includegraphics[scale=1]{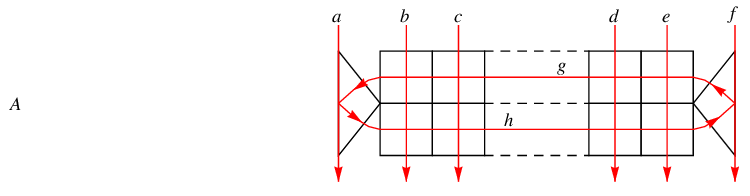}
	\end{minipage}
\end{equation}
%
By standard method using the Yang-Baxter equation and the reflection equations we can prove that such double row transfer matrices form a one parameter
family of commuting matrices \cite{sklyanin1988boundary}:
\begin{equation}
\label{eq:Tcom}
 \left[ T_L (u, z_0, z_1, \ldots, z_L, z_{L+1}), T_L (v, z_0, z_1, \ldots, z_L, z_{L+1}) \right] = 0 \; .
\end{equation}
At $n=1$ loop weight the model is stochastic, in analogy with the dense case \cite{pearce2002temperley}. A transfer matrix acts on a link pattern as a stochastic operator, sending it to a linear combination of other link patterns, 
weighted by Laurent polynomials in the rapidities $z_1, \ldots, z_L$ and boundary rapidities $z_0, z_{L+1}$. The weight is defined by the $R$ and $K$-matrices, constituting the $T$-matrix. A matrix element of the transfer matrix is the sum of the weights of all the path configurations which maps the preimage link pattern into the image link patter, possibly 0. 

Take a configuration with finite $L$ width and which is infinite upward, with an edge at the bottom. 
The edge at the bottom is a realization of $dLP_L$. Acting on this configuration with a $T$-matrix is
adding the $T$-matrix to the bottom edge, and consider the $dLP_L$ configuration on the new edge.
The probability distribution of the link pattern configurations is given by the ground state 
eigenvector of the $T$-matrix, which 
we will denote by $\ket{\Psi (z_0, z_1, \ldots, z_L, z_{L+1})}$. The existence and uniqueness of such a vector is provided by the Perron-Frobenius theorem:
\begin{equation}
 T \left( w, z_0, \ldots, z_{L+1} \right) \ket{\Psi \left( z_0, \ldots, z_{L+1} \right) } =
 N \left( w, z_0, \ldots, z_{L+1} \right) \ket{\Psi \left( z_0, \ldots, z_{L+1} \right) } \; ,
\end{equation}
where $N_L\left( w, z_0, \ldots, z_{L+1} \right)$ is the normalization of the $T$-matrix:
\begin{equation}
\begin{split}
\label{eq:Tnorm}
 N_L\left( w, z_0, \ldots, z_{L+1} \right)=&W_{K_l} (w, z_0) W_{K_l} (w^{-1}, -z_{L+1}) \prod_{i=1}^L W_R (w,z_i) W_R (z_i,w^{-1}) + \\
  & \tilde{W}_{K_l} (w, z_0) \tilde{W}_{K_l} (w^{-1}, -z_{L+1}) \prod_{i=1}^L \tilde{W}_R (w,z_i) \tilde{W}_R (z_i,w^{-1}) \; ,
\end{split}
\end{equation}
where  $W_{R}$ and $W_{K_l}$  are the normalization of the $R$ and $K_l$-matrices, and $\tilde{W}_{R}$ and $\tilde{W}_{K_l}$ are the following functions:
\begin{align}
 \tilde{W}_{R} (z_1,z_2)&= W_1 (z_1,z_2) - W_t (z_1,z_2) \; ,
 \\
 \tilde{W}_{K_l} (z, z_B)&=K_{id} (z,z_B) + K_m (z,z_B) - K_1 (z,z_B) \; .
\end{align}
%

The value of the normalization is non-trivial, since
 it is not the product of the normalization of the constituting $R$ and $K$-matrices.
The derivation for $N_L$ is in Appendix~\ref{app:Norm}.
The $K$-matrix on the right side of the $T$-matrix is computed by $K_L^{b.reversed}$. \\
Due to commutativity, that the  eigenvectors of the transfer matrix do not 
depend on $w$, which is only a spectral parameter and which we will call \emph{auxiliary rapidity}. 
The ground state vector is a vector of polynomials in $z_i$, in the basis $dLP_L$:
\begin{equation}
 \ket{\Psi_L (z_0, \ldots, z_{L+1})} = \sum_{\pi \in dLP_L} \psi_{\pi} (z_0, \ldots, z_{L+1}) \ket{\pi} \; .
\end{equation}
Here $\psi_{\pi} (z_0, \ldots, z_{L+1})$ is the polynomial coefficient of the basis element $\ket{\pi} \in dLP_L$. Sometimes we will
use the following notation: $\ket{\psi_{\pi}} = \psi_{\pi} (z_0, \ldots, z_{L+1}) \ket{\pi}$.
It is worth mentioning that even $z_0$ and $z_{L+1}$ have been introduced differently from the other rapidities, they behave similarly as all the other rapidities. The reason is that an open boundary $K$-matrix can be constructed from a closed boundary $KRR$ configuration, and the vertical rapidity of the $R$-matrices turn into the boundary rapidity (Appendix~\ref{app:KmxConstr}).

By introducing $dLP_L$ and $V_L$, we are able to compute quantities on the half-strip. In order to compute quantities on the full strip,
we introduce $dLP_L^{\ast}$ and $V^{\ast}_L$, dual to $dLP_L$ and $V_L$. The dual basis consist the link patterns in the downward direction. The dual space is spanned by the dual basis: $V_L^\ast = \text{span}(dLP_L^\ast)$. 
The scalar product of $\bra{\alpha} \in dLP_L^{\ast}$ and 
$\ket{\beta} \in dLP_L$:
\begin{equation}
 \langle \alpha | \beta \rangle = 
 \begin{cases}
   1 \quad \text{if the two link patterns match respecting the 
occupation}\\
	 0 \quad \text{otherwise}
 \end{cases}
\end{equation}
%
In an analogous fashion, we can built up the probabilistic picture for the dual vector space with a dual transfer-matrix. 

The transfer-matrices of $dLP_L$ and $dLP_L^{\ast}$ are related by the following relation:
	\begin{equation}
	\label{eq:TdualT}
  \begin{minipage}[c]{\linewidth}
      \centering
      \psfrag{a}{\tiny $z_0$}
	    \psfrag{b}{\tiny $z_1$}
	    \psfrag{c}{\tiny $z_2$}
	    \psfrag{d}{\tiny $z_{L-1}$}
	    \psfrag{e}{\tiny $z_L$}
	    \psfrag{f}{\tiny $z_{L+1}$}
	    \psfrag{g}{\tiny $w$}
	    \psfrag{h}{\tiny $w^{-1}$}
	    \psfrag{i}{\tiny $-w$}
	    \psfrag{j}{\tiny $-w^{-1}$}
	    \psfrag{A}{$=$}
	    \includegraphics[scale=1.1]{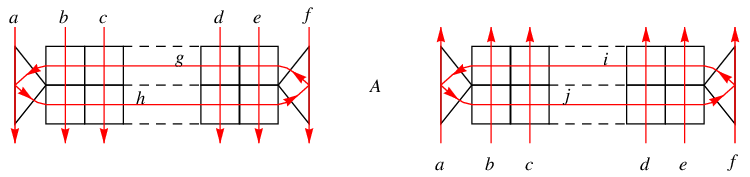}
			\vspace {5pt}
       \end{minipage}
\end{equation}
The transfer matrix of the dual space has a unique largest eigenvector:
\begin{equation}
\bra{\Psi} = \sum_{\pi \in dLP_L^\ast} \psi^\ast_\pi \bra{\pi} \equiv \sum_{\pi \in dLP_L^\ast} \bra{\psi_\pi} \, . 
\end{equation}
Eq.~\ref{eq:TdualT} defines the relation between the ground state elements:
\begin{equation}
\label{eq:DualEquiv}
 \psi_{\alpha} (z_0, \ldots, z_{L+1}) = \psi^\ast_{r(\alpha)} (z_{L+1}, \ldots, z_0) \; , 
\end{equation}
 where $\psi_\alpha \in dLP_L$ and $\psi^\ast_{r(\alpha)} \in dLP_L^{\ast}$ are related by a $180$ degree rotation, e.g.: \\
 $r(\ket{\bullet\bullet\bullet()(\bullet\bullet})=\bra{\bullet\bullet)()\bullet\bullet\bullet}$.

 With the ground state vector of both spaces, we can take scalar products of ground state elements:
\begin{subequations}
\begin{align}
 \langle \psi_{\alpha} | \psi_{\beta} \rangle &= \psi^\ast_\alpha (z_0, \ldots, z_{L+1}) \psi_\beta (z_0, \ldots , z_{L+1}) \langle \alpha | \beta \rangle \\ &= \psi_{r(\alpha)} (z_{L+1}, \ldots, z_0) \psi_\beta (z_0, \ldots , z_{L+1}) \langle \alpha | \beta \rangle,
\end{align}
\end{subequations}
which is equal to the probability of the full strip configuration. \\
For further calculation, we need to define three quantities: the empty element, the partition sum of the half strip, and the 
 partition sum of the full strip.
 The \emph{empty element}  is the link pattern with only unoccupied sites. 
 We denote by $ \psi_{EE}$ the weight of the empty element. 
 The \emph{partition sum of the half strip} is the sum of all ground state elements:
 \begin{equation}
  Z_{h.s.} (z_0, \ldots, z_{L+1}) = \sum_{\alpha \in dLP_L} \psi_{\alpha} (z_0, \ldots, z_{L+1}) \; .
 \end{equation}
The \emph{partition sum of the full strip} is the normalization of the probabilities on the full strip.
The partition sum of the full strip is 
\begin{equation}
\begin{split}
 Z_{f.s.} (z_0, \ldots, z_{L+1}) =& \braket{\Psi (z_0, \ldots, z_{L+1}) }{\Psi (z_0, \ldots, z_{L+1}) } = \\ 
=& \sum_{{\alpha \in dLP_L\atop r(\beta) \in dLP_L^{\ast}}} \psi_{\alpha} (z_0, \ldots, z_{L+1} ) \psi_{r(\beta)}^\ast (z_0, \ldots, z_{L+1}) =\\
=& \sum_{\alpha, \beta \in dLP_L} \psi_{\alpha} (z_0, \ldots, z_{L+1} ) \psi_{\beta} (z_{L+1}, \ldots, z_0) \; ,
\end{split}
\end{equation}
where we sum up to
$\alpha,\beta \in dLP_L$ link patterns with matching occupation.

Based on the properties of the $R$ and $K$-matrices, the transfer matrix satisfies the following 
conditions (with suppressing irrelevant notation):
\begin{subequations}
\begin{align}
\label{eq:RTeqTR}
 R_{i,i+1} (z_i, z_{i+1}) T \left(\ldots, z_i, z_{i+1}, \ldots \right) &= T \left(\ldots, z_{i+1}, z_{i} , \ldots \right)  R_{i,i+1} (z_i, z_{i+1}) \; ,\\
 K_l \left( z_1, z_0 \right) T \left( z_1, \ldots \right) &= T \left( z_1^{-1}, \ldots \right) K_l \left( z_1, z_0 \right) \; ,
\label{eq:KlTeqTKl}
\\
\label{eq:KrTeqTKr}
 K_r \left( z_L, z_{L+1} \right) T \left( \ldots, z_L \right) &= T \left( \ldots, z_L^{-1} \right) K_r \left( z_L, z_{L+1} \right) \; .
\end{align}
\end{subequations}
Acting with both sides on $\ket{\Psi_L}$, we derive the $q$-Knizhnik-Zamolodchikov equations \cite{di2005around, frenkel1992quantum, smirnov1992form} for the dilute $O(1)$ model,
 with open boundaries:
\begin{subequations}
 \begin{align}
 \label{eq:qKZeq}
 R_{i,i+1} (z_i, z_{i+1}) \ket{\Psi \left( \ldots, z_i, z_{i+1}, \ldots \right) } &=W_R  (z_i, z_{i+1}) \ket{\Psi \left( \ldots, z_{i+1}, z_{i}, \ldots \right) } \; ,\\
 \label{eq:leftbqKZeq}
 K_l \left( z_1, z_0 \right) \ket{\Psi \left(z_0, z_1, \ldots \right) } &=W_{K_l} \left( z_1, z_0 \right) \ket{\Psi \left(z_0, z_1^{-1}, \ldots \right) }  \; , \\
 \label{eq:rightbqKZeq}
K_r \left( z_L, z_{L+1} \right) \ket{\Psi \left( \ldots, z_L, z_{L+1}\right) } &=W_{K_r} \left( z_L, z_{L+1} \right) \ket{\Psi \left( \ldots, z_L^{-1}, z_{L+1} \right) } \; ,
\end{align}
\end{subequations}
where the prefactors $W$ follow from the stochasticity and the normalization of the $R$ and $K$-matrices. \\
The $q$KZ equations are the central tool to find the full ground state vector of the transfer matrix. The ground state elements are computed recursively. Starting from two specific elements, called fully nested elements (corresponding to the $\ket{))))\ldots )))}$ and the $\ket{((( \ldots ((}$ vectors), the full ground state is computable, by the repeated application of the $q$KZ equations~\cite{garbali2017dilute1, garbali2017dilute2, di2005around}. In practise, the computation is limited to small systems by computational power, in our case, the full ground state has been computed for $L=1,2,3$. \\
Explicit ground state elements and partition sum for $L=1$ are presented in Appendix~\ref{app:L1}.

\section{Definition of current, main results}
\label{sec:current}

\subsection{Boundary to boundary current}

In this section, we  give our main result, the exact finite size expression for the boundary to boundary current in the
dilute $O(1)$ model. \\
The spin-1 current, introduced in Eq.~(\ref{eq:Fdef}) is equal to the signed sum of all 
the possible plaquette configurations which have boundary to boundary path between the selected 
vertices. After Baxterization, $F^{\left( x_1, x_2 \right)}$ and $P(C)$
become a function of the rapidities. Because of the additivity of $F$, it is 
sufficient to concentrate on two cases, when $x_1$ and $x_2$ are on two adjacent sites. 
These markers can be separated by a horizontal lattice edge 
(Using horizontal and vertical lattice-indices for the coordinates $x_1$, $x_2$.):
\begin{equation}
 X^{\left( k \right)} = F^{\left( (k,i),(k+1,i) \right)} \; ,
\end{equation}
or a vertical one:
\begin{equation}
 Y^{\left( k \right)} = F^{\left( (k,j),(k,j+1) \right)} \; .
\end{equation}
After Baxterization, $X$ and $Y$ become functions of the rapidities and boundary rapidities. Because the $T$-matrices with different auxiliary rapidities commute, the $X$-current can not depend on them, but the $Y$-current still depends on the auxiliary rapidity between the points ($k,j$) and ($k,j$+1):
\begin{align}
 X^{(k)} &= X^{(k)} (z_0, \ldots, z_{L+1}) \; , \\
 Y^{(k)} &= Y^{(k)} (w, z_0, \ldots, z_{L+1}) \; .
\end{align}
The current is computable by the introduction of operators $\hat{X}^{(k)}$ and $\hat{Y}^{(k)}$, as the following expectation values:
\begin{align}
X^{(k)} (z_0, \ldots, z_{L+1}) &=\frac{\bra{\Psi (z_0, \ldots, z_{L+1})} \hat{X}^{(k)} \ket{ \Psi (z_0, \ldots, z_{L+1})}}{\braket{\Psi (z_0, \ldots, z_{L+1})}{\Psi (z_0, \ldots, z_{L+1})}} \; ,\\
Y^{(k)} (w, z_0, \ldots, z_{L+1}) &=\frac{ \bra{\Psi (z_0, \ldots, z_{L+1})} \hat{Y}^{(k)}(w, z_0, \ldots, z_{L+1}) \ket{ \Psi (z_0, \ldots, z_{L+1})} }{\bra{\Psi (z_0, \ldots, z_{L+1})} T(w, z_0, \ldots, z_{L+1})\ket{\Psi (z_0, \ldots, z_{L+1})}} \; .
\end{align}
Here $\hat{X}^{(k)}$ is a matrix, acting on $V_L$. The nonzero elements of $\hat{X}^{(k)}$ are $\pm 1$.  It has nonzero elements between $\bra{\alpha} \in dLP_L^\ast$ and $\ket{\beta} \in dLP_L$ if  in the $\braket{\alpha}{\beta}$ link pattern, a path is formed through the $k$th site, between the boundaries. The sign is chosen according to the spin-1 property. $\hat{Y}^{(k)}$ has a more complicated structure, as only the two link patterns do not tell information about the connectivity through a horizontal edge. $\hat{Y}^{(k)}$ is a modified $T$-matrix, where every $T$-matrix configuration is multiplied by $0$ or $\pm 1$, according to the current. \\
$\hat{X}^{(k)}$ does not depend on the auxiliary rapidity $w$, because $\hat{X}^{(k)}$ and $T$ commutes. However, this is not true for $\hat{Y}^{(k)}$, hence the explicit $w$ dependence. \\

\subsection{Main result}

Our main result is an explicit expression for both $X$ and $Y$ in an inhomogeneous dilute $O(1)$ model, defined on a strip,
infinite in the vertical direction (both upward and downward), with a finite width $L$. In order to present both expressions,
first we introduce some auxiliary functions. We  use the standard definition of the elementary symmetric functions:
\begin{equation}
  e_k (z_1, \ldots, z_n) = \sum_{1 \le i_1 < \ldots < i_k \le n} z_{i_1}\ldots z_{i_n}  \; \text{ for $1 \le k \le n$ , otherwise $0$ ,}
\end{equation}
and we introduce elementary symmetric functions over $z_i$ and 
$z_i^{-1}$:
\begin{equation}
 E_i(z_0, z_1, \ldots, z_L, z_{L+1}) =e_i \left( z_0, z_1, \ldots, z_L, z_{L+1}, z_0^{-1}, z_1^{-1}, \ldots, z_L^{-1}, z_{L+1}^{-1} \right) \; .
\end{equation}
These symmetric polynomials take the same value for indices $i$ and $L+1-i$:
\begin{equation}
 E_{i} (z_0, z_1, \ldots, z_L, z_{L+1}) = E_{L+1-i} (z_0, z_1, \ldots, z_L, z_{L+1}) \; .
\end{equation}
%
%
%

Define an auxiliary polynomial:
\begin{multline}
 P^p_L (z_0, z_1, \ldots ,z_{L+1}) = -\frac{\ii^{L+1}}{2 (\omega - \omega^{-1})} \left( \prod_{j=0}^{L+1} \frac{(\omega z_j + \ii)(\omega^{-1} z_j - \ii)}{z_j} \; - \right. \\ \left. \prod_{j=0}^{L+1} \frac{(\omega^{-1} z_j + \ii)(\omega z_j - \ii)}{z_j} \right) \; .
\end{multline}
With these definitions, Garbali and Nienhuis found closed expressions for the empty element and the partition sums (in \cite{garbali2017dilute1, garbali2017dilute2}, where alternative definitions are also presented):
\begin{align}
 \psi_{L,EE} (z_0, \ldots, z_{L+1}) &=\frac{\det_{1 \leq i,j \leq L+1} \left( E_{3j-2i}-E_{3j+2i-4(L+2)} \right) }{P_L^p (z_0, \ldots, z_{L+1})} \; ,\\
 \label{eq:Zhs}
  Z_{L,h.s.} (z_0, \ldots, z_{L+1}) &=2^L \psi_{L,EE} (z_0, \ldots, z_{L+1}) \; ,\\
  Z_{L,f.s.} (z_0, \ldots, z_{L+1}) &=2^L \left( \psi_{L,EE} (z_0, \ldots, z_{L+1}) \right)^2 \; .
\end{align}
In this paper, under some technical assumption, we will prove that, on an inhomogeneous lattice of width $L$:
\begin{align}
 X_L^{(i)} (z_0, \ldots, z_{L+1}) &= \frac{1- 2\omega}{2} \left( z_i - \frac{1}{z_i} \right) \frac{1}{E_1 (z_0, \ldots, z_{L+1})} \; ,
 \label{eq:X}\\ \nonumber \\
 Y_L^{(i)} (w, z_0, \ldots, z_{L+1}) &=3\; (-1)^{L+1} \left( w - \frac{1}{w} \right)^2  \times \nonumber \\ 
 \label{eq:Y}
 &\times \frac{{w^{2(L+2)}}}{ W_Y  (w, z_0, \ldots, z_{L+1})} \; \frac{\psi_{L+2,EE} (w,-w, z_0, \ldots, z_{L+1})}{E_1 (z_0, \ldots, z_{L+1}) \; \psi_{L,EE} (z_0, \ldots, z_{L+1}) }  
 \; .
 \end{align}
%
%
Note that $Y_L^{(i)}$ is in fact independent of the index $i$. Here $W_Y (w, z_0, \ldots, z_{L+1})$ is an auxiliary function:
\begin{equation}
\label{eq:WYdef}
W_Y  (w, z_0, \ldots, z_{L+1}) =\prod_{i=0}^{L+1} W_R (z_i,w) W_R (w^{-1},z_i)+\prod_{i=0}^{L+1} \tilde{W}_R (z_i,w) \tilde{W}_R (w^{-1},z_i) \; .
\end{equation}
Notice that $W_Y$ is a symmetrized version of the $T$-matrix normalization $N_L$. Instead of the $K$-matrix normalization, it contains $R$-matrix normalization, belonging to the boundary rapidities. 
$X$ can be written in the following form:
\begin{equation}
 X_L^{(i)} (z_0, \ldots, z_{L+1}) =\frac{1-2 \omega}{2} z_i \frac{\partial}{\partial z_i} \log E_1 (z_0, \ldots, z_{L+1}) \; .
\end{equation}
%

\section{Recursion relations}
\label{sec:RS1}

In this section, we derive recursion equations, relating
systems with sizes $L$ and $L-1$. 
For a special ratio of the rapidities of two consecutive sites, the possible configurations of the two sites are restricted in such a way that they act as a single site. 
The same mechanism works on the boundary, by setting the boundary rapidity and the first (last) rapidity 
to the special ratio. This allows us to relate systems of different sizes $L$ and $L-1$ to each other by the recursion relations. Similar equations hold for the dense loop model \cite{qasimi2017skein}. \\
The polynomial weights of the configurations become $0$, if the configuration is not allowed, and 
factorize into a product of a symmetric prefactor and the polynomial weight of the smaller configuration, if 
it is allowed.\\
This situation is well known in IQFT literature, usually referred as fusion equation and boundary fusion equation \cite{ghoshal1994boundary}.

\subsection{Fusion equation}
\label{sec:RS1bulk}

The $R$-matrix factorizes into the product of two 'triangle operators', 
if we set the two  variables to $z \omega^{-1}$ and $z \omega$:
\begin{multline}
 R \left( z \omega^{-1}, z \omega \right)= (-1-\omega) M \cdot S = \\
 (-1-\omega) \left( \Mmxe + \Mmxlr + \Mmxtl + \Mmxtr \right) \cdot \left( \Smxe + \Smxlr + \Smxlb + \Smxrb \right) 
\end{multline}
  \par
  \begin{minipage}{\linewidth}
            \centering
           \psfrag{a}{$z \omega$}
	    \psfrag{b}{$\!\!\!\!\!\! z \omega^{-1}$}
	    \psfrag{c}{$z$}
	    \psfrag{A}{$=$}
	    \includegraphics[scale=0.5]{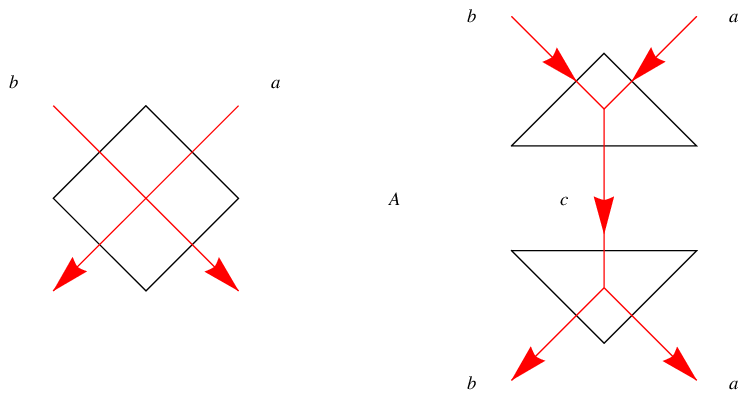}
       \end{minipage}

With the help of $M$, the fusion equation holds:
\begin{equation}
\label{eq:fusion}
 R_i (z \omega, w) R_{i+1} (z \omega^{-1}, w) M_i = 2 \frac{(w - z) (w + z)}{z^2} M_i R_i (z, w) \; .
\end{equation}
  \par
  \begin{minipage}{\linewidth}
            \centering
           \psfrag{a}{$z $}
	    \psfrag{b}{$\! w$}
	    \psfrag{c}{$z \omega$}
	    \psfrag{d}{$z \omega^{-1}$}
	    \psfrag{A}{$=$}
	    \includegraphics[scale=0.5]{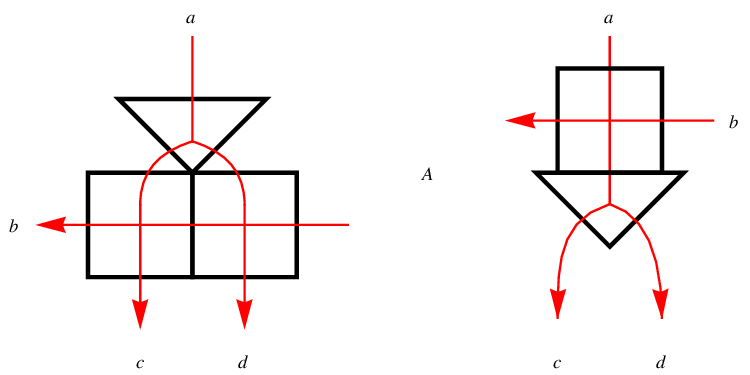}
       \end{minipage}

This is derived in \cite{garbali2017dilute1, garbali2017dilute2}, and we give a detailed proof in Appendix~\ref{app:RecRelProof}.
Setting $z_i=z \omega$ and $z_{i+1}=z \omega^{-1}$ means that we can use the fusion equation from row to row, which
effectively decreases the system size by one. In fact, the 'triangle operators' are intertwiners between $dLP_{L}$ and $dLP_{L-1}$. This relates the transfer-matrices:
\begin{equation}
\label{eq:YenBTmx}
 M_i T_L (\ldots,z_{i-1}, z \omega, z \omega^{-1},z_{i+2}, \ldots)= T_{L-1} (\ldots,z_{i-1}, z,z_{i+2}, \ldots) M_i \; .
\end{equation}
%
%
Act by both sides on $\ket{\Psi_L (\ldots, z \omega, z \omega^{-1}, \ldots)}$ and using that 
$\ket{\Psi_L}$ is the eigenvector of $T_L$:
\begin{equation}
 N M_i \ket{\Psi_L  (\ldots,  z \omega, z \omega^{-1}, \ldots) } = T_{L-1} (\ldots, z, \ldots) \left( M_i \ket{\Psi_L (\ldots, z \omega, z \omega^{-1}, \ldots) } \right) \; .
\end{equation}
Hence, using the uniqueness of the eigenvector:
\begin{multline}
 M_i \ket{\Psi_L (\ldots,z_{i-1},  z \omega, z \omega^{-1},z_{i+2}, \ldots)} = \\=F(z;z_0, \ldots, z_{i-1}, z_{i+2}, \ldots, z_{L+1}) \ket{\Psi_{L-1} (\ldots,z_{i-1}, z,z_{i+2}, \ldots)} \; ,
\end{multline}
where $F(z;z_0, \ldots, z_{i-1}, z_{i+2}, \ldots, z_{L+1})$ is a proportionality factor \cite{garbali2017dilute1}:
\begin{equation}
\label{eq:Fpropfact}
 F(z; z_1, \ldots, z_n) = \prod_{j=1}^n E_1(z,z_j) = \prod_{j=1}^n \frac{(1+z_j z) (z_j + z)}{z_j z} \; .
\end{equation}
How $M_i$ maps from $dLP_L$ to $dLP_{L-1}$ depends on the link pattern on the two sites 
(Here '$|$' denotes '(' or ')' without specification.): 
\begin{itemize}
 \item Mapping to empty site: 
 \begin{align*}
  M:\quad &\ket{\ldots()\ldots} \rightarrow \ket{\ldots \bullet \ldots} \\
	        &\ket{\ldots\bullet\bullet \ldots} \rightarrow \ket{\ldots\bullet \ldots }	
 \end{align*}

 \item Mapping to one occupied, one empty site: 
  \begin{align*}
  M:\quad &\ket{\ldots\bullet| \ldots} \rightarrow \ket{\ldots| \ldots} \\
	        &\ket{\ldots|\bullet \ldots} \rightarrow \ket{\ldots| \ldots}	
 \end{align*}
 \item Disappearing elements: 
  \begin{align*}
  M:\quad &\ket{\ldots )( \ldots} \rightarrow 0 \\
  M:\quad &\ket{\ldots )) \ldots} \rightarrow 0 \\
  M:\quad &\ket{\ldots (( \ldots} \rightarrow 0
 \end{align*}
\end{itemize}
This recursion relation can be understood by considering the following $q$KZ equation:
\begin{equation}
 R_{i,i+1} \left( z \omega^{-1}, z \omega \right) \ket{\Psi_L \left( \ldots  z \omega^{-1}, z \omega \ldots \right)}=
 \ket{\Psi_L \left( \ldots  z \omega, z \omega^{-1} \ldots \right)} \; .
\end{equation}
Since $W_2 \left( z \omega^{-1}, z \omega \right) = 0 $, it is 
impossible to have two not connected line at sites $i, i+1$, effectively decreasing the size by one.
\subsection{Boundary recursion relations}

Based on a very similar argument, as in Section~\ref{sec:RS1bulk}, we have a recursion relation involving the $K$-matrix.
Setting $z_0= z \omega$, $z_1=z \omega^{-1}$, or $z_L= z \omega$, $z_{L+1}= z \omega^{-1}$ 
effectively decreases the size of the system by one.
The reasoning is basically identical for the left and right boundary, so here we present only the one for the 
left side.\\
Setting $z_0= z \omega$, $z_1=z \omega^{-1}$, the left $K$-matrix factorizes into an upper and a lower triangle:
\begin{multline}
 K_l \left( z \omega^{-1}, z \omega \right) = -\frac{-1+2 \omega + z^2 + \omega z^2}{-1+ \omega + z^2} L_l \cdot U_l = \\
 = -\frac{-1+2 \omega + z^2 + \omega z^2}{-1+ \omega + z^2} \left(\, \Lmxe + \Lmxl \,\right)  \cdot \left(\, \Umxe + \Umxl \,\right)  
\end{multline}
%

  \par
  \begin{minipage}{\linewidth}
            \centering
           \psfrag{a}{$z \omega$}
	    \psfrag{b}{$z \omega^{-1}$}
	    \psfrag{c}{$z^{-1} \omega$}
	    \psfrag{d}{$z$}
	    \psfrag{e}{$z$}
	    \psfrag{A}{$=$}
	    \includegraphics[scale=0.5]{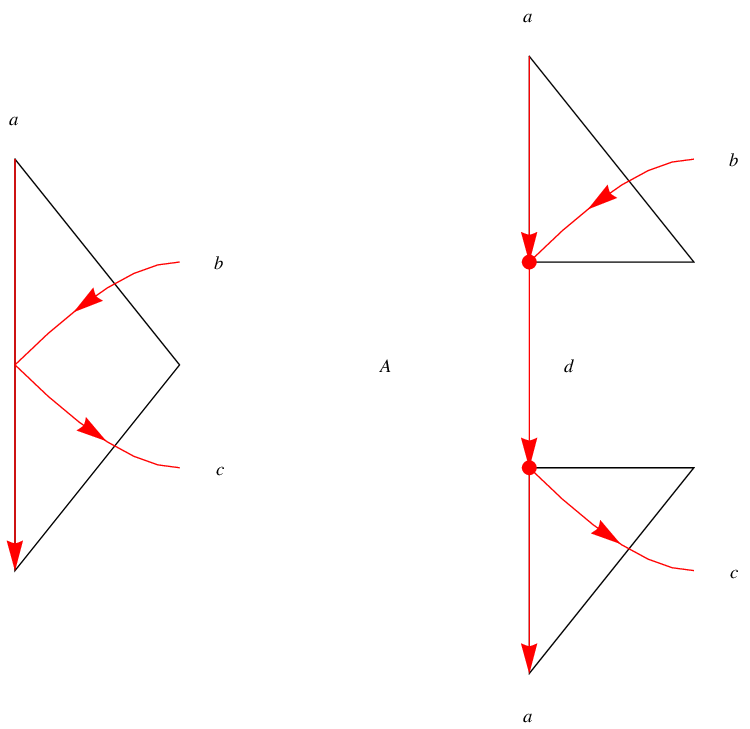}
       \end{minipage}

Using the operator $U$, the boundary reflection holds:
\begin{multline}
 U_l R_1 \left( w^{-1}, z \omega^{-1} \right) K_l (w,z \omega) R_1 \left(z \omega^{-1}, w \right) =\\ =
 \frac{\omega (w + z) (-1 + w z) w^2 (1 + z^2)^2}{z^2 (-w + \omega w - \omega z + \omega w^2 z - w z^2)} K_l (w, z) U_l
\end{multline}

  \par
  \begin{minipage}{\linewidth}
            \centering
           \psfrag{a}{$z \omega$}
	    \psfrag{b}{$w$}
	    \psfrag{c}{$w^{-1}$}
	    \psfrag{d}{$z \omega^{-1}$}
	    \psfrag{e}{$z$}
	    \psfrag{A}{$=$}
	    \includegraphics[scale=0.5]{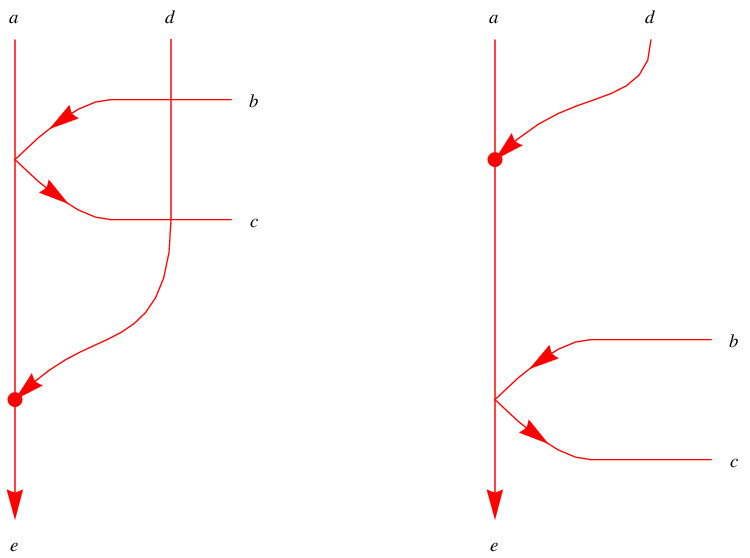}
       \end{minipage}

This can be proved by combining the fusion equation and the open boundary $K$-matrix construction described in Appendix~\ref{app:KmxConstr}. 
Thus, the following intertwining equation holds:
\begin{equation}
\label{eq:bYenBTmx}
 L_l T_L (w, z \omega, z \omega^{-1},z_2, \ldots)= T_{L-1} (w, z,z_2  \ldots) L_l \; .
\end{equation}
%
%
Similarly, as before, $L_l$ effectively decreases the system size by one:
\begin{equation}
 L_l \ket{\Psi_L ( z \omega, z \omega^{-1}, z_2, \ldots)} =F(z,z_2, \ldots, z_{L+1}) \ket{\Psi_{L-1} (z, z_2 \ldots)} \; ,
\end{equation}
%
%
where $F(z,z_0, \ldots, z_{i-1}, z_{i+2}, \ldots, z_{L+1})$ is the same proportionality factor (Eq.~(\ref{eq:Fpropfact})).
This recursion relation decreases the system size by effectively erasing the first site. The rule for the mapping is the following:
\begin{itemize}
	\item Elements with erased first site:
	 \begin{align*}
    L_l:\quad &\ket{\bullet \ldots} \rightarrow \ket{\ldots} \\
	            &\ket{)\ldots} \rightarrow \ket{\ldots}	
 \end{align*}
	\item Disappearing elements:
	 \begin{align*}
	  L_l:\quad &\ket{( \ldots} \rightarrow 0 
	 \end{align*}
\end{itemize}
A completely analogous derivation holds for the right hand side. This means that based on the fusion and the boundary fusion equations,
concerning the recursion, we can treat all the rapidities and boundary rapidities on the same footing, and in the following, we
do not have to distinguish them.

Based on these results, the following recursion relations can be derived for the partition sums:
\begin{align}
 Z_{L,h.s.} \left( \ldots z_{i-1},z \omega , z_ \omega^{-1}, z_{i+2} \ldots \right) &= 2 \left (\prod_{k \neq i,i+1} E_1 (z, z_k) \right) Z_{L-1,h.s}  \left( \ldots z_{i-1}, z, z_{i+2}, \ldots \right) \; , \\
 \label{eq:ZfsRS1}
 Z_{L,f.s.} \left( \ldots z_{i-1}, z \omega , z \omega^{-1}, z_{i+2} \ldots \right) &= 2 \left( \prod_{k \neq i,i+1} E_1^2 (z, z_k) \right) Z_{L-1,f.s} \left( \ldots z_{i-1}, z, z_{i+2}, \ldots \right) \; .
\end{align}
The factor $2$ is coming from the property that every smaller ground state element has two possible source in the larger system. 
The $F^2$ proportionality factor for $Z_{L,f.s.}$ is a result of it being a sum of products of two ground state elements. The recursion acts on the dual space $dLP_L^\ast$ in the flipped order, hence the recursion only takes place after acting with $M$ to project out the extra site. 

%
%
%
%

\section{Symmetries}
\label{sec:Sym}

The expectation values $Y^{(k)}_L$ and $X^{(k)}_L$ exhibit a number of symmetries, most arising from the unitary relations and the $q$KZ and boundary $q$KZ equations. $X^{(k)}$ and $Y^{(k)}$ are symmetric under $z_i \leftrightarrow z_j$ (for $X^{(k)}$, $i,j \neq k$), and under $z_i \rightarrow z_i^{-1}$ (for $X^{(k)}$, $i \neq k$). \\
 We prove these symmetries in the following steps: \\
(i) First, we prove that $X^{(k)}$ and $Y^{(k)}_L$ are symmetric under $z_i \leftrightarrow z_j$ for two disjoint sets of indices, namely
\begin{align}
 \nonumber X^{(k)} :& \quad 1 \leq i,j \leq k-1 \text{ or}\\
 \nonumber         &  \quad k+1 \leq i,j \leq L \\
 \nonumber \\
 \nonumber Y^{(k)} :& \quad 1 \leq i,j \leq k-1 \text{ or}\\
 \nonumber         &  \quad k \leq i,j \leq L
\end{align}
We did not find a way to expand the symmetry to the boundary rapidities with indices $0$ and $L+1$. We checked these symmetries numerically and analytically for small systems. \\
(ii) Using this, we prove the $z_i \rightarrow z_i^{-1}$ symmetry, for $X^{(k)}$, $0 \leq i \leq L+1, \, i \neq k$, and for $Y^{(k)}$, $0 \leq i \leq L+1$. \\
(iii) We use the $z_i \rightarrow z_i^{-1}$ inversion symmetry as an ingredient to prove the full symmetry of $Y^{(k)}_L$ under $z_i \leftrightarrow z_j$ without restriction on the indices. \\
As we did not find a proof for the symmetry in the bulk and boundary rapidity, and the full symmetry of $X^{(k)}_L$, we use these as necessary assumptions  during the proof of the main result for $X$ and $Y$.

\subsection{Partial symmetries of $X_L$ and $Y_L$}
\label{sec:XYsymPartial}

The symmetry of $X^{(k)}$ and $Y^{(k)}$ under $z_i \leftrightarrow z_j$ on the two separate side of $k$ is based on the unitarity of the $R$-matrix (Eq.~\ref{eq:RReqid}), and the $q$KZ equation (Eq.~\ref{eq:qKZeq}). The proof is the same for both expressions, so here we present the one for $X$. Use the operator form of $X$ (suppressing irrelevant notations and normalization):
\begin{equation}
\begin{split}
 X^{(k)} ( z_i, z_{i+1} ) 
\simeq& \bra{\Psi(z_i, z_{i+1})} \hat{X}^{(k)} \ket{\Psi(z_i, z_{i+1})} \simeq \\ 
\simeq&  \bra{\Psi(z_i, z_{i+1})} \hat{X}^{(k)} R_{i,i+1} (z_{i+1}, z_i) R_{i,i+1} (z_i, z_{i+1}) \ket{\Psi(z_i, z_{i+1})} \simeq \\
\simeq& \bra{\Psi(z_i, z_{i+1})} R_{i,i+1} (z_{i+1}, z_i) \hat{X}^{(k)} \ket{\Psi(z_{i+1}, z_i)} \simeq \\
\simeq& \bra{\Psi(z_{i+1}, z_i)} \hat{X}^{(k)} \ket{\Psi(z_{i+1}, z_i)} \simeq  X^{(k)} ( z_{i+1}, z_i ) \; .
\end{split}
\end{equation}
The $R$-matrix and the $\hat{X}^{(k)}$ operators commute, if $i+1 < k$ or $k<i$. By repeating this procedure, we extend the symmetry to ${z_1, \ldots, z_{k-1}}$ and ${z_{k+1}, \ldots, z_{L}}$. We did not find a way to extend the proof to the $z_0$, $z_{L+1}$ boundary rapidities. 
The same proof with the same flaw stands for $Y^{(k)}$, the only difference is in the restriction for the indices, i.e. for $z_i \leftrightarrow z_{i+1}$, $i+1 \leq k$ or $k <i$. \\
To give evidence of the symmetry involving the bulk rapidities, we made the following analytic and numerical check: We checked analytically the expression for $X$ for $L=1,2,3$. We checked analytically the expression for $Y$ for $L=0,1$. For larger system sizes $L=2, 3$, we made numerical checks. Due to the largely increasing terms --especially in $Y$-- in these expressions, these checks are strong evidences supporting our assumption. \\
The symmetry of $X$ and $Y$ under $z_i \rightarrow z_i^{-1}$ is based on the unitarity of the $K$-matrix (Eq.~(\ref{eq:leftKKid},~\ref{eq:rightKKid})) and on the boundary $q$KZ equations (Eq.~(\ref{eq:leftbqKZeq},~\ref{eq:rightbqKZeq})). The proof is the same for $X^{(k)}$ and $Y^{(k)}$, so we present it for $X^{(k)}$. Derive the inversion symmetry for $z_1$ , using the unitarity of the left $K$-matrix (assume $k \neq 1$):
\begin{equation}
\begin{split}
  X^{(k)} &( z_1 ) 
\simeq \bra{\Psi(z_1)} \hat{X}^{(k)} \ket{\Psi(z_1)}  
\simeq  \bra{\Psi(z_1)} \hat{X}^{(k)} K_l (z_1^{-1}, z_0) K_l (z_1, z_0) \ket{\Psi(z_1)} \\
\simeq& \bra{\Psi(z_1)} K_l (z_1^{-1}, z_0) \hat{X}^{(k)} \ket{\Psi(z_1^{-1})} 
\simeq \bra{\Psi(z_1^{-1})} \hat{X}^{(k)} \ket{\Psi(z_1^{-1})} \simeq  X^{(k)} ( z_1^{-1} ) \; .
\end{split}
\end{equation}
The same way, we can prove the $z_L \rightarrow z_L^{-1}$ symmetry, using the unitarity of the right $K$-matrix. The inversion symmetry can be extended to the other rapidities on the two sides of the position of the current by the consecutive application of $z_i \leftrightarrow z_j$ symmetry. By this we can extend the inversion symmetry to $0 \leq i \leq k-1$ and $k+1 \leq i \leq L+1$ for $X^{(k)}$, and for all $0 \leq i \leq L+1$ for $Y^{(k)}$.
\\
Using the additivity property around an elementary plaquette, 
\begin{equation}
X^{(i)\text{(mid)}} - Y^{(i+1)}-X^{(i)\text{(top)}}+Y^{(i)}=0 \; .
\end{equation}
Since $Y$ is position independent (as we will prove in the next section), this implies that $X$ in the middle of the $T$-matrix has the same properties, as the one
on the edge, which has been explicitly computed for $L=1,\,2$.


\subsection{Full symmetry of $Y_L$}
\label{sec:Ysym}

\begin{figure}[htb]
\begin{center}
\includegraphics[scale=0.6]{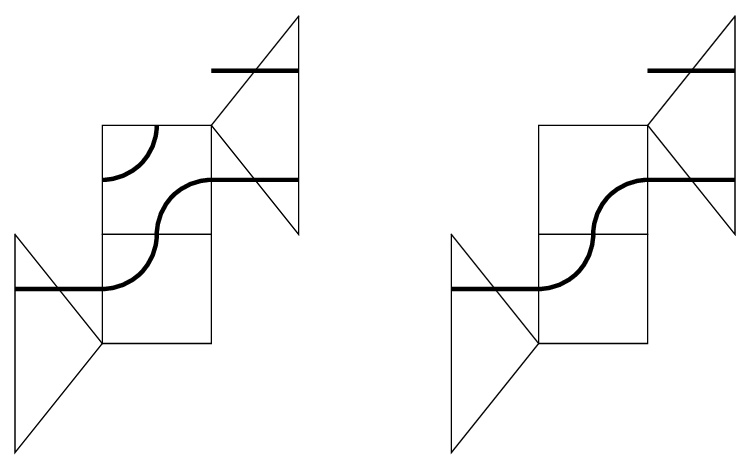}
\caption{These two paths are not equivalent.}
\label{fig:NonEqvPaths}
\end{center}

\end{figure}

In this section, we will show that $Y_L^{(k)}$ is independent of the position
$k$, and symmetric in the variables $z_0, \ldots, z_{L+1}$.\\
Define $p$ as a path going from the left boundary to the right boundary. 
In this section, the path is defined as a set of $K$ and $R$-matrices which constitute the actual path connecting the two boundaries.
Regard two paths to be different, if the actual path connecting the two boundaries are the same, 
but there is difference between the content of the $K$ and $R$-matrices (E.g. in Fig.~\ref{fig:NonEqvPaths}).
By this definition, we identify configurations which only differ in their position, and which are related by a vertical translation.

\begin{figure}[htb]
\begin{center}
\includegraphics[scale=0.6]{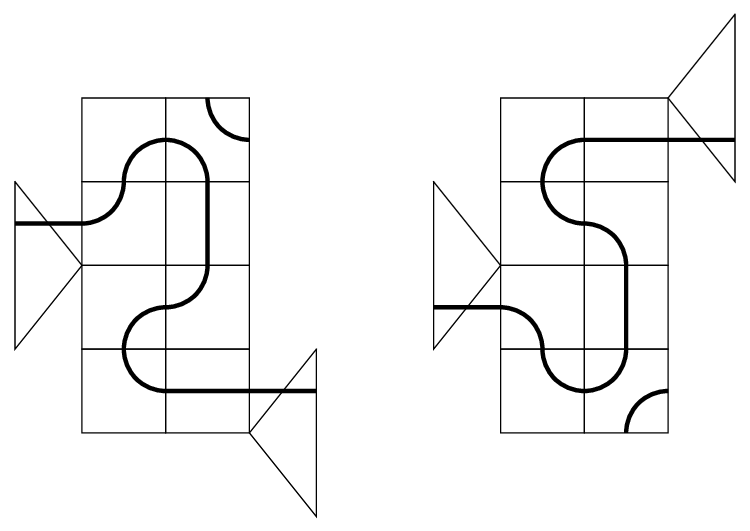}
\caption{Path $p$ and the mirrored path $\tilde{p}$.}
\label{fig:MirroredPaths}
\end{center}
\end{figure}

The weight of the $p$ path is the weight of the constituting matrices. Since $Y_L^{(k)}$ depends only on one auxiliary rapidity, we set 
all the auxiliary rapidities to the same value $w$. Denote the weight of $p$ by $\Omega_p (w,z_0 \ldots z_{L+1})$.
\\
The set of all paths, $P$ is a union of two disjoint sets, $P_T$ and $P_B$. $P_T$ contains the paths starting from the top of the left $K$-matrix, 
$P_B$ contains the ones starting from the bottom of it. 
Every path, $p \in P_T$ is in bijection with a path $\tilde{p} \in P_B$, by a horizontal mirroring (as in Fig.~\ref{fig:MirroredPaths}).
By the properties of the $R$ and $K$-matrices, 
it is easy to see that $\Omega_p (w,z_0 \ldots z_{L+1})=\Omega_{\tilde{p}} (w,z_0^{-1} \ldots z_{L+1}^{-1})$.

\begin{figure}[htb]
\begin{center}
\includegraphics[scale=0.6]{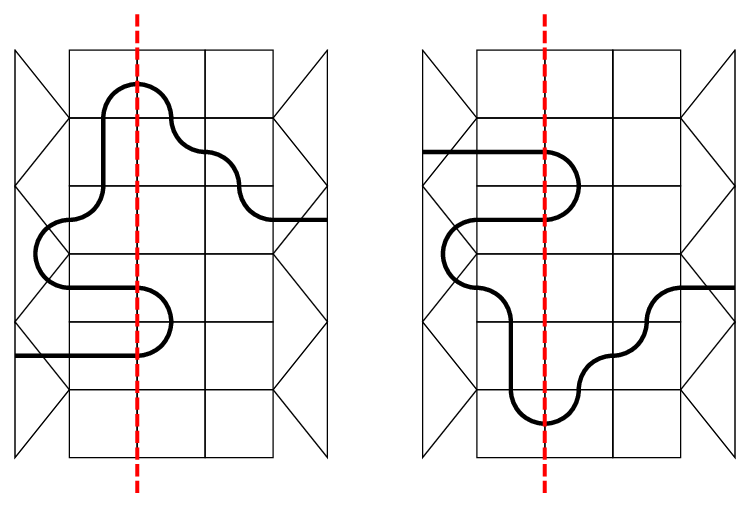}
\caption{$m_{p,k=1,T}=2$, $m_{p,k=1,B}=-1$, $m_{\tilde{p},k=1,T}=-1$, $m_{\tilde{p},k=1,B}=2$}
\label{fig:PathSumRule}
\end{center}
\end{figure}

Introduce $m_{p,k,x}$, where $p$ denotes the path, $0 \le k \le L$ denotes the horizontal position of $Y_L^{(k)}$, 
and $x \in {T,B}$ denotes 'top' or 'bottom', respectively.
Define $m_{p,k,x}$ as the signed crossing of the path $p$ at horizontal line $k$ on the 'top' or 'bottom' section, 
i.e. at the top or bottom of the double row transfer-matrix (See in Fig.~\ref{fig:PathSumRule}).
Signed crossing means that if the line crosses from left to right, it counts as $1$, if  from right to left, it counts as $-1$.
Since a path is crossing once more from left to right then to right to left, $m_{p,k,T}+m_{p,k,B}=1$ (E.g. in Fig.~\ref{fig:PathSumRule}).

By the mirroring, the crossing from the top of the path $p$ maps to the crossing to the bottom of the path $\tilde{p}$, and vice versa.
This means that $m_{p,k,x}=m_{\tilde{p},k,\bar{x}}$, where $\bar{T}=B$, $\bar{B}=T$. 
Combining these features, it is clear that $m_{p,k,x}+m_{\tilde{p},k,x}=1$.
Denote by $Y_L^{(k,T)}$ and $Y_L^{(k,B)}$ the current through the top and the bottom of the $T$-matrix, respectively.
By this definitions, the $Y$ current is given by:
\begin{equation}
\begin{split}
 Y_L^{(k,x)} &(w,z_0, \ldots z_{L+1}) = \\
 =&\sum_{p \in P_T} m_{p,k,x} \Omega_p \left( w,z_0, \ldots z_{L+1} \right) + 
 \sum_{\tilde{p} \in P_B} m_{\tilde{p},k,x} \Omega_{\tilde{p}} \left( w,z_0, \ldots z_{L+1} \right) =\\
 =&\sum_{p \in P_T} \left( m_{p,k,x} \Omega_p \left( w, z_0 , \ldots z_{L+1} \right) + 
  m_{\tilde{p},k,x} \Omega_{p} \left( w, z_0^{-1}, \ldots z_{L+1}^{-1}\right) \right) \; .
\end{split}
\end{equation}
Using that $Y$ is symmetric under $z_i \rightarrow z_i^{-1}$, we consider the following construction:
\begin{equation}
\begin{split}
 Y_L^{(k,x)} & (w,z_0, \ldots z_{L+1}) =
 \frac{1}{2} \left( Y_L^{(k,x)} \left( w, \left\lbrace z_i \right\rbrace \right) +
 Y_L^{(k,x)} \left( w, \left\lbrace z_i^{-1} \right\rbrace \right) \right) = \\ =&
 \frac{1}{2} \left( \sum_{p \in P_T} m_{p,k,x} \Omega_p \left( w, \left\lbrace z_i \right\rbrace \right)+ 
 m_{\tilde{p},k,x} \Omega_p \left( w, \left\lbrace z_i^{-1} \right\rbrace \right)  +   \right. \\
  & \qquad \left.  +
 \sum_{p \in P_T} m_{p,k,x} \Omega_p \left( w, \left\lbrace z_i^{-1} \right\rbrace \right)+ 
 m_{\tilde{p},k,x} \Omega_p \left( w, \left\lbrace z_i \right\rbrace \right)
 \right) = \\ 
 =& \frac{1}{2} \left( \sum_{p \in P_T} (m_{p,k,x} + m_{\tilde{p},k,x}) \Omega_p \left( w, \left\lbrace z_i \right\rbrace \right) + \sum_{p \in P_T} (m_{p,k,x} + m_{\tilde{p},k,x}) \Omega_p \left( w, \left\lbrace z_i^{-1} \right\rbrace \right) 
 \right) = \\
 =&\frac{1}{2} \sum_{p \in P_T} \Omega_p \left( w, \left\lbrace z_i \right\rbrace \right) + 
 \Omega_p \left( w, \left\lbrace z_i^{-1} \right\rbrace \right) \; .
\end{split}
\end{equation}
It thus follows that each path in $P_T$ contribute to $Y$, by the average of the weight of $p$ and $\tilde{p}$. 
It is also clear from this reasoning that $Y^{(k)}$ is independent of $k$. 
This means, there is no further restriction on its symmetries, so it is symmetric in $z_i$, and under $z_i \rightarrow z_i^{-1}$, $\forall i$.

%
%
%
%
%
\section{Proof of the main result}
\label{sec:Proof}
The $X$ and $Y$ currents satisfy the following recursion relations:
\begin{subequations}
\begin{align}
\label{eq:XRS1}
 X_L^{(i)} \left(\ldots, z_{j-1}, z \omega, z \omega^{-1}, z_{j+2}, \ldots\right) & =X_{L-1}^{(i)} (\ldots, z_{j-1}, z, z_{j+2}, \ldots), \quad \forall i \neq j,j+1 \; , \\
 \label{eq:YRS1}
 Y_L \left(\ldots, z_{j-1}, z \omega, z \omega^{-1}, z_{j+2}, \ldots\right)  & =Y_{L-1} (\ldots, z_{j-1}, z, z_{j+2}, \ldots), \quad \forall i \neq j,j+1 \; .
\end{align}
\end{subequations}
%
Note that depending on if $j$ smaller or larger then $i$, the actual position of the current might change, however,
we misuse $X^{(i)}$ to denote both cases. 
Because of the symmetry properties of $X$ and $Y$, the recursion relations extend  to non-adjacent rapidities. 
Our strategy in both cases is the following: First we list the recursion relation for the \emph{unnormalized} expressions $X_{u.n.}$ and 
$Y_{u.n.}$, and based on that, we prove the recursion relation for the normalized ones. \\

\subsection{Proof for the $Y$ current}
\label{sec:Yproof}

In Section~(\ref{sec:Ysym}) we have seen that $Y_L (w, z_0, \ldots , z_{L+1})$ symmetric in all the $z_i$'s and under $z_i \rightarrow z_i^{-1}$.
The unnormalized version of $Y$ is computable in the following way:
\begin{equation}
 Y_L^{(i),u.n.}=\sum_{\alpha, \beta, \gamma} (-1)^{\text{sign}(\alpha, \beta, \gamma)} \psi_{\alpha}^\ast T_{\beta} \psi_{\gamma} \; .
\end{equation}
Here, $\alpha \in dLP_L^\ast$, $\gamma \in dLP_L$ and $T_{\beta}$ is a $T$-matrix configuration which provides the necessary plaquettes
to form a path through the top row of the $T$-matrix, at position $i$, $\text{sign}(\alpha, \beta, \gamma)= \pm 1$ according to the direction.
The relation between the normalized and unnormalized $Y$ current is
\begin{equation}
 Y_L^{(i)}=\frac{Y_L^{(i),u.n.}}{Z_{L,f.s.} N_{L}} \; .
\end{equation}
Here $N_{L}$ is the normalization of the $T$-matrix.\\
Since the recursion relation is known for $Y_L^{(i)}$ (Eq.~\ref{eq:YRS1}) and $Z_{L,f.s}$ (Eq.~\ref{eq:ZfsRS1}):
\begin{equation}
 \left. \frac{Y_L^{(i),u.n.}}{W_{T,L}} \right|_{z_j=z \omega, z_{j+1}=z \omega^{-1}} =2 \left( \prod_{k \neq j,j+1} E_1^2 \left( z,z_k \right) \right)
 \frac{Y_{L-1}^{(i),u.n.}}{W_{T,L-1}} \; .
\end{equation}
Since $Y$ is fully symmetric and $N_L$ is not fully symmetric, $Y^{u.n.}$ is not symmetric.
To exploit the symmetric properties, we introduce auxiliary Laurent polynomial function which is symmetric by construction:
\begin{equation}
\label{eq:tildeYdef}
  \tilde{Y}^{u.n.}_L=Y^{u.n.}_L \frac{W_{Y,L}}{N_{L}} \; .
\end{equation}
$W_Y$ is a predefined quantity (Eq.~\ref{eq:WYdef}) with the following recursion relation (which can be checked by simple calculation):
\begin{align}
  W_{Y,L} (\ldots,z \omega, z \omega^{-1},\ldots) = - w^2 E_1 (w, z) E_1 (-w1, z) W_{Y,L-1} (\ldots,z, \ldots) \; .
\end{align}
Now, we can write up the recursion relation for $\tilde{Y}^{u.n.}$, based on the previous equations:
%
\begin{equation}
 \tilde{Y}^{u.n.}_L(z_i=z \omega \ldots z_j=z \omega^{-1}) = -2 w^2 E_1(w,z) E_1(-w,z) \prod_{k \neq i,j} E_1^2 (z,z_k) \tilde{Y}^{u.n.}_{L-1}(z) \; .
\end{equation}
Since $Y$ is fully symmetric, the two rapidities set to the recursion ratio is not restricted to adjacent rapidities. Because of the symmetry 
under $z_i \rightarrow z_i^{-1}$, the recursion relation holds for two additional values. The full set of recursion relations involving
two chosen variables $z_i$, $z_j$:
\begin{subequations}
\begin{align}
 Y_L (z_i= z \omega \ldots z_j = z \omega^{-1}) &= Y_{L-1} (z) \; , \\
 Y_L (z_i= z \omega^{-1} \ldots z_j = z \omega) &= Y_{L-1} (z) \; , \\
 Y_L (z_i=( z \omega )^{-1} \ldots z_j = z^{-1} \omega) &= Y_{L-1} (z) \; , \\
 Y_L (z_i= z^{-1} \omega \ldots z_j = ( z \omega )^{-1}) &= Y_{L-1} (z) \; .
\end{align}
\end{subequations}
There are four recursion relations involving two chosen variables. Consequently, fully exploiting symmetry and inversion symmetry, there are $4 (L+1)$ recursion relations with one chosen variable $z_k$,  relating $Y_L$ and $Y_{L-1}$. \\
$\tilde{Y}^{u.n.}$ has been calculated for $L=0,1$, giving starting element for the following form:
\begin{align}
 \nonumber
 \tilde{Y}^{u.n.} (w, z_0, \ldots, z_{L+1}) &= 2^{L}\; 3 (-1)^{L+1}\; w^{2(L+2)}  \left( w-\frac{1}{w} \right)^2 \times \\
 & \times \frac{\psi_{L,EE} (z_0 \ldots, z_{L+1}) \psi_{L+2,EE} (w,-w,z_0 \ldots, z_{L+1})}{E_1 (z_0, \ldots, z_{L+1})} \; .
\end{align}
The degree width of this expression is $4(L+1)-1$ for system size $L$, which means that the recursion fully fixes $\tilde{Y}^{u.n.}$ for any
system size, as a Lagrange interpolation. From this, based on Eq.~(\ref{eq:tildeYdef}), we have $Y$ for any system size. 
Since $Y_L=\frac{\tilde{Y}^{u.n.}_L}{Z_{L,f.s.}W_{Y,L}}$, the $Y$ current has the proposed form of Eq. (\ref{eq:Y}):
\allowdisplaybreaks[0]
\begin{multline*}
Y_L^{(i)} (w, z_0, \ldots, z_{L+1}) =3 (-1)^{L+1} \left( w-\frac{1}{w} \right)^2 \times \\
\times \frac{{w^{2(L+2)}}}{ W_Y  (w, z_0, \ldots, z_{L+1})} \frac{\psi_{L+2,EE} (w,-w, z_0, \ldots, z_{L+1})}{E_1 (z_0, \ldots, z_{L+1}) \psi_{L,EE} (z_0, \ldots, z_{L+1}) }  
  \; .
\end{multline*}
By this, based on the recursion relation, under the assumption of the symmetry of bulk and boundary rapidities, we proved that, the unique solution which satisfies Eq. (\ref{eq:YRS1}) with the computed
starting element, is indeed Eq.~(\ref{eq:Y}).

\subsection{Proof for the $X$ current}

As an unnormalized quantity, we can compute $X$ in the following way:
\begin{equation}
 X^{(i)u.n.}_L= \bra{\Psi_L} \hat{X}^{(i)}_L \ket{\Psi_L} = \sum_{\alpha, \beta} (-1)^{\text{sign}(\alpha, \beta)} \psi_{\alpha}^\ast \psi_{\beta} \; ,
\end{equation}
 where $\alpha \in dLP_L^\ast$, $\beta \in dLP_L$, and $\alpha$ and $\beta$ are chosen such that they form a boundary to 
 boundary path through the site $i$, and $\text{sign}(\alpha, \beta)= \pm 1$ according to the direction. 
 The relation between the normalized and unnormalized $X$ current is:
 \begin{equation}
  X^{(i)}_L = \frac{X^{(i)u.n.}_L}{Z_{L,f.s.}} \; .
 \end{equation}
 Since $Z_{L,f.s.}$ is fully symmetric, $X^{(i)}_L$ and $X^{(i)u.n.}$ share the same symmetry properties.
 The recursion relation for $X^{(i)u.n.}$:
\begin{equation}
\label{eq:XunRS1} 
  X^{(i)u.n.}_L (z_j = z \omega \ldots z_k= z \omega^{-1}) = 2 \prod_{n \neq j,k} E_1^2 (z,z_n) X^{(i)u.n.}_{L-1} (z) \; .
\end{equation}
This quantity has been computed explicitly for $L=1,2,3$, and give hint for the following conjectured form:
\begin{equation}
 X^{(i)u.n.}_L (z_0, \ldots, z_{L+1}) = (2-\omega) 2^{L-1} \left(z_i - \frac{1}{z_i} \right)  \frac{\psi_{L,EE}^2 (z_0, \ldots, z_{L+1})}{E_1 (z_0, \ldots, z_L)} \; .
\end{equation}
The computation for $L=1$ is presented in Appendix~\ref{app:L1}. 
Based on the recursion properties of $\psi_{L,EE}$ and $E_1$, this form clearly satisfies Eq.~(\ref{eq:XRS1}). 
To prove the uniqueness of the solution, we will assume the symmetry in all the variables expect $z_i$.
The degree width of $X^{(i)u.n.}$ in any $z_j \neq z_i$ rapidity is $4L-1$. Based on a similar counting as in the $Y$, 
there are $4L$ recursion relations relating systems $L$ and $L-1$. All the arguments for $Y$ hold for $X$, only the number of variables is smaller by one 
(since $X^{(i)}$ is not symmetric in $z_i$, but all the other $z$'s.).
By this, we see that under the aforementioned assumption, we have found the unique solution for $X$ described in
Eq. (\ref{eq:X}):
\begin{equation*}
  X_L^{(i)} (z_0, \ldots, z_{L+1}) = \frac{(1- 2\omega)}{2} \left(z_i - \frac{1}{z_i} \right) \frac{1}{E_1 (z_0, \ldots, z_{L+1})} \; .
\end{equation*}
\section{Conclusion}

We have computed the spin-1 boundary to boundary current for the dilute O($n$=1) loop model, on a strip, with finite width and infinite height, with open boundary conditions. We have conducted the computation for the inhomogeneous case, i.e. our expressions are symmetric rational functions in the rapidities and boundary rapidities of the model. 

Our model has open boundaries, which we have constructed from the closed boundary case, with the insertion of a rapidity line. We have proved the fusion equation of the model too, and we got the boundary fusion equations as corollary. 

\section*{Acknowledgment}

GF is grateful to Alexandr Garbali, Kayed Al Qasimi, Jasper Stokman, Jan de Gier and Paul Zinn-Justin for useful discussion. \\
The final phase of the work of GF was supported by the BME-Nanotechnology FIKP grant of EMMI (BME FIKP-NAT) and by the National Research Development and Innovation Office (NKFIH) under the KH-17 grant no. 125567. 

 \appendix
 \addcontentsline{toc}{section}{APPENDIX}
 
 \section{The dilute O($n$=1) model and the site percolation on a triangular lattice}
 \label{app:Perco}

\begin{figure}
        \centering
        \begin{subfigure}[b]{0.475\textwidth}
            \centering
            \includegraphics[width=\textwidth]{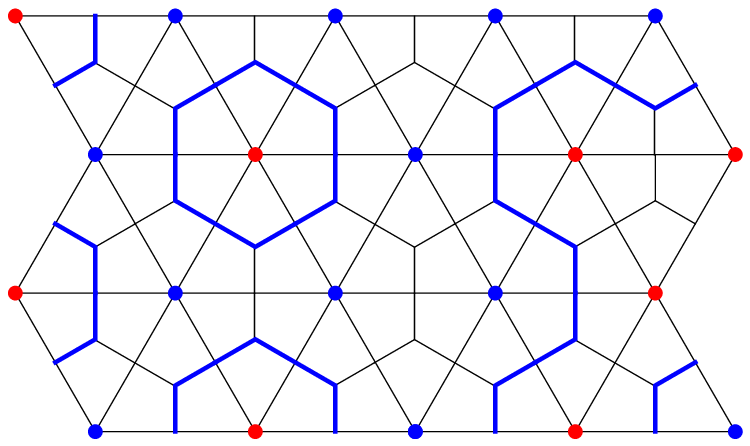}
            \caption[]%
            {{\small The site percolation on a triangular lattice, with the domain walls}}    
            \label{fig:PercoToLoop1}
        \end{subfigure}
        \hfill
        \begin{subfigure}[b]{0.475\textwidth}  
            \centering 
            \includegraphics[width=\textwidth]{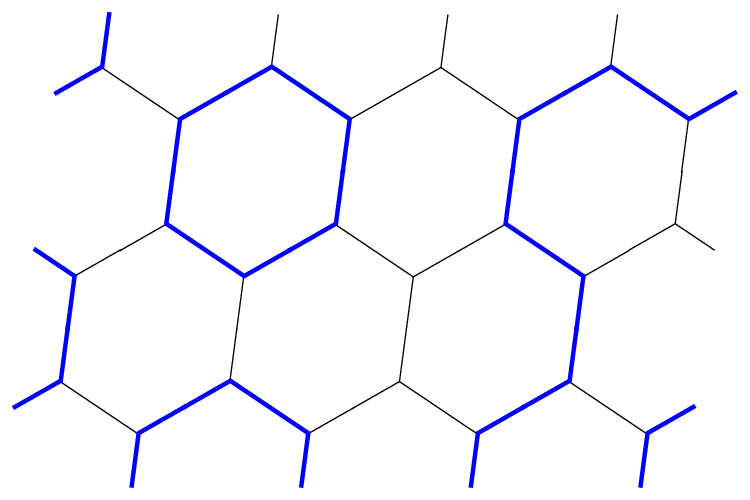}
            \caption[]%
            {{\small The domain walls form loops on a honeycomb lattice}}    
            \label{fig:PercoToLoop2}
        \end{subfigure}
        \vskip\baselineskip
        \begin{subfigure}[b]{0.475\textwidth}   
            \centering 
            \includegraphics[width=\textwidth]{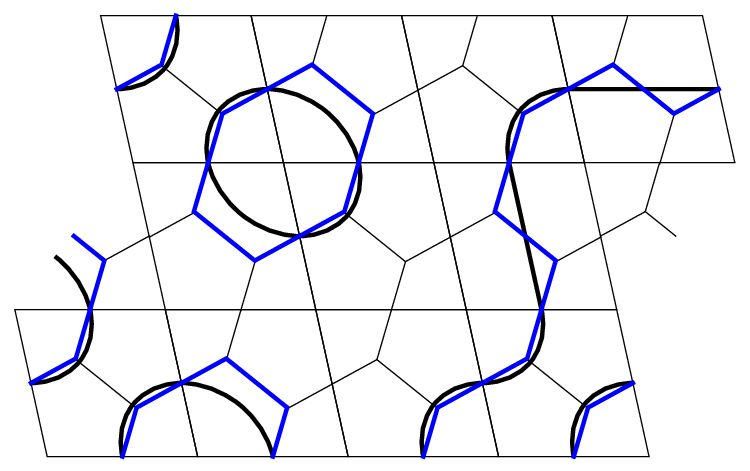}
            \caption[]%
            {{\small This maps to the dilute loops in the appropriate limit}}    
            \label{fig:PercoToLoop3}
        \end{subfigure}
        \quad
        \begin{subfigure}[b]{0.475\textwidth}   
            \centering 
            \includegraphics[width=\textwidth]{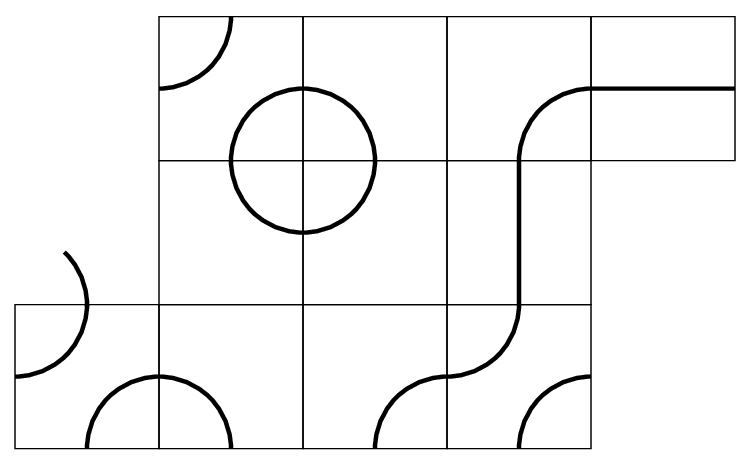}
            \caption[]%
            {{\small The corresponding loop configuration}}    
            \label{fig:PercoToLoop4}
        \end{subfigure}
        \caption[]
        {The mapping of a site percolation configuration on a triangular lattice to a dilute loop configuration on the square lattice. For convenience, the figures are distorted, in order to have the usual triangular and square lattice on the two sides of the mapping.} 
        \label{fig:PercoToLoop}
    \end{figure}

 The dilute O($n$=1) loop model on a rectangular lattice maps to the critical site percolation on a triangular lattice. The site percolation is defined as such: Consider a triangular lattice, where every site is either red or blue, both with $p_c=\frac12$ probability. The loop model in a certain homogeneous limit maps to this percolation model. \\
The mapping is depicted in Fig.~\ref{fig:PercoToLoop}. In the mapping, the loops became the domain walls of the percolation problem. Since the domain walls live on the dual lattice of the triangular lattice, they live on a honeycomb lattice. The mapping takes place in a way that out of the 9 possible plaquettes of the loop model one has to be zero, and all the others have to have equal weights. Since a plaquette belongs to 4 sites of the site percolation, if all the remaining 8 plaquette configurations have equal weights, this gives back independent site probabilities, as the 16 possible 4 site percolation configurations has equal weight, after a factorization of the two colors. Since one of the weights disappears, and the others have equal weight, this corresponds to the loop model after the fusion limit, and the factorization of the $R$-matrix elements. The mapping can be realized by two different way, setting either the weight of $\Rmxtlbr$ or $\Rmxtrbl$ to $0$. This can be realized to set $W_2 (z, w) = 0$ or $W_m(z,w)=0$, which means $w= z q^2$ or $w= z q^{-2}$ respectively. In order to realize this situation on all sites of the lattice model, the rapidities are set to this ratio on all sites. In the figure, the later realization is depicted. Either way, the remaining 8 configurations have equal weights, hence the independent probabilities of the percolation model are guaranteed.

 \section{Normalization of the transfer-matrix}
 \label{app:Norm}
 
 In order to prove the form of the transfer matrix normalization (Eq.~\ref{eq:Tnorm}), first, we map the dilute $O(1)$ loop model
 to a site percolation model, and compute the normalization in the percolation model, using the all-$1$ left eigenvector.\\
 There is a mapping between the dilute $O(1)$ loop model, and an unusual site percolation model on the square lattice, different from the one discussed in the previous section. \\
  The site percolation model is built up from randomly distributed spins taking the values $s = \pm 1$ on the vertices of the tiles.
 The mapping takes place as the paths of the loop model are mapped to the domain walls of the site percolation.
 To implement the $R$ and $K$-matrix weight, we introduce the following plaquette-interactions:
 \begin{itemize}
  \item For the $R$-matrix: $R=a+b s_1 s_2 s_3 s_4$
  \item For the $K$-matrix: $K=A+B s_1 s_3$
 \end{itemize}
 Here $s_1,s_2,s_3,s_4$ are the four spins in the corners of the $R$-matrix, and $s_1,s_3$ are the spins in the vertices in the upper
 and lowermost corner of the $K$-matrix.
 The aforementioned definitions coincide with the plaquette weights, if
\begin{subequations}
 \begin{align}
  a &=\tfrac{1}{2} \left( W_1+W_{t} \right) \equiv \tfrac{1}{2} W_R \; , \\
  b &=\tfrac{1}{2} \left( W_1-W_{t} \right) \equiv \tfrac{1}{2} \tilde{W}_R \; , \\
  A &=\tfrac{1}{2} \left( K_{id}+K_m+K_1 \right) \equiv \tfrac{1}{2} W_K \; , \\
  B &=\tfrac{1}{2} \left( K_{id}+K_m-K_1 \right) \equiv \tfrac{1}{2} \tilde{W}_K \; .
 \end{align}
\end{subequations}
We define a percolation state, as a sequence of spins along the bottom edge of the $T$-matrix.
The percolation state is equal to the sum of loop states which locally realize the required domain walls, irrespectively of the connectivity. This mapping leaves the overall $\mathbb{Z}_2$ symmetry of the percolation state undecided, hence we sum up to it too. E.g.: $\ket{1,-1,1,1}_{\text{perco}} + \ket{-1,1,-1,-1}_{\text{perco}} = \ket{))\bullet} + \ket{)(\bullet} + \ket{()\bullet} + \ket{((\bullet} $. 
Even the states are not in bijection -after summing up to the $\mathbb{Z}_2$ symmetry of the percolation- the $T$-matrix configurations are, consequently the normalization for both $T$-matrices are the same.

\begin{figure}
    \begin{center}
   \includegraphics[scale=0.6]{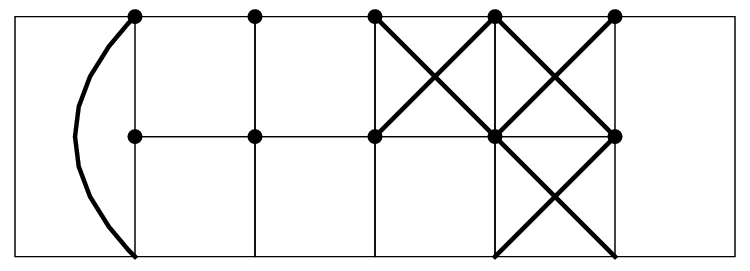}
   \caption{Graphical representation of a summand in the normalization of the percolation transfer-matrix. 
   For better understanding, we denote the $K$-matrices by rectangulars, instead of triangles. The spins --we sum up to-- are emphasized by dots.}
   \label{fig:PercoSum}
   \end{center}
  \end{figure}

 By definition, the $T$-matrix is a left stochastic matrix, so all the columns of it sum to $N_L$. 
 Consequently, the left eigenvector is the $(1,1, \ldots , 1)$ vector. 
 The corresponding $T$-matrix normalization is proportional to the weight of summing over all inner configurations of the $T$-matrix. Summing over all inner configurations means summing over all spins of the $T$-matrix,
 with the exception of the bottom edge (In Fig.~\ref{fig:PercoSum}, the summed over spins are marked by $\bullet$). 
 The weight of a $T$-matrix is $\prod R \prod K$, and the normalization is 
\begin{equation}
  N=\frac{\sum_{\text{all config.}} \prod R \prod K}{\sum_{\text{all config.}} 1}=2^{-2 (L+1)}\sum_{\text{all config.}} \prod_{<i,j,k,l>} (a+b s_i s_j s_k s_l) \prod_{<i,j>} (A+B s_i s_j) \; ,
 \end{equation}
 where $<i,j,k,l>$ and $<i,j>$ denote spins in the four corners of the $R$-matrix, and the two corners of the $K$-matrix, respectively. Expanding the products, the $\prod (a+b s_i s_j s_k s_l )\prod (A + B s_i s_j )$ summands of $N_L$ are polynomials in $s_m$, and because $s_m$ is summed over $+1$ and $-1$, if at least one $s_m$ has odd power,
 the contribution  of that summand cancels out as we sum over all the configurations.  
 It is easy to see that all the summand has at least one odd-powered $s_m$, 
 with the exception of $\prod a \prod A$ and $\prod b \prod B$ by the following argument:
 If we represent the $b s_1 s_2 s_3 s_4$ term by a cross at the given square, and $B s_1 s_3$ by a line connecting $s_1$ and
 $s_3$, a given summand is a partial filling of the $T$-matrix with crosses and lines (Fig.~\ref{fig:PercoSum}), and the power of a spin is equal to the lines starting from that vertex. Clearly, in case of the empty filling, all the spins have even power (and particularly all of them are 0), resulting in $\prod a \prod A$. 
 By putting somewhere a cross or a line, it is clear that the full $T$-matrix has to be filled in order to avoid any vertices with odd number of lines connected to it (during which procedure, we disregard the spins on the bottom edge).
By this, we see that
 the only non-vanishing contributions are $\prod a \prod A$ (the ``empty'' $T$-matrix) and $\prod b \prod B$ (the ``completely filled'' $T$-matrix).
 Consequently, the normalization is 
 $N=2^{-2(L+1)}\sum_{\text{all config.}}\prod a \prod A + \prod b \prod B=\prod a \prod A + \prod b \prod B$. The $2^{-2(L+1)}$ prefactor is canceled by the summation over all configurations. Including inhomogeneous weights, we get the following expression for the normalization:
\begin{equation}
\begin{split}
 N\left( w, z_0, \ldots, z_{L+1} \right) =& W_{K_l} (w, z_0) W_{K_l} (w^{-1}, -z_{L+1}) \prod_{i=1}^L W_R (w,z_i) W_R (z_i,w^{-1}) + \\
  &+\tilde{W}_{K_l} (w, z_0) \tilde{W}_{K_l} (w^{-1}, -z_{L+1}) \prod_{i=1}^L \tilde{W}_R (w,z_i) \tilde{W}_R (z_i,w^{-1}) \; .
  \end{split}
\end{equation}

\section{$L=1$ ground state elements and $X$ current}
\label{app:L1}

As an example, here we present the $L=1$ ground state elements, and the computation of the $X$ current for this case:
\begin{subequations}
\begin{align}
 \psi_{\bullet} (z_0, z_1, z_2) &\equiv \psi_{EE,L=1} (z_0, z_1, z_2) = z_0 + z_1 + z_2 + z_0^{-1} + z_1^{-1} + z_2^{-1} =\\ \nonumber & =\frac{z_0 z_1 + z_0 z_2 + z_1 z_2 + z_0^2 z_1 z_2 + z_0 z_1^2 z_2 + z_0 z_1 z_2^2 }{z_0 z_1 z_2} \\
 \psi_{(} (z_0, z_1, z_2) &= \frac{(\omega z_0 + z_1)(\omega + z_0 z_1)}{\omega  z_0 z_1} \\
 \psi_{)} (z_0, z_1, z_2) &= \frac{(\omega z_1 + z_2)(1 +\omega z_1 z_2)}{\omega  z_1 z_2}.
\end{align}
\end{subequations}
The normalization of the full strip follows by summing up for all the allowed connectivity:
\begin{equation}
\begin{split}
Z_{f.s.,L=1}& (z_0, z_1, z_2) = \psi_\bullet (z_0, z_1, z_2) \psi^\ast_\bullet (z_0, z_1, z_2) +\\
 &+ \left( \psi_( (z_0, z_1, z_2) +\psi_) (z_0, z_1, z_2) \right) \left( \psi^\ast_( (z_0, z_1, z_2) + \psi^\ast_) (z_0, z_1, z_2)  \right) =\\ =&\psi_\bullet (z_0, z_1, z_2) \psi_\bullet (z_2, z_1, z_0) + \left( \psi_( (z_0, z_1, z_2) +\psi_) (z_0, z_1, z_2) \right) \\ & \left( \psi_) (z_2, z_1, z_0) + \psi_) (z_2, z_1, z_0)  \right) = \\
=&\frac{2 (z_0 z_1 + z_0 z_2 + z_1 z_2 + z_0^2 z_1 z_2 + z_0 z_1^2 z_2 + 
   z_0 z_1 z_2^2)^2}{z_0^2 z_1^2 z_2^2}.
\end{split}
\end{equation}
 To compute the $X$ current, we utilize Eq.~\ref{eq:DualEquiv}. The unnormalized current:
\begin{equation}
\begin{split}
 X^{(1)u.n.}_{L=1} (z_0, & z_1, z_2) = \psi_{)}(z_0, z_1, z_2) \psi^\ast_{(}(z_0, z_1, z_2) - \psi_{(}(z_0, z_1, z_2) \psi^\ast_{)}(z_0, z_1, z_2) =\\=& \psi_{)}(z_0, z_1, z_2) \psi_{)}(z_2, z_1, z_0) - \psi_{(}(z_0, z_1, z_2) \psi_{(}(z_2, z_1, z_0) = \\
=& (1-2\omega) \; \frac{(z_1^2-1)(z_0 z_1 + z_0 z_2 + z_1 z_2 + z_0^2 z_1 z_2 + z_0 z_1^2 z_2 + z_0 z_1 z_2^2)}{ z_0 z_1^2 z_2}.
\end{split}
\end{equation}
The normalized current follows after dividing by the partition sum of the full strip:
\begin{equation}
\begin{split}
 X^{(1)}_{L=1} (z_0, z_1, z_2) =& \frac{1-2\omega}2 \left( z_i - \frac{1}{z_i} \right) \; \frac{1}{E_1 (z_0, z_1, z_2)} =\\=&  \frac{1-2\omega}2 \frac{z_1^2-1}{z_1} \; \frac{1}{z_0+z_1+z_2+z_0^{-1}+z_1^{-1}+z_2^{-1}} = \\
=& \frac{1-2\omega}2 \frac{z_1^2-1}{z_1} \; \frac{z_0 z_1 z_2}{z_0 z_1 + z_0 z_2 + z_1 z_2 + z_0^2 z_1 z_2 + z_0 z_1^2 z_2 + z_0 z_1 z_2^2}.
\end{split}
\end{equation}

\section{Proof of the fusion equation}
\label{app:RecRelProof}

In this section, we prove the~(\ref{eq:fusion}) fusion equation:
\begin{equation*}
 R_i (z \omega, w) R_{i+1} (z \omega^{-1}, w) M_i = 2 \frac{(w - z) (w + z)}{z^2} M_i R_i (z, w) \; .
\end{equation*}
  \par
  \begin{minipage}{\linewidth}
            \centering
           \psfrag{a}{$z $}
	    \psfrag{b}{$w$}
	    \psfrag{c}{$z \omega$}
	    \psfrag{d}{$z \omega^{-1}$}
	    \psfrag{A}{$=$}
	    \includegraphics[scale=0.5]{YenBaxterEq.eps}
			\vspace{11pt}
       \end{minipage}
If such an equation holds, every $RR$ configuration of the l.h.s. belongs to one of the nine possible faces of the $R$-matrix. In other words, we should be able to group the $RR$ configurations such a way that the sum of their weight at the fusion values ($z \omega$, $z \omega^{-1}$) are  proportional to the corresponding $R$-matrix weight. First, on the l.h.s. of the fusion equation, we group the $RR$ configurations according to their external connectivity on the five external sides of the l.h.s of Eq.~(\ref{eq:fusion}). This classification puts the possible 41 $RR$ configurations in 21 sets. At the value $z_1 = z \omega$, $z_2 = z \omega^{-1}$, 3 of these sets have vanishing weights. The remaining 18 sets can be grouped into the expected 9 groups, according to the connectivity on the bottom, i.e. two empty sites or the two sites connected to each other turn into an empty site, one empty site and one occupied site turn into an occupied site, connected to the original connection of the occupied site.The grouping is exactly the same, as the elements of $dLP_L$ maps to $dLP_{L-1}$ under the recursion. \\
The disappearing elements are exactly the ones with two not linked lines on the bottom, i.e. which cannot be mapped to a proper one site. \\
The classification is depicted in Tab.~\ref{tab:ClassificationRR}. The first column is the corresponding $R$-matrix (the r.h.s. of the equation), the second and third are the corresponding $RR$ configurations (where we keep the original classification to 18 sets according to the external connectivity on the five external sides of the l.h.s. of Eq.~(\ref{eq:fusion})). The top triangle operator is not shown, as it is uniquely defined by the depicted top two sides. As the triangles have equal weights, we do not have to take them into account at the weight counting. The proportionality factor $2 \frac{(z-w)(z+w)}{z^2}$ is exposed in Eq.~(\ref{eq:fusion}). It is even true that the 18 sets independently proportional to their corresponding $R$-matrix weight, with factor $\frac{(z-w)(z+w)}{z^2}$. \\
All the aforementioned statements are computed directly.

\begin{table}
\begin{tabular}{ l l l }
  $R$- matrix & $RR$ config.  & $RR$ config.\\
  element & $\bullet\bullet \rightarrow \bullet$, $|\bullet \rightarrow | $ & $() \rightarrow \bullet$, $\bullet| \rightarrow | $\\
  \includegraphics[height=2em]{R_e}  & \includegraphics[height=2em]{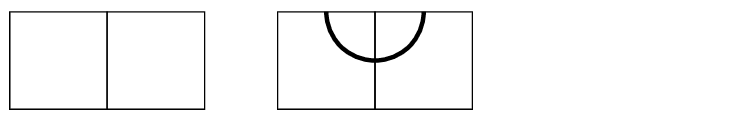} & \includegraphics[height=2em]{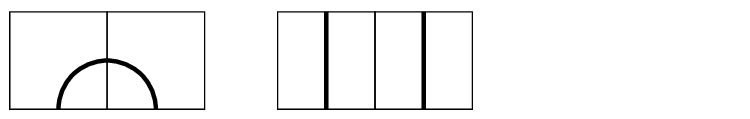} \\
  \includegraphics[height=2em]{R_tl}  & \includegraphics[height=2em]{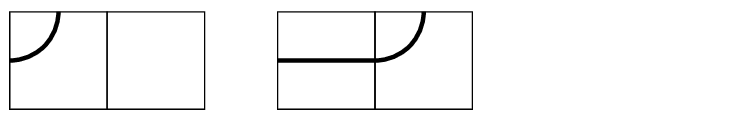} & \includegraphics[height=2em]{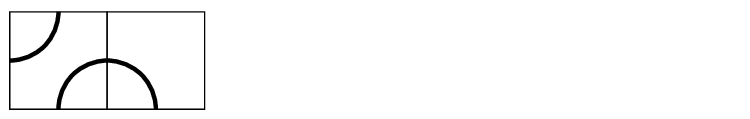} \\
  \includegraphics[height=2em]{R_bl}  & \includegraphics[height=2em]{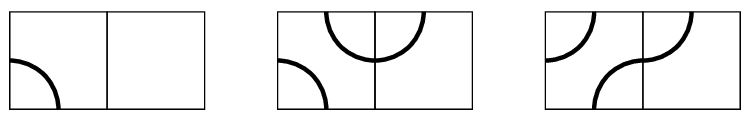} & \includegraphics[height=2em]{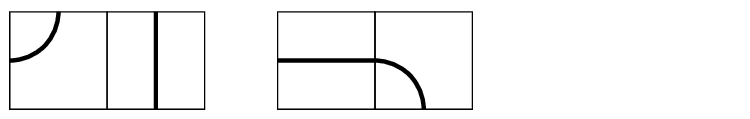} \\
  \includegraphics[height=2em]{R_br}  & \includegraphics[height=2em]{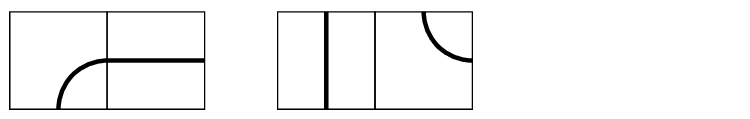} & \includegraphics[height=2em]{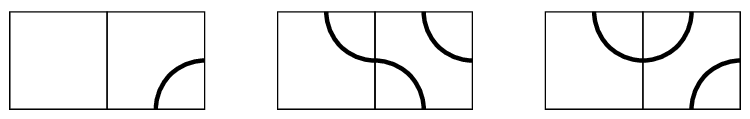} \\
  \includegraphics[height=2em]{R_tr}  & \includegraphics[height=2em]{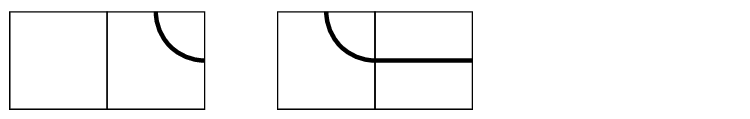} & \includegraphics[height=2em]{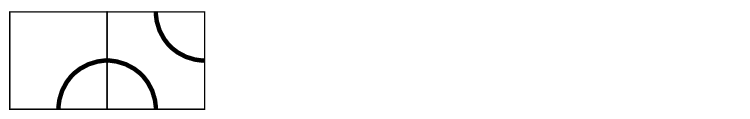} \\
	\includegraphics[height=2em]{R_lr}  & \includegraphics[height=2em]{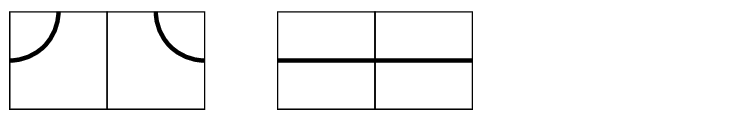} & \includegraphics[height=2em]{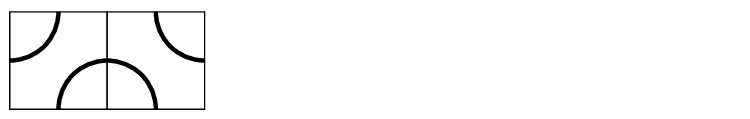} \\
	\includegraphics[height=2em]{R_tb}  & \includegraphics[height=2em]{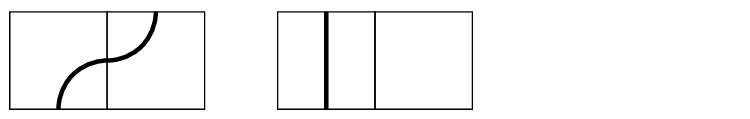} & \includegraphics[height=2em]{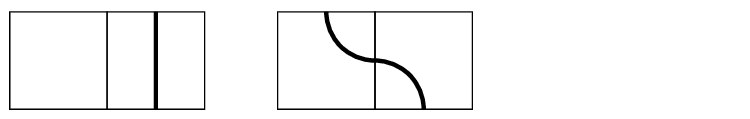} \\
	\includegraphics[height=2em]{R_tlbr}  & \includegraphics[height=2em]{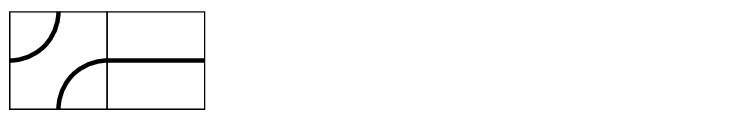} & \includegraphics[height=2em]{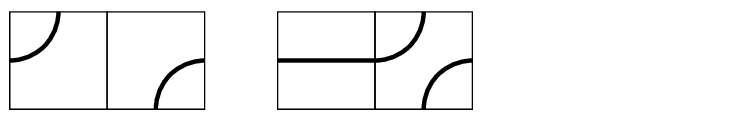} \\
	\includegraphics[height=2em]{R_trbl}  & \includegraphics[height=2em]{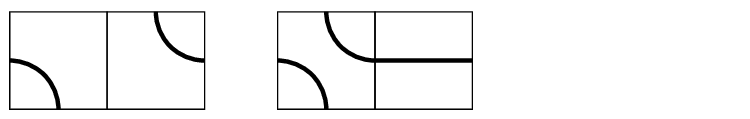} & \includegraphics[height=2em]{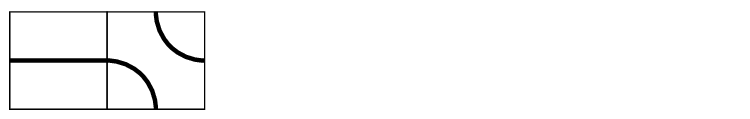} \\
	$0$  & \includegraphics[height=2em]{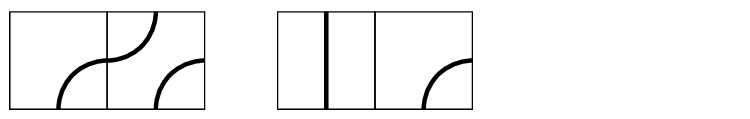} &  \\
	$0$  & \includegraphics[height=2em]{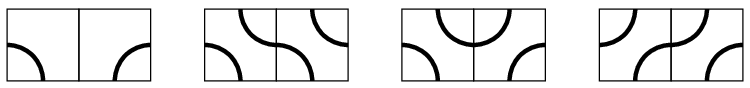} &  \\
	$0$  & \includegraphics[height=2em]{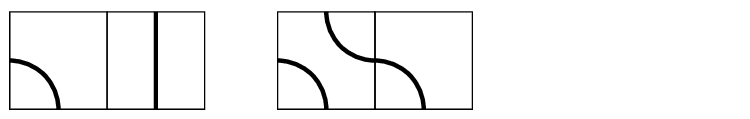} &  \\
\end{tabular}
   \caption{The classification of $RR$ configurations in the fusion equation. The first column contains the resulting $R$-matrix element, the second and third column the corresponding $RR$-configurations, grouped according to their external connectivity. The connectivity is taken on the five external sides of the l.h.s. of the fusion equation. The top triangle is not drawn, as it is determined by the top two sites of the $RR$ configuration.} 
   \label{tab:ClassificationRR}
\end{table}

\section{Construction of the open boundary $K$-matrix from the closed boundary $K$-matrix via insertion of a rapidity line}
\label{app:KmxConstr}

In this section, we show the construction of the open boundary $K$-matrix weights from the closed boundary case, by the well
known method of insertion of a line. The main advantage of this description of the open boundary $K$-matrix, 
that for certain quantities we can extend the symmetry arguments to the boundary rapidities, and also using the fusion equation, 
we get the boundary fusion equation as a corollary. \\
The closed left boundary $K$-matrix consist two elements, with identical weights, and satisfies the reflection equation:
\begin{equation}
 K_{\text{cbc}}=\Klmxe + \Klmxl
\end{equation}
\begin{equation}
 K_{\text{cbc}} R\left(v^{-1},u \right) K_{\text{cbc}} R \left( u,v \right) = R \left( v^{-1}, u^{-1} \right) K_{\text{cbc}} R \left( u^{-1}, v\right) K_{\text{cbc}} \; .
\end{equation}

  \begin{figure}
    \begin{center}
    \psfrag{a}{(a)}
	    \psfrag{b}{(b)}
	    \psfrag{A}{$=$}
   \includegraphics[scale=0.6]{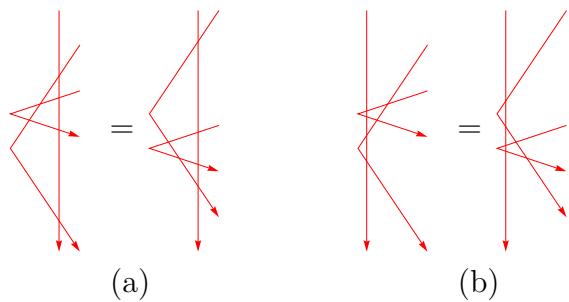}
   \caption{The construction of a new $K$-matrix via insertion of a line. In this figure --for convenience-- we use different style, then in the previous equation. Straight lines represent the rapidity-lines, 
a crossing of two rapidity lines is an $R$-matrix, a cusp in a line is a $K$-matrix.}
   \label{fig:InsertionOfLine}
   \end{center}
  \end{figure}

\begin{figure}
    \begin{center}
   \includegraphics[scale=0.2]{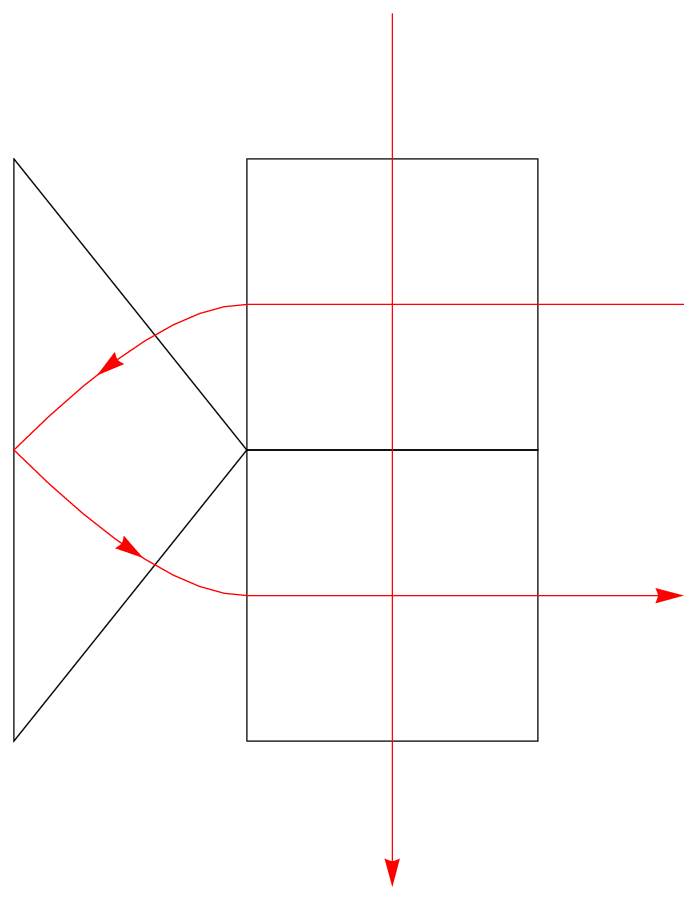}
      \caption{A $KRR$ block. The vertical rapidity is the boundary rapidity of the open boundary condition $K$-matrix, the reflecting rapidity is the normal one.}
   \label{fig:KRRblock}
   \end{center}
  \end{figure}

The idea of the insertion of a line is as in Fig.~\ref{fig:InsertionOfLine}. 
Multiply the closed boundary reflection equation from the right with a column of four $R$-matrices (Fig.~\ref{fig:InsertionOfLine}, (a)), 
and by the means of the Yang-Baxter equation, move the $R$-matrices inside (Fig.~\ref{fig:InsertionOfLine}, (b)). 
In this configuration the $KRR$ blocks can be regarded as the elements of the new $K$-matrix, 
and the weight of the new $K$-matrix is equal to the sum of the weights of the corresponding $KRR$ blocks (Fig.~\ref{fig:KRRblock}).
Our aim is to follow this procedure to create the open boundary (left) $K$-matrix. (The procedure is the same for the right boundary.) \\
Since we want to create \emph{independent} weights, and the possible $KRR$ configurations depend on if a path or an empty site enters
on the top of the top $R$-matrix, we can elaborate our procedure. For every open boundary $K$-matrix element, we want to have two groups of $KRR$ 
configurations, one with an entering path on the top, one without. We expect the sum of these weights to be equal, in order to produce independent 
open $K$-matrix weights. \\
Since we expect the right sides of the two $R$-matrices to be the top and bottom half of the open boundary $K$-matrix, 
the occupancy on the left and on the top already defines the six groups associated with the empty, the top, and the bottom type $K$-matrix.
(First three row of Table~\ref{tab:KRRgrouping}.) \\
\begin{table}
\begin{tabular}{ l l l }
  Open BC $K$- & $KRR$ config. with & $KRR$ config. with\\
  matrix element & entering line & empty top\\
  \includegraphics[scale=0.1]{Kl_t}  & \includegraphics[scale=0.45]{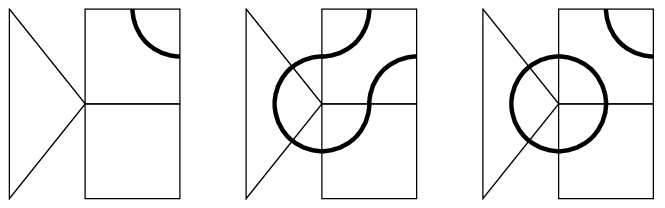} & \includegraphics[scale=0.45]{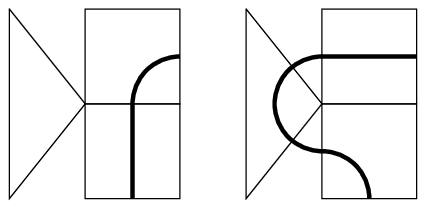} \\
  \includegraphics[scale=0.1]{Kl_b}  & \includegraphics[scale=0.45]{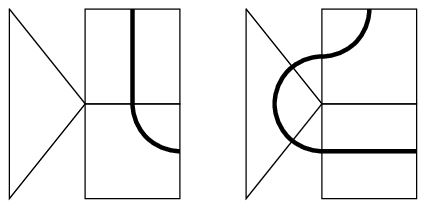} & \includegraphics[scale=0.45]{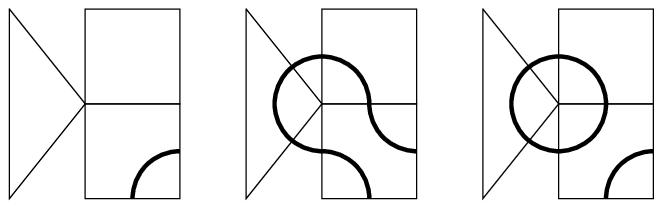} \\
  \includegraphics[scale=0.1]{Kl_e}  & \includegraphics[scale=0.45]{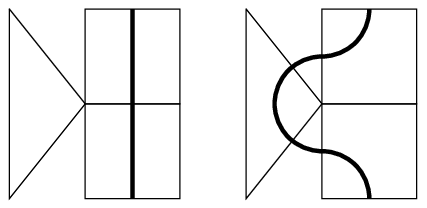} & \includegraphics[scale=0.45]{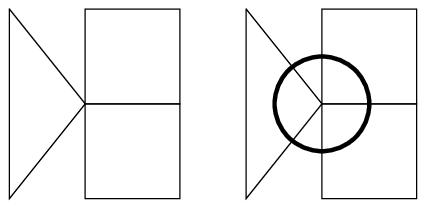} \\
  \includegraphics[scale=0.1]{Kl_m}  & \includegraphics[scale=0.45]{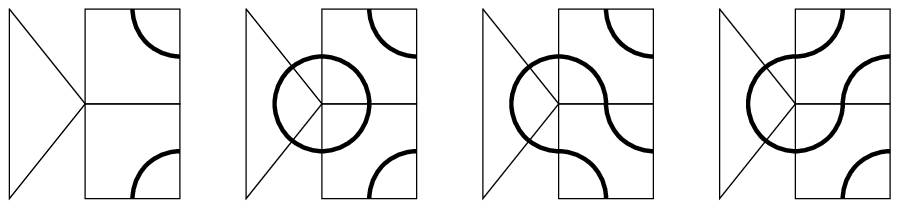} & \psfrag{t}{$\frac{3}{4}$} \includegraphics[scale=0.45]{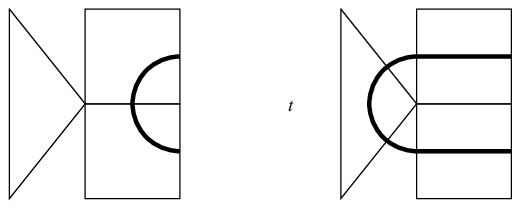} \\
  \includegraphics[scale=0.1]{Kl_l}  & \includegraphics[scale=0.45]{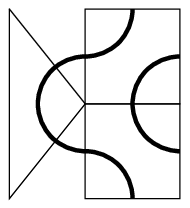} & \psfrag{o}{$\frac{1}{4}$} \includegraphics[scale=0.45]{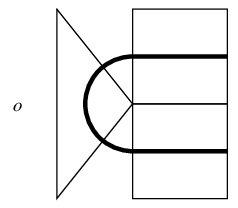} \\
\end{tabular}
   \caption{Open boundary condition $K$-matrix elements expressed by closed B.C. $KRR$ configurations. Note the $\frac{3}{4}$ and $\frac{1}{4}$ factors: In order to reproduce the
  open boundary condition weights, we have to split the configuration into two part, with certain probabilities. These probabilities goes into $\frac{3}{4}$ and $\frac{1}{4}$ in the 
  homogenous limit. For the general expression, consult the text.} 
   \label{tab:KRRgrouping}
\end{table}
Distinguishing the two remaining elements (the 'line': $\Klmxl$ and the 'monoid': $\Klmxm$) is a bit more tricky, and can be done in the following way: 
We look at configurations with a line entering, and we group them according to their connectivity on the left: If the two left side are connected, 
they belong to the 'line', if not, they belong to the 'monoid'. Now we have to choose the other two groups according to the criteria that with and without 
the entering path, the weights should be the same. Based on this criteria, we can uniquely make the choice, however, there is one $KRR$ configuration which has to be 'divided'
between the line and the monoid. Not considering these divided case, the following statement holds for the weight of the open boundary condition $K$-matrix and the weight 
of the $KRR$ configuration:

\begin{equation}
  \frac{\left( 1+z_B^2 \right)^2 z_1^2}{zB^2} W_{K_{obc}} (z_1, z_B) = \sum_{i \in G_{K_{obc}}} R_{i,\text{top}}(z_B,z_1) R_{i,\text{bottom}} (z_1^{-1}, z_B) \; .
\end{equation}

Here $W_{K_{obc}}$ denotes the weigh of a specific open boundary conditions $K$-matrix element, and the sum on the other side runs over the $KRR$ 
configurations which contribute to the given open boundary conditions. $K$-matrix element (As given in Table~\ref{tab:KRRgrouping}).
The divided cases have the prefactors $\frac{3}{4}$ and $\frac{1}{4}$ in the homogeneous case, in the inhomogeneous case, the 
following relations hold:

\begin{align}
 \frac{\left( 1+z_B^2 \right)^2 z_1^2}{z_B^2} W^l_m (z_1, z_B) &= W_t (z_B, z_1) W_t \left( z_1^{-1}, z_B \right) + \nonumber\\
     +  W_1 (z_B, z_1) W_1  \left( z_1^{-1}, z_B \right) & \left( \frac{1}{f\left( z_1^{-1}, z_B \right) + 1} + \frac{1}{\left( f\left(z_B, z_1 \right)+1 \right) \left( f\left( z_B, z_1^{-1} \right)+1 \right)} \right) \; , \\
     \nonumber \\
 \frac{\left( 1+z_B^2 \right)^2 z_1^2}{zB^2} W^l_id (z_1, z_B) &= \frac{f\left( z_B, z_1^{-1} \right)+1}{f\left( z_1, z_B \right)+1} 
     W_1 (z_B, z_1) W_1 \left( z_1^{-1}, z_B \right) \; .
\end{align}

Here $f$ is defined as:

\begin{equation}
 f(z_1,z_2) = \frac{W_2 (z_1,z_2)}{W_m (z_1,z_2)} \; ,
\end{equation}

with the property: $f^{-1} (z_1,z_2) = f(z_2,z_1)$. The prefactors, involving the $f$'s equal to $\frac{3}{4}$ and 
$\frac{1}{4}$, respectively, if the rapidities are equal to $1$.\\
An intuitive understanding of the divination of this $KRR$ configuration is missing, however, the aforementioned 
relations have been thoroughly checked analytically. \\
Since the vertical rapidity becomes the boundary rapidity, the previous argument about the symmetry in the rapidities for certain quantities 
extends to the boundary rapidities too. The extension depends on if the considered quantity commutes with the construction of the open boundary $K$-matrix. E.g. for the partition sum of the half and the full strip, we can use this construction to prove the symmetry in the boundary rapidity. However, for the considered spin-1 current, the construction does not commute with the operators $\hat{X}$ and $\hat{Y}$. \\
It is easy to prove the boundary fusion relation, based on this construction and the fusion relation. If we extend the $KRR$ 
configuration into a $KRRRR$ configuration, and we apply the recursion relation on the four $R$-matrix, we get the boundary fusion 
relation, as a corollary.

 

\bibliographystyle{hunsrt}
\bibliography{bibliography_bibtex}

\end{document}